\documentclass[11pt,a4paper,twoside,fleqn]{article}
\usepackage{color,graphicx,natbib,lscape,here,geometry,multirow,enumitem}
\usepackage[dvipsnames]{xcolor}
\usepackage{graphicx}
\usepackage{setspace}
\usepackage{authblk}
\usepackage{silence}
\usepackage{rotating}
\usepackage{multirow}

\usepackage{booktabs}
\usepackage{diagbox}
\usepackage{color}
\usepackage{algorithm2e}
\usepackage{enumitem}
\usepackage{tikz}

\usepackage{pdflscape}
\usepackage{afterpage}
\usepackage{comment}
\usepackage{mathtools}
\usepackage[hang,flushmargin]{footmisc} 
\usepackage[british]{babel}
\usepackage{bm}
\usepackage{caption}
\usepackage{subcaption}
\usepackage{colortbl}
\providecommand{\shadeRow}{\rowcolor[rgb]{0.9, 0.9, 0.9}}
\providecommand{\shadeBench}{\rowcolor[rgb]{0.95, 0.3, 0.3}}
\usepackage{longtable}
\usepackage{array}

\definecolor{nblue}{HTML}{000660}
\usepackage[colorlinks=true,urlcolor=nblue,linkcolor=nblue,citecolor=nblue]{hyperref}
\usepackage{bigstrut}

\usepackage{tabularx}
\usepackage{threeparttable}
\usepackage{dcolumn}
\newcolumntype{d}[1]{D{.}{.}{#1}}

\usepackage{etoolbox} 
\makeatletter
\patchcmd{\BR@backref}{\newblock}{\newblock[}{}{}
\patchcmd{\BR@backref}{\par}{]\par}{}{}
\makeatother

\newcolumntype{C}[1]{>{\centering\arraybackslash}p{#1}}

\usepackage[hang]{footmisc}
\pdfoutput=1
 \usepackage{float}
 \restylefloat{table}

\usepackage[title,titletoc]{appendix}
\makeatletter

\renewenvironment{appendices}{%
    \begin{oldappendices}%
    \renewcommand{\thefigure}{\ifnum \c@section>\z@ \thesection.\fi\@arabic\c@figure}%
    \@addtoreset{figure}{section}%
    \renewcommand{\thetable}{\ifnum \c@section>\z@ \thesection.\fi\@arabic\c@table}%
    \@addtoreset{table}{section}}{%
    \end{oldappendices}%
}\makeatother

\usepackage{titlesec} 
\titleformat{\section}[block]{\large}{\thesection. }{0em}{\MakeUppercase} 
\titleformat{\subsection}[block]{\large}{\thesubsection. }{0em}{\itshape} 
\titleformat{\subsubsection}[block]{\large}{}{0em}{\itshape} 

\let\natbibcitet\citet
\renewcommand\citet{\bibpunct{(}{)}{,}{a}{,}{,}\natbibcitet}
\let\natbibcitep\citep
\renewcommand\citep{\bibpunct{(}{)}{;}{a}{,}{;}\natbibcitep}
\newcommand{\bi}{\begin{itemize}}
\newcommand{\ei}{\end{itemize}}
\newcommand{\be}{\begin{equation}}
\newcommand{\ee}{\end{equation}}
\defcitealias{ieo14}{IEO, 2014}

\long\def\symbolfootnote[#1]#2{\begingroup%
\def\thefootnote{\fnsymbol{footnote}}\footnote[#1]{#2}\endgroup}

\makeatletter
\def\ubar#1{\underline{\sbox\tw@{$#1$}\dp\tw@\z@\box\tw@}}
\def\obar#1{\overline{\sbox\tw@{$#1$}\dp\tw@\z@\box\tw@}}
\makeatother

\makeatletter\let\p@subfigure\thefigure\makeatother

\floatstyle{plaintop}
\restylefloat{table}

\usepackage{fancyhdr} 
\usepackage{cleveref}
\crefname{chapter}{Chapter}{Chapters}
\crefname{section}{Section}{Sections}
\crefname{subsection}{Section}{Sections}
\crefname{subsubsection}{Section}{Sections}
\crefname{figure}{Figure}{Figures}
\crefname{table}{Table}{Tables}
\crefname{equation}{Equation}{Equations}
\crefname{appendix}{Appendix}{Appendices}
\crefname{appendices}{Appendix}{Appendices}
\crefname{appsec}{Appendix}{Appendices}

\usepackage{catoptions}
\makeatletter

\def\Autoref#1{%
  \begingroup
  \edef\reserved@a{\cpttrimspaces{#1}}%
  \ifcsndefTF{r@#1}{%
    \xaftercsname{\expandafter\testreftype\@fourthoffive}
      {r@\reserved@a}.\\{#1}%
  }{%
    \ref{#1}%
  }%
  \endgroup
}
\def\testreftype#1.#2\\#3{%
  \ifcsndefTF{#1autorefname}{%
    \def\reserved@a##1##2\@nil{%
      \uppercase{\def\ref@name{##1}}%
      \csn@edef{#1autorefname}{\ref@name##2}%
      \autoref{#3}%
    }%
    \reserved@a#1\@nil
  }{%
    \autoref{#3}%
  }%
}
\makeatother

\title{\LARGE{Flexible Mixture Priors for Large Time-varying Parameter Models}}
\author{\large{
\uppercase{Niko Hauzenberger}}\thanks{
\noindent Salzburg Centre of European Union Studies, University of Salzburg. Address: M\"{o}nchsberg 2a, 5020 Salzburg, Austria. Email: \href{mailto:niko.hauzenberger@sbg.ac.at}{niko.hauzenberger@sbg.ac.at}. The author gratefully acknowledges financial support from the Austrian Science Fund (FWF, grant no. ZK 35) and the Oesterreichische Nationalbank (OeNB, Anniversary Fund, project no. 18127). I thank Sylvia Fr{\"u}hwirth-Schnatter, Paul Hofmarcher, Florian Huber, Karin Klieber, Gary Koop, Luca Onorante, Michael Pfarrhofer and Anna Stelzer for valuable comments and suggestions. 
}
\\\vspace*{-0.5em}
\textit{University of Salzburg}}
\date{}


\def\equationautorefname~#1\null{%
  Eq.~(#1)\null
}
\def\equationautorefname~#1\null{
Eq.~(#1)\null
}
\usepackage[utf8, latin1]{inputenc}                 

\begin{document}
\maketitle\thispagestyle{empty}\normalsize\vspace*{-2em}\small

\begin{center}
\begin{minipage}{0.8\textwidth}
\noindent\small Time-varying parameter (TVP) models often assume that the TVPs evolve according to a random walk. This assumption, however, might be questionable since it implies that coefficients change smoothly and in an unbounded manner. In this paper, we relax this assumption by proposing a flexible law of motion for the TVPs in large-scale vector autoregressions (VARs). Instead of imposing a restrictive random walk evolution of the latent states, we carefully design hierarchical mixture priors on the coefficients in the state equation. These priors effectively allow for discriminating between periods where coefficients evolve according to a random walk and times where the TVPs are better characterized by a stationary stochastic process. Moreover, this approach is capable of introducing dynamic sparsity by pushing small parameter changes towards zero if necessary. The merits of the model are illustrated by means of two applications. Using synthetic data we show that our approach yields precise parameter estimates. When applied to US data, the model reveals interesting patterns of low-frequency dynamics in coefficients and forecasts well relative to a wide range of competing models.
\end{minipage}
\end{center}

\onehalfspacing\normalsize\renewcommand{\thepage}{\arabic{page}}
\newpage\setcounter{page}{1}
\begin{spacing}{1.5}
\section{Introduction}\label{sec:intro}
A growing number of papers introduces time-varying parameters (TVP) in econometric models for capturing structural breaks in relations across macroeconomic fundamentals \citep[see, for example,][]{cogley2005drifts, primiceri2005, sims2006were, korobilis2013assessing, eickmeier2015classical, mumtaz2018changing, paul2019time} and to achieve more accurate macroeconomic forecasts \citep[see, for instance,][]{koop2012forecasting, koop2013large, giannone2013,groen2013real, bauwens2015contribution, hhko2019,hko2020, hkp2020}. 

In this paper, we focus on estimating TVP vector autoregressive (VAR) models with a large number of endogenous variables. Due to severe overfitting issues in large TVP-VARs, special emphasis is paid to important modeling decisions, such as whether coefficients evolve gradually, change abruptly or remain constant for subset of periods. In macroeconomic applications, it is common to assume that coefficients evolve according to a random walk, implying that parameters change smoothly over time. As noted by the recent literature \citep[see, for example,][]{lopes2016parsimony, hhko2019}, however, this assumption might be overly simplistic and lead to model misspecification.  

In large TVP-VARs it is often reasonable to assume that most parameters remain constant over time, while only few vary. To capture this behaviour, the Bayesian literature frequently uses shrinkage priors on the state innovation variances to sufficiently push them towards zero \citep{sfs_wagner2010, belmonte2014hierarchical}. A severe drawback of this strategy is that it only accounts for the case that a given coefficient is constant for all points in time (labeled static sparsity). 

Another common situation faced by researchers is that coefficients change only at certain points in time (this is referred to as dynamic sparsity). 
Using a mixture distribution on the innovation variances, for example, allows to push small parameter changes towards zero \citep[see, inter alia,][]{mcculloch1993bayesian, gerlach2000efficient, giordani2008efficient, koop2009evolution, huber2019should}.\footnote{Other dynamic sparsification techniques include different forms of dynamic shrinkage processes \citep[see, inter alia,][]{kg2014, uribe2017dynamic, rockova2018, kowal2019dynamic, hhk2020}, latent threshold models \citep{nakajima2013bayesian} or dynamic model selection techniques \citep{chan2012time, koop2013large}.} Alternatively, \cite{hhko2019} introduce a more flexible law of motion by assuming a conjugate hierarchical location mixture prior directly on the time-varying part of the coefficients. This location mixture allows for a dynamically adjusting the prior mean on the TVPs to capture situations with a low, moderate or even large number of structural breaks in the coefficients. However, both techniques come with drawbacks. For instance, the mixture innovation model of \cite{huber2019should}, equipped with a latent threshold mechanism, discriminates between a high and a low innovation variance state. However, the authors do not discard the random walk law of motion, which might be too restrictive. \cite{hhko2019} use either a conjugate g-prior \citep{Zellnergprior} or a conjugate Minnesota prior \citep{doan1984forecasting, litterman1986forecasting}, potentially lacking flexibility to disentangle abrupt from gradual changes.

In this paper, we carefully design suitable mixture priors for the state equation. In a first variant, a mixture prior is not only introduced on the state innovations, but also on the autoregressive coefficients in the state equation to obtain sufficient flexibility. To achieve parsimony in large models, a latent binary indicator determines the law of motion for the TVPs and detect periods where coefficients evolve according to a random walk and times where the TVPs are better characterized by a stationary stochastic process. Combined with a mixture on the innovation volatilites and suitable shrinkage priors, this approach is capable of automatically capturing a wide range of typical parameter changes. In a second variant, the sparse finite location mixture model of \cite{hhko2019} is extended by considering non-conjugate shrinkage priors and by replacing the location mixture with a location-scale mixture. Here, an additional mixture on the state variances captures the notion that structural breaks in coefficients happen infrequently (with potentially large TVP innovations), while most of the time coefficients are constant (with TVP innovations pushed towards zero), similar to mixture innovation models. 
 
In the previous paragraphs we repeatedly stated that our techniques are well suited to handle overfitting issues in large TVP-VARs. But large TVP models also raise the question of computational feasibility. In this contribution, computational complexity is reduced by using recent advances in estimating large-scale TVP regression \citep[see][]{chan2009efficient, mccausland2011simulation, hhk2020}. These are based on rewriting the TVP model in its static regression form. In this representation, the TVP model is treated as a very big regression model and the techniques proposed in \cite{bhattacharya2016fast} can be used. Since these algorithms are designed for single equation models, we estimate the VAR model using its structural representation and thus estimate a set of unrelated TVP regressions \citep[see][]{carriero2019large}.

Based on two applications we investigate the merits of the techniques developed in the paper. 
First, in an application using synthetic data we illustrate that the proposed methods work well in detecting small and large structural breaks in coefficients. Second, we employ a large US macroeconomic dataset for an empirical application. Our proposed methods reveal interesting patterns in the low-frequency relationship between unemployment and inflation. Moreover, to evaluate predictive performance of our approach, we perform a comprehensive forecasting exercise. This forecasting horse race shows that the proposed framework works well relative to a wide range of competing models. Even for large TVP-VARs introducing flexible mixture priors in the state equation tends to improve forecast accuracy. 

The remainder of the paper is structured as follows. Section \ref{sec:econ} introduces a TVP regression model with flexible mixture priors and sketches the main contributions of the paper. 
Section \ref{sec:infer} generally outlines inference in these models, while Section \ref{sec:postTVP} discusses the posterior sampling algorithm of \cite{bhattacharya2016fast}, when applied to non-centerred TVP regressions. Section \ref{sec::sim} and Section \ref{sec:real} show the results for artificial data and US data, respectively. 
Finally, Section \ref{sec:conclusions} summarizes and concludes.

\section{Econometric Framework}\label{sec:econ}
\subsection{A TVP Regression}\label{ssec:tvp}
Let $y_t$ denote a scalar time series and $\bm x_t$ refer to a $K$-dimensional vector of predictors, then the observation equation for a TVP regression can be written as:
\begin{equation}
\begin{aligned}
		y_t = \bm x_t'\bm \alpha_t + \varepsilon_t, \quad \varepsilon_t \sim \mathcal{N}(0, \sigma^2_t).
\end{aligned}\label{eq:obs}
\end{equation}
Here, $\bm \alpha_t$ is a $K$-dimensional vector of TVPs that relates $\bm x_t$ to the quantity of interest and $\varepsilon_t$ denotes the measurement error with mean zero and time-varying variance $\sigma^2_t$. For the state equation of $\sigma^2_t$, we assume a stochastic volatility (SV) specification and refer to Appendix \ref{app:sv} and \cite{kastner2014ancillarity} for details. 

Typically, $\bm \alpha_t$ is assumed to evolve according to a random walk (RW). In this paper, interest centers on relaxing this assumption. In the following, to achieve both sufficient flexibility and model parsimony, we use two different hierarchical mixture specifications for $\bm \alpha_t$.  
In the first variant, we assume that coefficients evolve according to a mixture of a random walk and a white noise process. In the second variant, interest centers on extending the methods proposed in \cite{hhko2019}. 

\subsection{A Hierarchical Mixture between a Random Walk and a White Noise Process} 
For a mixture between a random walk and white noise process we assume that the evolution of $\bm \alpha_t$ is given by: 
\begin{equation}
\bm \alpha_t = \bm \alpha_0 + \bm \phi_t(\bm \alpha_{t-1} - \bm \alpha_0) + \bm \varsigma_t, \quad  \bm \varsigma_t \sim \mathcal{N}(\bm 0, \bm \Psi_t) \label{eq:v1}
\end{equation}
with $\bm \alpha_0$ denoting a $K$-dimensional intercept vector, $\bm \phi_t$ being a $K$-dimensional diagonal autoregressive coefficient matrix and $\bm \varsigma_t$ denoting a $K$-dimensional vector of state innovations, which are centered on zero and feature a $K \times K$-dimensional variance-covariance matrix $\bm \Psi_t$.  
Moreover, we assume $\bm \phi_t$ and $\bm \Psi_t$ to evolve according to a regime-switching process:
\begin{equation}\label{eq:phi}
\bm \phi_t = \bm S_t 
\end{equation}
and 
\begin{equation}\label{eq:psi}
\bm \Psi_t = \bm S_t \bar{\bm \Psi}_1 + (\bm I_K - \bm S_t) \bar{\bm \Psi}_0,
\end{equation}
with $\bm S_t= \text{diag}~(s_{1t}, \dots, s_{Kt})$ denoting a binary indicator matrix with $\{s_{it}\}_{i = 1}^{K}$ being either zero or one, $\bm I_K$ being a $K$-dimensional identity matrix while $\bar{\bm \Psi}_1 = \text{diag}~(\bar{\psi}_{11}, \dots, \bar{\psi}_{K1})$ and $\bar{\bm \Psi}_0 = \text{diag}~(\bar{\psi}_{10}, \dots, \bar{\psi}_{K0})$ denote $K$-dimensional diagonal matrices. 
\autoref{eq:phi} assumes that coefficients evolve according to a mixture of a random walk and a white noise process, while \autoref{eq:psi} ensures sufficient flexibility of the state innovations, respectively. For example, if the covariate-specific indicator $s_{it} = 1$ in the $t^{th}$ period, the $i^{th}$ covariate follows a random walk with state innovation variance $\psi_{it} = \bar{\psi}_{i1}$, while if $s_{it} = 0$ in the $t^{th}$ period, it follows a stochastic process with variance $\psi_{it} = \bar{\psi}_{i0}$. 

This specification (henceforth labeled as \texttt{TVP-MIX}) nests a wide variety of popular TVP models, such as standard RW state equations and mixture innovation models.\footnote{In the empirical application, these are considered as important benchmarks.} 
A standard random walk evolution is trivially obtained by setting $\bm S_t = \bm I_K$. A so-called mixture innovation model assumes $\bm \phi_t = \bm I_K$ and specifies $\bm \Psi_t$ similar to \autoref{eq:psi} \citep{gerlach2000efficient, giordani2008efficient,koop2009evolution, huber2019should}. Additionally, mixture innovation specifications restrict $\bar{\bm \Psi}_{0} = \kappa \hat{\bm \Psi}_{0}$ with $\kappa$ being a small value close to zero and $\hat{\bm \Psi}_{0}$ being a diagonal matrix collecting variable specific scaling parameters.\footnote{Related to literature on variable selection \citep{george1993variable, george1997approaches}, here $\bar{\bm \Psi}_{1}$ is commonly referred to as slab component and $\bar{\bm \Psi}_{0}$ as spike component \citep[see, for example,][]{huber2019should}.} 

Apart from discussing the relation to other popular TVP models, it is also worth highlighting additional features of the model proposed in (\ref{eq:v1}) to (\ref{eq:psi}). If a parameter is almost constant, but also features larger abrupt changes for some periods, we would expect that $\bar{\psi}_{i0} > \bar{\psi}_{i1}$. This case is of particular interest, when compared to a standard mixture innovation model with random walk state equation.
Conversely, if a coefficient features large, more persistent swings, but also some periods of parameter stability, we would expect $\bar{\psi}_{i0} < \bar{\psi}_{i1}$. Intuitively, the relative proportions of $\bar{\psi}_{i0}$ and $\bar{\psi}_{i1}$ depend mainly on the nature of coefficient changes. Alternatively, if the $i^{th}$ coefficient is constant or negligible (static sparsity), this can be achieved with $ \bar{\psi}_{i1}$ and/or $\bar{\psi}_{i0}$ close to zero \citep{lopes2016parsimony}. Note that in the special case of constant coefficients, the proposed specification is not identified. We address this issue in the context of interpreting the state indicators $\bm S_t$. 
 
\subsection{A Hierarchical Pooling Specification}
For a hierarchical pooling specification, we follow \cite{hhko2019} and assume that the time-varying part of $\bm \alpha_t$ follows a sparse finite mixture in the spirit of \cite{malsiner2016model}. 

The specification of the state equation $\bm \alpha_t$ (labeled as \texttt{TVP-POOL}) reads as:
\begin{equation}\label{eq:v2}
\bm \alpha_t = \bm \alpha_0 + \bm \gamma_t.
\end{equation}
Here, $\bm \alpha_0$ denotes a $K$-dimensional constant coefficient vector and $\bm \gamma_t$ is assumed to be a $K$-dimensional vector of random coefficients featuring a specific structure. That is, conditional on latent group indicators $\theta_t$ that takes a value $n \in \{1, \dots, N\}$, $\bm \gamma_t$ follows a multivariate Gaussian distribution:
\begin{equation}\label{eq:mixprior}
\bm \gamma_t = \bm \mu_n + \bm \varsigma_{t}, \quad  \bm \varsigma_{t} \sim \mathcal{N}(\bm 0, \bm \Psi_t), \quad \text{if} \quad \theta_t = n, 
\end{equation}
where $\bm \mu_n$ refers to the group-specific mean and $\bm \Psi_t$ denotes the variance-covariance matrix. 
It is also worth noting that $\theta_t$ serves as group indicator for $\bm \gamma_t$. The probability that $\bm \gamma_t$ is assigned to cluster $n$ is defined as $P(\theta_t = n) = \omega_n$. 

This structure is closely related to the setup of \cite{hhko2019}. In the following, we extend their location mixture prior to a location-scale mixture prior by introducing a regime-switching specification on $\bm \Psi_t$ similar to \autoref{eq:psi}.
That is:
\begin{equation}\label{eq:psi_v2}
\bm \Psi_t = \bm S_t \bar{\bm \Psi}_1 + (\bm I_K - \bm S_t) \bar{\bm \Psi}_0,
\end{equation}
with both $\bar{\bm \Psi}_0$ and $\bar{\bm \Psi}_1$ being diagonal matrices and $\bm S_t$ denoting a binary indicator matrix. Similar to standard mixture innovation models one component serves to detect larger breaks, while a second component handles dynamic sparsity. We therefore discard the conjugate prior assumption of \cite{hhko2019} and instead assume non-conjugate shrinkage priors on both state variances (described in more detail in Subsection \ref{ssec:nc}). 

Before proceeding, it is also worth sketching the general idea of this random coefficient specification. This model can be seen as a stochastic variant of multiple break point specifications \citep{koop2007estimation}, which is capable of capturing situations with a low, moderate or even large number of structural breaks. To estimate the number of regimes, we follow \cite{malsiner2016model} and \cite{hhko2019} and specify an ``overfitting'' model by setting $N$ to a large integer (i.e. consider many regimes a priori). To achieve parsimony, we come up with an estimate for the number of clusters $\hat{N}$ (usually $\hat{N} < N$) by specifying a shrinkage prior on both the mixture weights and the component means. Thus, overall shrinkage is determined between two interacting objectives: we aim at eliminating irrelevant clusters, while at the same time attempting to avoid highly overlapping component means. 

At this stage one might ask, why we do not assume $N$ different state innovation variances (i.e. using the group indicators $\theta_t$ for both $\gamma_t$ and $\bm \Psi_t$)? Here, it is worth discussing two important considerations. First, $N$ denotes a large integer and might lead to overfitting issues without assuming additional hierarchical shrinkage/pooling priors on the state innovation variances. Second, covariate-specific binary indicators ($\bm S_t$) for the scales already render the model highly flexible and it allows to introduce shrinkage on the state innovation variances in a simpler way. Moreover, the two-state mixture on the state variances (see \autoref{eq:psi_v2}) is designed to support inference about the locations $\gamma_n$, for $n = \{1, \dots, N\}$. We expect that many elements in $\{\bm \gamma_t\}_{t = 1}^{T}$ cluster around zero (i.e. coefficients are constant with $\bm \Psi_t$ close to zero), while occasionally there are structural breaks in some coefficients (requires relatively large values in $\bm \Psi_t$). Especially we aim to detect these two extremes (changes/no changes in $\bm \alpha_t$) with $\bm \gamma_t$. 

\subsection{The Latent State Indicator Matrix}
Sofar we remained silent on the evolution of $\bm S_t$. There are many different possibilities how the binary indicators $s_{it}$, for $i = \{1, \dots, K\}$ evolve over time. In the following, we assume two laws of motion: 
\begin{enumerate}
\item \textbf{Pooled Markov-switching process:}
When assuming a first-order Markov process for each $s_{it}$ independently, sampling the state indicators can be computationally cumbersome, especially if $K$ is large. Since one has to rely on forward filtering backward sampling algorithms, computation time quickly adds up. Therefore, we replace $\bm S_t$ with $s_t \bm I_K$. In the following, $s_t$ is assumed to be common to all $K$ covariates in period $t$ and governed by a joint Markov process.\footnote{Alternatively, \cite{koop2009evolution} group coefficients and assume class-specific indicators.} 
This process is driven by a transition probability matrix given by:
\begin{equation*}
P = \begin{pmatrix}
p_{00} & 1-p_{11} \\
1 - p_{00} & p_{11}
\end{pmatrix},
\end{equation*}
with transition probabilities from state $k$ to $l$ denoted by $p_{kl}$ and following a Beta distribution $p_{kk} \sim \mathcal{B}(c_{0k},c_{1k})$, for $k = \{0, 1\}$ \citep[see][]{uribe2017dynamic}. 

\item \textbf{Independent over time and covariate-specific indicators:} 
The assumption that a joint indicator governs the evolution of large number of coefficients might be too inflexible in certain cases. For this reason, we also specify covariate-specific indicators, coupled with independent mixture priors \citep[see][]{lopes2016parsimony}. In contrast to covariate-specific Markov processes, mixture priors are assumed to be independent over time and thus do not involve computationally demanding forward filtering backward sampling algorithms. In the following, $s_{it}$ is assumed to follow an independent Bernoulli distribution with $P(s_{it} = 1) = p_i$ and $p_i$ being Beta distributed, i.e. $p_i \sim  \mathcal{B}(c_{i,0},  c_{i,1})$.
\end{enumerate}

Moreover, it should be noted that the prior choice on the binary indicators is quite influential. For the random walk/white noise mixture (\texttt{TVP-MIX}), the hyperparameters are chosen in such a way that it is more likely that gradual changes have a higher (unconditional) expected duration (with $s_t = 1)$ than abrupt changes (with $s_t = 0$). In the empirical application we therefore set $c_{00} = 0.3 , c_{01} = 30, c_{10} = 30$, $c_{11} = 0.3$ for the Markov-switching process and $c_{i,0} = 0.3$, $c_{i,1} = 30$, $i = \{1, \dots, K\}$, for the independent mixture distribution. For the location-scale mixture (\texttt{TVP-POOL} with $\bm S_t$ solely governing the state innovation variances, we take a more agnostic approach by assuming $c_{00} = 0.3 = c_{01} =  c_{i,0} = 0.3$ and $c_{10} = c_{11} = c_{i,1} = 3$. 

\section{Bayesian Inference}\label{sec:infer}
To discuss inference for both variants outlined in Section \ref{sec:econ}, we introduce a very general state equation for $\bm \alpha_t$: 
\begin{equation}
\begin{aligned}
\bm \alpha_t = \bm \alpha_0 + \bm \gamma_t +  \bm \phi_t(\bm \alpha_{t-1} - \bm \alpha_0) + \bm \varsigma_t, \quad  \bm \varsigma_t \sim \mathcal{N}(\bm 0, \bm \Psi_t).
\end{aligned}\label{eq:state}
\end{equation}
\autoref{eq:state} nests both approaches with the first variant (\texttt{TVP-MIX}) being  obtained by setting $\bm \gamma_t = \bm 0_{K \times 1}$, while the second approach (\texttt{TVP-POOL}) is given by defining $\bm \phi_t = \bm 0_{K \times K}$ and $\bm \gamma_t = \bm \mu_n$, if $\theta_t = n$.

\subsection{The Non-Centered Parameterization}\label{ssec:nc}
In this subsection we exploit the non-centered parameterization to write $\bar{\bm \Psi}_0$ and $\bar{\bm \Psi}_1$ as part of the observation equation, enabling shrinkage on the regime-switching state innovation volatilities \citep{sfs_wagner2010}. 

We therefore recast the model as follows:
\begin{equation}
\begin{aligned}
		y_t =& \bm x_t'\underbrace{(\bm \alpha_0 + \sqrt{\bm \Psi}_t\tilde{\bm \alpha}_t)}_{\bm \alpha_t} + \sigma_t \epsilon_t, \quad \epsilon_t \sim \mathcal{N}(0,1), \\
		\tilde{\bm \alpha}_t =& \tilde{\bm \gamma}_t +  \bm \phi_t \tilde{\bm \alpha}_{t-1} + \bm \eta_t, \quad  \bm \eta_t \sim \mathcal{N}(\bm 0, \bm I_K), \quad \tilde{\bm \alpha}_0 = \bm 0, \bm \phi_1 = \bm I_K. 
\end{aligned}\label{eq:nc}
\end{equation}
Here, $\tilde{\bm \alpha}_t$ is a $K$-dimensional vector of normalized states, defined as $\tilde{\bm \alpha}_t = (\sqrt{\bm \Psi}_t)^{-1}(\bm \alpha_t - \bm \alpha_0)$ 
and $\tilde{\bm \gamma}_t = (\sqrt{\bm \Psi}_t)^{-1} \bm \gamma_t$ with $\sqrt{\bm \Psi}_t = \text{diag}~(\sqrt{\psi}_{1t}, \dots, \sqrt{\psi}_{Kt})$ denoting the (matrix) square-root of $\bm \Psi_t$. 
Using the definition of $\bm \Psi_t$ in \autoref{eq:psi} (or \autoref{eq:psi_v2}) the observation equation in \autoref{eq:nc} can be rewritten as: 
\begin{equation*}
y_t = \bm x_t'(\bm \alpha_0 + \bm S_t \sqrt{\bar{\bm \Psi}}_1 \tilde{\bm \alpha}_t + (\bm I_K -\bm S_t) \sqrt{\bar{\bm \Psi}}_0 \tilde{\bm \alpha}_t) + \sigma_t \epsilon_t,
\end{equation*}
and, more compactly, as a standard regression model: 
\begin{equation*}
y_t = \hat{\bm x}_t'\hat{\bm \alpha} + \sigma_t \epsilon_t.
\end{equation*}
Here, $\hat{\bm x}_t = (\bm x'_t, (\bm S_t \bm x_t \odot \tilde{\bm \alpha}_t)', ((\bm I_K - \bm S_t) \bm x_t \odot \tilde{\bm \alpha}_t)')'$ denotes a $3K$-dimensional covariate vector with $\odot$ referring to the dot product and $\hat{\bm \alpha} = (\bm \alpha'_0, \sqrt{\bar{\psi}}_{11}, \dots, \sqrt{\bar{\psi}}_{K1}, \sqrt{\bar{\psi}}_{10}, \dots, \sqrt{\bar{\psi}}_{K0})'$ being a $3K$-dimensional coefficient vector. 

On the time-invariant $\hat{\bm \alpha}$ we use a hierarchical global-local shrinkage prior \citep[see][]{polson2010shrink}:
\begin{equation*}
\hat{\alpha}_j \sim \mathcal{N}(0, \tau_j), \quad \tau_j|\lambda \sim f, \quad \lambda \sim g, \quad \text{for} \quad j = 1, \dots 3K,
\end{equation*}
where $\hat{\alpha}_j$ refers to the $j^{th}$ element in $\hat{\bm \alpha}$, $\tau_j$ induces local shrinkage with mixing density $f$ and $\lambda$ denotes a global shrinkage parameter with density $g$. In the empirical application, we focus on the Normal-Gamma \citep{griffin2010inference} shrinkage prior and choose $f$ and $g$ accordingly. 
This shrinkage prior has been proven to be successful in macroeconomic and financial application \citep[see, for example,][]{huber2017adaptive} and is quite common in the literature.\footnote{It is worth noting that any global-local shrinkage prior might be used. Other popular choices are the SSVS prior \citep{george1993variable, george1997approaches}, the Horseshoe prior \citep{carvalho2010horseshoe}, the Bayesian Lasso \citep{park2008bayesian} or the Triple-Gamma prior \citep{cadonna2019triple}. See also \cite{hko2020}, \cite{huberkastner2020} and \cite{cross2020macroeconomic} for thorough studies of global-local shrinkage priors in macroeconomic applications.} The exact prior specification is outlined in Appendix \ref{app:ng}.

\subsection{The Static Representation}\label{ssec:static}
If interest centers on estimating the latent states $\{\tilde{\bm \alpha}_t\}_{t=1}^{T}$, we can straightforwardly recast \autoref{eq:nc} in a static regression form by conditioning on $\bm \alpha_0$, the state innovation volatilities $\{\sqrt{\bm \Psi}_t\}_{t = 1}^{T}$ and the stochastic volatilities in $\bm \Sigma = \text{diag}~(\sigma_1, \dots, \sigma_T)$. We define $\bm y$ as a $T$-dimensional vector, $\bm X$ as a $T \times K$-dimensional matrix and $\bm \epsilon$ as a $T$-dimensional vector with  $y_t$, $\bm x_t'$ and $\epsilon_t$ on the $t^{th}$ position, respectively. Then, the static form of \autoref{eq:nc} is:  
\begin{equation*}
\begin{aligned}
		\bm y =& \bm X \bm \alpha_0 + \bm W \tilde{\bm \alpha}+ \bm \Sigma \bm \epsilon, \quad \bm \epsilon \sim \mathcal{N}(0,\bm I_T), \\
		\bm \Phi \tilde{\bm \alpha} =& \tilde{\bm \gamma} + \bm \eta, \quad  \bm \eta \sim \mathcal{N}(\bm 0, \bm I_{\nu}).
\end{aligned}
\end{equation*}
Here, $\tilde{\bm \alpha} = (\tilde{\bm \alpha}_1', \dots, \tilde{\bm \alpha}_T')'$ is a $\nu~(=TK)$-dimensional latent state vector, $\tilde{\bm \gamma} = (\tilde{\bm \gamma}'_1, \dots, \tilde{\bm \gamma}'_T)'$ is a $\nu$-dimensional intercept vector and $\bm \eta$ is a $\nu$-dimensional shock vector.  
After defining $\tilde{\bm x}_t' = \bm x_t' \sqrt{\bm \Psi}_t$, the precise structure of $\bm W$ and $\bm \Phi$ is given by:
\begin{equation*} 
\bm W = 
\begin{pmatrix} 
  \tilde{\bm x}_1'              & \bm 0_{K \times 1}' & \dots  & \bm 0_{K \times 1}' \\
  \bm 0_{K \times 1}'    & \tilde{\bm x}_{2}'           & \dots  & \bm 0_{K \times 1}' \\ 
  \vdots                   & \vdots                & \ddots & \vdots                \\
  \bm 0_{K \times 1}'    & \bm 0_{K \times 1}' & \dots  & \bm \tilde{\bm x}_{T}'
\end{pmatrix}, \quad \text{and }\quad
\bm \Phi = 
\begin{pmatrix} 
  \bm I_{K}  & \bm 0_{K \times K}   & \dots  & \bm 0_{K \times K} & \bm 0_{K \times K} \\
  -\bm \phi_2  & \bm I_{K}   & \dots  & \bm 0_{K \times K} & \bm 0_{K \times K} \\
  \vdots    & \vdots   & \ddots & \vdots& \vdots \\
  \bm 0_{K \times K} & \bm 0_{K \times K}   & \dots & - \bm \phi_T & \bm I_{K}
\end{pmatrix},
\end{equation*}
with $\bm 0_{K \times K}$ denoting $K \times K$-dimensional zero matrix. Solving for $\tilde{\bm \alpha}$ yields:  
\begin{equation*}
\tilde{\bm \alpha} = \bm \Phi^{-1}(\tilde{\bm \gamma} + \bm  \eta),
\end{equation*} 
implying that $\tilde{\bm \alpha} \sim \mathcal{N}\left(\bm a_0, \bm \Omega_0 \right)$ with  prior mean $\bm a_0 = \bm \Phi^{-1} \tilde{\bm \gamma}$ and prior variance-covariance matrix $\bm \Omega_0 = (\bm  \Phi'\bm \Phi)^{-1}$  of $\tilde{\bm \alpha}$ \citep[see, for instance,][]{chan2009efficient, chanstrachan2020statespace}. In the special case of $\bm \phi_t = \bm 0_{K \times K}$, for all $t$, $\bm \Phi$ (and thus $\bm \Omega_0$) reduces to an identity matrix, while $\bm \phi_t \neq \bm 0_{K \times K}$, for any $t$, induces a (specific) banded lower-triangular (block diagonal) structure of $\bm \Phi$ ($\bm \Omega_0$).\footnote{Note that $\bm \phi_t = \bm 0_{K \times K}$, for $t = \{1, \dots, T\}$, is always true for the \texttt{TVP-POOL} model, but not ruled out for the \texttt{TVP-MIX} specification. Moreover, if $\bm \Omega_0 = \bm I_{\nu}$, $\bm a_0 = \tilde{\bm \gamma}$.} Here it is worth emphasizing, that the prior variance-covariance matrix $\bm \Omega_0$ solely depends on state indicators $\bm S_t$. 

Moreover, the prior mean $\bm a_0$ also depends on the structure of $\bm \Phi$ and $\tilde{\bm \gamma}$. The simplest thing is to set $\bm a_0$ to a zero vector, which we implicitly assume for the \texttt{TVP-MIX} variants. For the \texttt{TVP-POOL} approach we use a hierarchical mixture prior on $\bm a_0$, described in detail next. 
 
\subsection{A Hierarchical Prior Mean}\label{ssec::mixprior}
The model outlined in (\ref{eq:v2}) to (\ref{eq:psi_v2}) denotes a sparse finite location-scale mixture. After recasting the model in the non-centered parameterization, we are able to replace the location-scale mixture prior on $\bm \alpha_t$  (outlined in \autoref{eq:mixprior}) with a location mixture prior on the normalized latent states $\tilde{\bm \alpha}_t$, since the scales of \autoref{eq:psi_v2} ($\bar{\bm \Psi}_0$ and $\bar{\bm \Psi}_1$) are now part of the observation equation.
That is:
\begin{equation}\label{eq:mixprior2}
\tilde{\bm \alpha}_t|\theta_t = n \sim \mathcal{N}(\tilde{\bm \mu}_n, \bm I_K).
\end{equation}
with group-specific mean $\tilde{\bm \mu}_n$, for $n = \{1, \dots, N\}$ and variance-covariance matrix $\bm I_K$.
In the following, the prior mean is defined as $\bm a_0 = (\bm a'_{01}, \dots, \bm a'_{0T})'$, with $\bm a_{0t} = \tilde{\bm \mu}_n$ if $\theta_t = n$. 

The sparse finite location mixture in \autoref{eq:mixprior2} allows us to use a similar prior setup as proposed in \cite{malsiner2016model} and \cite{hhko2019}. To ensure model parsimony we use a Dirichlet prior on $\bm \omega = (\omega_1, \dots, \omega_N)'$:
\begin{equation*}
\bm \omega|\xi \sim \text{Dir}(\xi, \dots, \xi),
\end{equation*}
with $\xi$ referring to an intensity parameter. 
The prior on the intensity parameter is specified as:  
\begin{equation*}
\xi \sim \mathcal{G}(d_0, d_0N),
\end{equation*}
with $d_0 = 10$ in the empirical application. 
Here, we closely follow \cite{malsiner2016model}, who show that this prior choice is successful in detecting superfluous components and obtaining a parsimonious mixture representation. 

Moreover, on the group means we specify the following shrinkage prior:   
\begin{equation*}
\tilde{\bm \mu}_n|\tilde{\bm \alpha} \sim \mathcal{N}(\bm 0_{K \times 1}, \bm \Lambda_0),
\end{equation*}
with $\tilde{\bm \mu}_n$ being centered on zero and prior variance-covariance matrix $\bm \Lambda_0 = \bm L \bm R \bm L$. Here, $\bm L = \text{diag}~(\sqrt{l_1}, \dots, \sqrt{l_K})$ and $\bm R = \text{diag}(r_1^2, \dots, r_K^2)$ with $r_i$ denoting the range of $\tilde{\bm \alpha}_{i}= (\tilde{\alpha}_{i1}, \dots,\tilde{\alpha}_{iT})'$. Moreover, we specify a Gamma prior on the elements in $\bm L$:  
\begin{equation*}
l_i \sim \mathcal{G}(e_0, e_1). 
\end{equation*}
Following \cite{yau2011hierarchical} and \cite{malsiner2016model}, we define $e_0 = e_1 = 0.6$ to push the group-specific prior means towards zero.

\section{Posterior computation}\label{sec:postTVP}
In this subsection, we outline the MCMC sampling step for $\tilde{\bm\alpha}$. We stress that drawing $\hat{\bm\alpha}$ is computationally fast, also for relatively large $K$, but sampling the $\nu$-dimensional vector $\tilde{\bm \alpha}$ is computationally demanding \citep{hhk2020}. Thus for $\hat{\bm\alpha}$ (and the remaining parameters) we use standard MCMC techniques with sampling steps and conditional posteriors outlined in Appendix \ref{app:post}. 

For $\tilde{\bm\alpha}$, irrespectively of the structure of $\bm a_0$ and $\bm \Omega_0$, we obtain standard conditional Gaussian posterior quantities with $\tilde{\bm y} = \bm \Sigma^{-1}(\bm y- \bm X\bm \alpha_0)$ and $\tilde{\bm W} = \bm \Sigma^{-1}\bm W$:
\begin{equation*}
\begin{aligned}
\tilde{\bm \alpha}| \tilde{\bm y}, \tilde{\bm W}, \bm a_0, \bm \Omega_0 \sim& \mathcal{N}(\bm a_1, \bm \Omega_1) \quad \text{with} \\
\bm \Omega_1^{-1} = \underbrace{(\tilde{\bm W}' \tilde{\bm W} + \bm \Omega_0^{-1})}_{\nu \times \nu}  \quad \text{and}& \quad \bm a_1 = \bm \Omega_1 (\tilde{\bm W}' \tilde{\bm y}+\bm \Omega_0^{-1} \bm a_0).
\end{aligned}
\end{equation*}
The main issue, however, is that the inversion of $\bm \Omega_1^{-1}$ is computationally costly, since it is $\nu \times \nu$-dimensional matrix with $\nu = TK$ and $\{T, K\}$ being potentially large integers. 

Thus, to avoid high-dimensional full matrix inversions and Cholesky decompositions for drawing the normalized latent states $\tilde{\bm \alpha}$, we rely on the algorithms proposed in \cite{bhattacharya2016fast}, applied to TVP models in \cite{hhk2020}. This method involves the following steps: 
\begin{enumerate}
\item Draw a $\nu$-dimensional vector $\bm u \sim \mathcal{N}(\bm 0, \bm I_{\nu}).$  
\item Sample a $T$-dimensional vector $\bm v \sim \mathcal{N}(\bm 0, \bm I_T).$
\item Define $\bm q = \bm a_0 + \bm \Phi^{-1}\bm u$, with $\bm \Phi^{-1}$ denoting the lower Cholesky factor of $\bm \Omega_0$, and $\bm r = \tilde{\bm W} \bm q + \bm v.$
\item Compute $\tilde{\bm \Omega}_1 = (\bm I_T + \tilde{\bm W} \bm \Omega_0 \tilde{\bm W}')^{-1}.$ 
\item Set $\bm f = \tilde{\bm \Omega}_1(\tilde{\bm y} - \bm r).$ 
\item  Obtain a draw for $\tilde{\bm \alpha} = (\bm \Omega_0 \tilde{\bm W}' \bm f) + \bm q.$ 
\end{enumerate}

Moreover, using the static representation for a TVP regression the involved matrices are sparse, which can be exploited to achieve additional computational gains \citep[see][]{chan2009efficient, hhko2019, hhk2020}. 
Depending on the structure of $\bm \Phi$ there a two extreme cases as briefly discussed in Subsection \ref{ssec:static}. Computationally the most expensive case is a random walk state equation ($\bm \phi_t = \bm I_K$,  $\forall$ $t$), while having no autoregressive structure in the state equation ($\bm \phi_t = \bm 0_{K \times K}$, $\forall$ $t$) it is computationally less demanding.\footnote{See \cite{hhk2020} for a comparison between the two extremes.} Recall, the former $\bm \Phi$ has a specific lower triangular structure (rendering $\bm \Omega_0$ block diagonal) and in the latter both $\bm \Phi$ and $\bm \Omega_0$ are diagonal. Thus, even for a random walk state equation (the most dense case), using sparse algorithms pays off in terms of computation.
Moreover, if $\bm \phi_t = \bm S_t$, for some $t$, we have to account for an intermediate computational burden lying between the two extremes that eventually depend on the exact structure of $\bm \Phi$ ($\bm \Omega_0$). 

\subsection{Equation-wise estimation for a TVP-VAR}\label{ssec:var}
The methods outlined in the previous subsection are designed for single equation models.
To use these algorithms also for posterior inference in TVP-VARs, we rewrite the multivariate model as a set of unrelated TVP regressions \citep[see][]{carriero2019large}. 

This can be done by using the structural form of the TVP-VAR: 
\begin{equation}
	\bm Y_t = \bm B_{0t} \bm Y_t + \sum_{i = 1}^{p} \bm B_{it} \bm Y_{t-i} +  \bm C_t +  \bm \epsilon_t, \quad \bm \epsilon_t \sim \mathcal{N}(\bm 0,\bm \Sigma_t).\label{eq:strVAR}
\end{equation}	
Here, $\bm Y_{t} = (Y_{1t}, \dots, Y_{mt})'$ denotes an $m$-dimensional vector of endogenous variables with $\bm B_{0t}$ being an $m \times m$-dimensional strictly lower-triangular matrix (with zero main diagonal) defining contemporaneous relationships between the elements of $\bm Y_t$. Moreover, $\bm B_{it}$, for $i = 1, \dots, p$, denotes an $m \times m$-dimensional time-varying coefficient matrix, $\bm C_t$ is an $m$-dimensional intercept vector and $\bm \epsilon_t$ refers to an $m$-dimensional Gaussian distributed error vector, centered on zero and with time-varying $m$-dimensional diagonal variance-covariance matrix $\bm \Sigma_t = \text{diag}~(\sigma^2_{1t}, \dots, \sigma^2_{mt})$. 
Before proceeding, it is convenient to define $\bm B_{t} = (\bm B_{t}, \dots, \bm B_{pt})$.

In the following, for $i= 2, \dots, m$, the $i^{th}$ equation of $\bm Y_t$ is given by: 
\begin{equation*}
y_{it} =  \bm x_{it}'\underbrace{(\bm \alpha_{i0} + \tilde{\bm \alpha}_{it})}_{\bm \alpha_{it}} + \epsilon_{it}, \quad \epsilon_{it} \sim \mathcal{N}(0, \sigma^2_{it}),
\end{equation*}
with $\bm x_{it}$ denoting a $K_i (= mp+i)$-dimensional covariate vector with $\bm x_{it} = (\{y_{jt}\}_{j=1}^{i-1}, \bm Y_{t-1}', \dots, \bm Y_{t-p}', 1)'$ and $\bm \alpha_{it} = (\{b_{ij,0t}\}_{j =1}^{i-1}, \bm B_{i \bullet, t}, c_{it})'$ a $K_i$-dimensional vector of time-varying coefficients. Here $b_{ij,0t}$ refers to the $(i,j)^{th}$ element of $\bm B_{0t}$, $\bm B_{i\bullet, t}$ denotes the $i^{th}$ row of $\bm B_t$ and $c_{it}$ the $i^{th}$ element of $\bm C_t$.  Moreover, for the first equation ($i = 1$) we have $\bm x_{1t} = (\bm Y'_{t-1}, \dots, \bm Y'_{t-p}, 1)'$ and $\bm \alpha_{1t} = (\bm B_{1\bullet,t}, c_{1t})'$.

\section{Simulation study}
\label{sec::sim}
In this section we use synthetic data to illustrate the features of the proposed mixture variants. For the data generating process (DGP) we assume that the number of observations is $T = 100$ and the number of covariates is given by $K = 5$. The covariates are  simulated with $\bm X_j \sim \mathcal{N}(\bm 0, \bm I_{T})$ for $j= 1,\dots, (K-1)$ and $\bm X_K = \bm \iota_T$ with $\bm \iota_T$ being a $T$-dimensional vector of ones. 
For the error variance $\sigma^2_t$, we assume an SV specification with $\log(\sigma^2_t) = h_t$ following a random walk process. That is, $h_t = h_{t-1} + \vartheta_t$ with $\vartheta_t \sim \mathcal{N}(0, 0.1)$ and $h_0 = \log(0.1)$. For the time-varying parameters $\bm \alpha_t$ we assume quite specific laws of motion. We define $\bm \alpha_0 =  (-4,3, -2, 2, 0)'$ as initial level and assume that both regime-switching autoregressive parameters and regime-switching variances in the state equation are governed by a joint Markov process $s_t$. Here, we let $\bm S_t = s_t \bm I_K$, $\bm \phi_t = \bm S_t$, $\bm \Psi_t = \bm S_t \bar{\bm \Psi}_1 + (\bm I_K - \bm S_t) \bar{\bm \Psi}_0$ with $\bar{\bm \Psi}_0 = \text{diag}~(10^{-10}, 0.5, 0.1, 10^{-10}, 1)$ and $\bar{\bm \Psi}_1 = \text{diag}~(1, 0.1, 0.5, 10^{-10}, 10^{-10})$. The joint indicator $s_t$ is simulated with transition probabilities $p_{00} = 0.6$ and $p_{11} = 0.95$, effectively leading to a higher unconditional probability that $\bm \alpha_t$ follows a random walk evolution. 

In particular, the first coefficient features larger, more persistent parameter changes (with $\psi_{11} = 1$), while for a small number of periods $\bm \alpha_{1t}$ is basically constant (achieved through the white noise state equation with $\psi_{10}$ close to zero). The second parameter features small gradual changes over time ($\psi_{21} = 0.1$), but also larger abrupt breaks ($\psi_{20} = 0.5$). The third parameter is similar to the first coefficient, but assumes $\psi_{30} > \psi_{10}$. The fourth coefficient is assumed to be constant over time ($\psi_{04}$ and $\psi_{14}$ are both close to zero) and, finally, the fifth parameter features some extremely large breaks (with $\psi_{05} = 1$), while it is otherwise assumed to be constant (here achieved through the random walk state equation with $\psi_{51}$ close to zero).   

To assess the flexibility of our approaches we compare them to models assuming a standard random walk evolution of coefficients and to those assuming constant coefficients. Moreover, we consider a typical mixture innovation model as an important benchmark. For each model we use a Normal-Gamma \citep{griffin2010inference} prior on the constant part and allow for SV in the measurement error variance. Furthermore, each TVP model features a Normal-Gamma prior on the square root of the state innovation variances, which are potentially regime-switching (see \autoref{eq:nc}). 

\begin{landscape}
\begin{figure}
\centering
\begin{minipage}[b]{1.6\textwidth}
\begin{minipage}[t]{0.19\textwidth}
\centering
\scriptsize
$1^{st}$ coefficient with \textit{true} parameters:\\
\hspace{30pt}
\tiny
$\alpha_{10} = -4, \bar{\psi}_{10} \approx 0, \bar{\psi}_{11} =1$ 
\end{minipage}
\begin{minipage}[t]{0.19\textwidth}
\centering
\scriptsize
$2^{nd}$ coefficient with \textit{true} parameters:\\
\hspace{30pt}
\tiny
$\alpha_{20} = 3, \bar{\psi}_{20} = 0.5, \bar{\psi}_{21} = 0.1$  
\end{minipage}
\begin{minipage}[t]{0.19\textwidth}
\centering
\scriptsize
$3^{rd}$ coefficient with \textit{true} parameters:\\
\hspace{30pt}
\tiny
$\alpha_{30} = -2, \bar{\psi}_{30} = 0.1, \bar{\psi}_{31} = 0.5$  
\end{minipage}
\begin{minipage}[t]{0.19\textwidth}
\centering
\scriptsize
$4^{th}$ coefficient with \textit{true} parameters:\\
\hspace{30pt}
\tiny
$\alpha_{40} = 2, \bar{\psi}_{40} \approx 0, \bar{\psi}_{41} \approx 0$   
\end{minipage}
\begin{minipage}[t]{0.19\textwidth}
\centering
\scriptsize
$5^{th}$ coefficient with \textit{true} parameters:\\
\hspace{30pt}
\tiny
$\alpha_{50} = 0, \bar{\psi}_{50} = 1, \bar{\psi}_{51} \approx 0$  
\hspace{500pt}
\end{minipage}
\end{minipage}\vfill

\begin{minipage}[b]{\textwidth}
\centering
(a) \small{\textit{\texttt{TVP-MIX} with flexible state variances (\texttt{FLEX}) and $\bm S_t = s_t \bm I_K$ (\texttt{MS})}:}
\hspace{5pt}
\end{minipage}\vfill

\begin{minipage}[b]{1.6\textwidth}
\begin{minipage}[t]{0.19\textwidth}
\centering
\includegraphics[scale=.17]{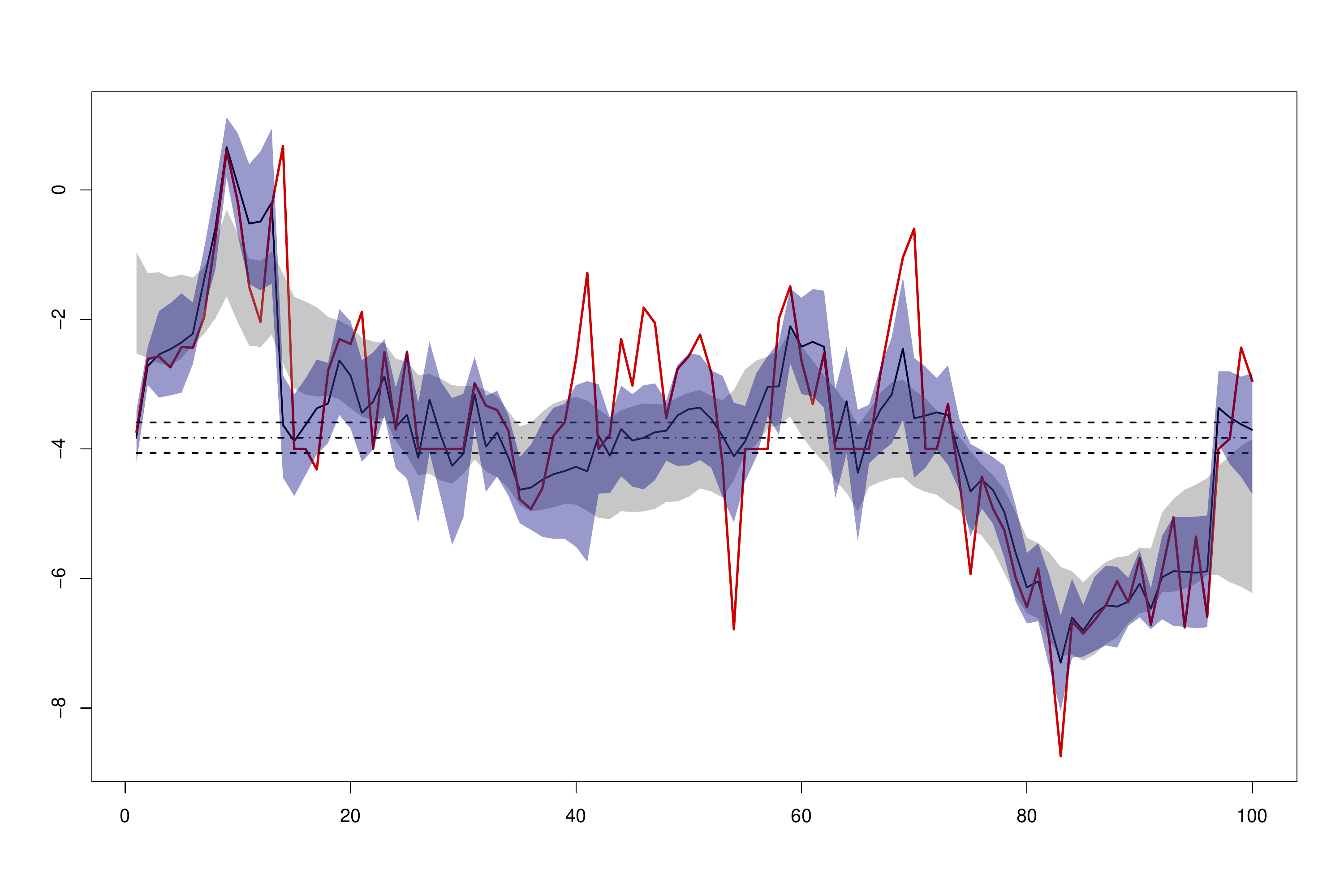}
\end{minipage}
\begin{minipage}[t]{0.19\textwidth}
\centering
\includegraphics[scale=.17]{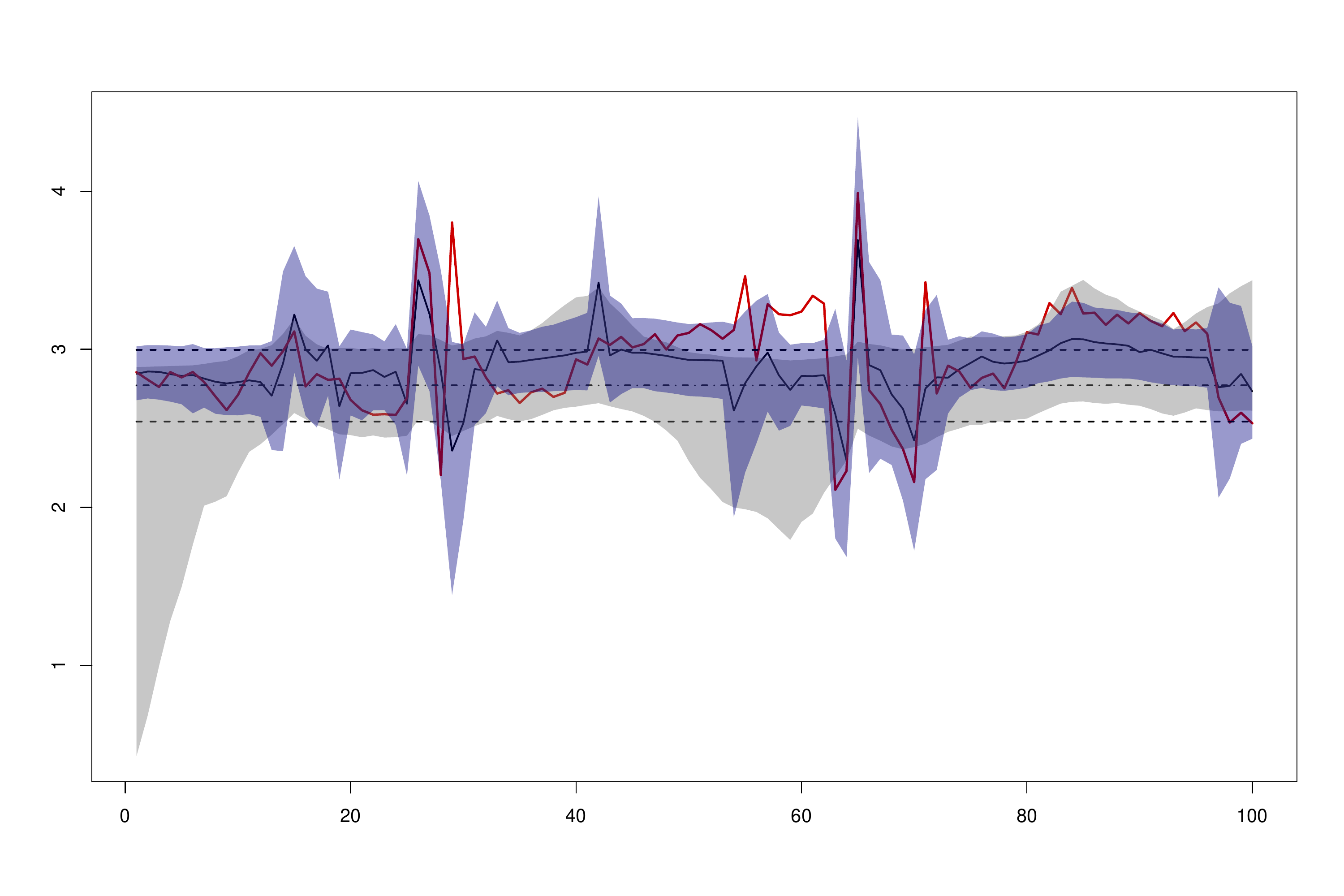}
\end{minipage}
\begin{minipage}[t]{0.19\textwidth}
\centering
\includegraphics[scale=.17]{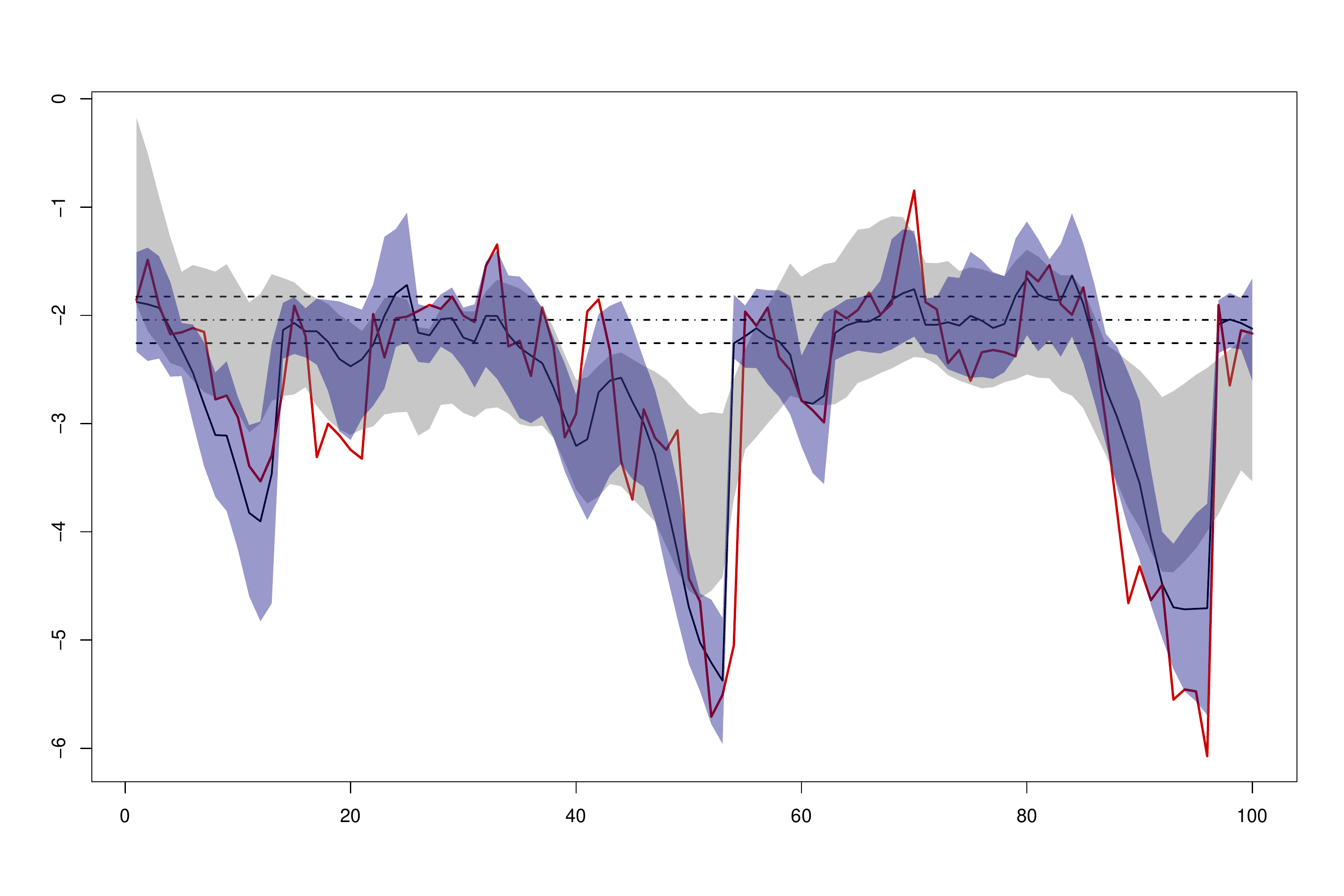}
\end{minipage}
\begin{minipage}[t]{0.19\textwidth}
\centering
\includegraphics[scale=.17]{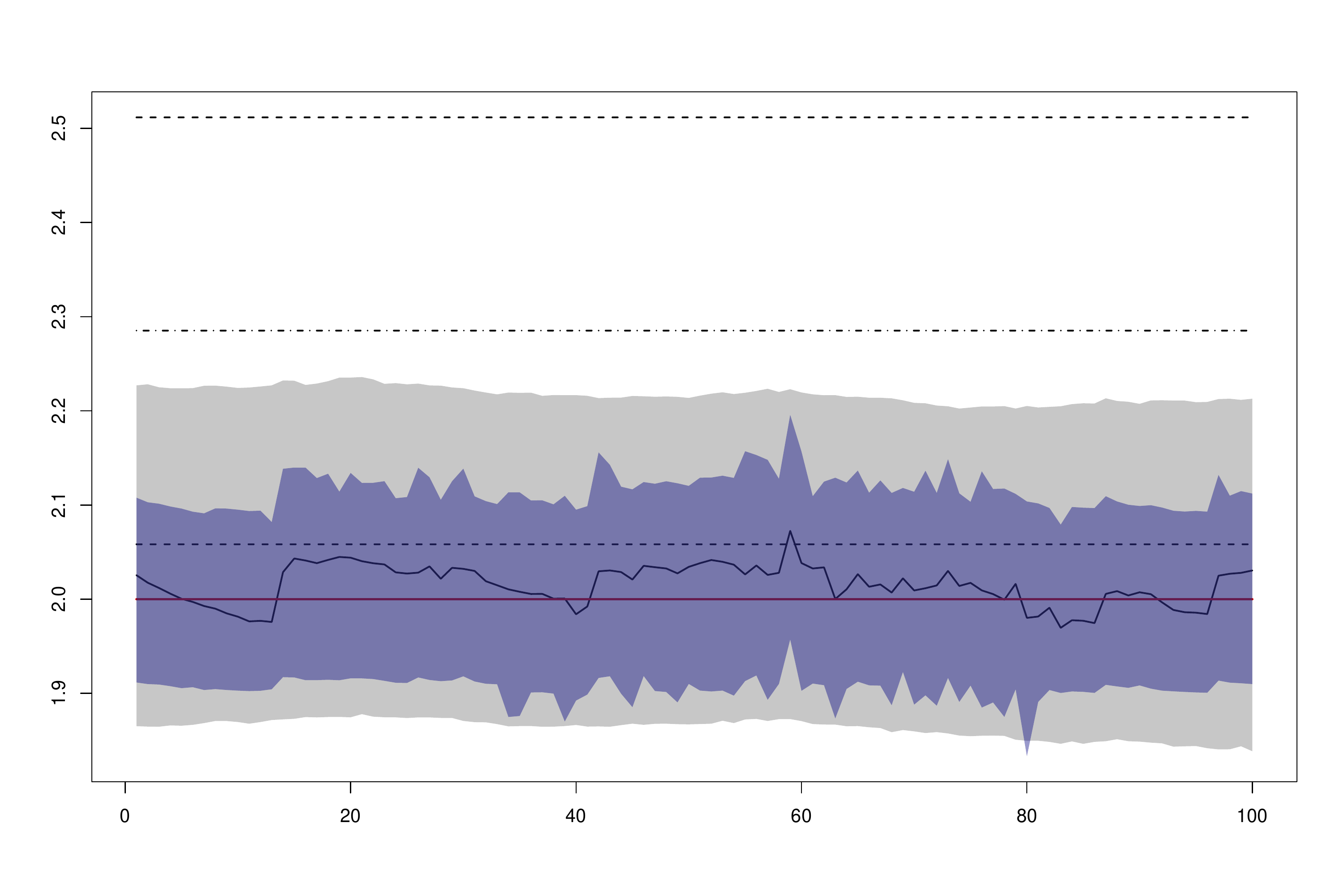}
\end{minipage}
\begin{minipage}[t]{0.19\textwidth}
\centering
\includegraphics[scale=.17]{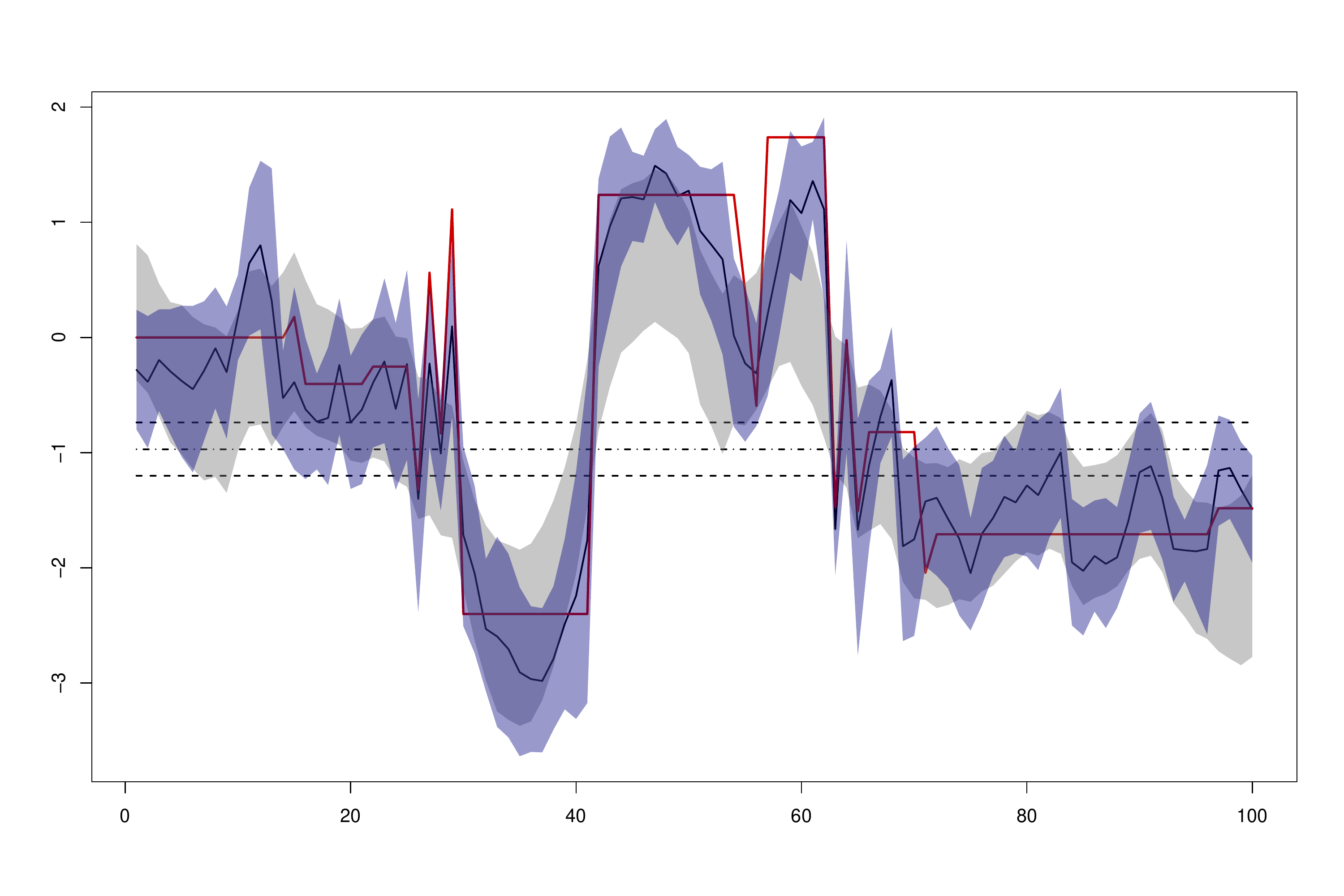}
\end{minipage}
\end{minipage}\vfill

\begin{minipage}[b]{\textwidth}
\centering
(b) \small{\textit{\texttt{TVP-MIX} with flexible state variances (\texttt{FLEX}) and covariate-specific indicators (\texttt{MIX})}:}
\hspace{5pt}
\end{minipage}\vfill

\begin{minipage}[b]{1.6\textwidth}
\begin{minipage}[t]{0.19\textwidth}
\centering
\includegraphics[scale=.17]{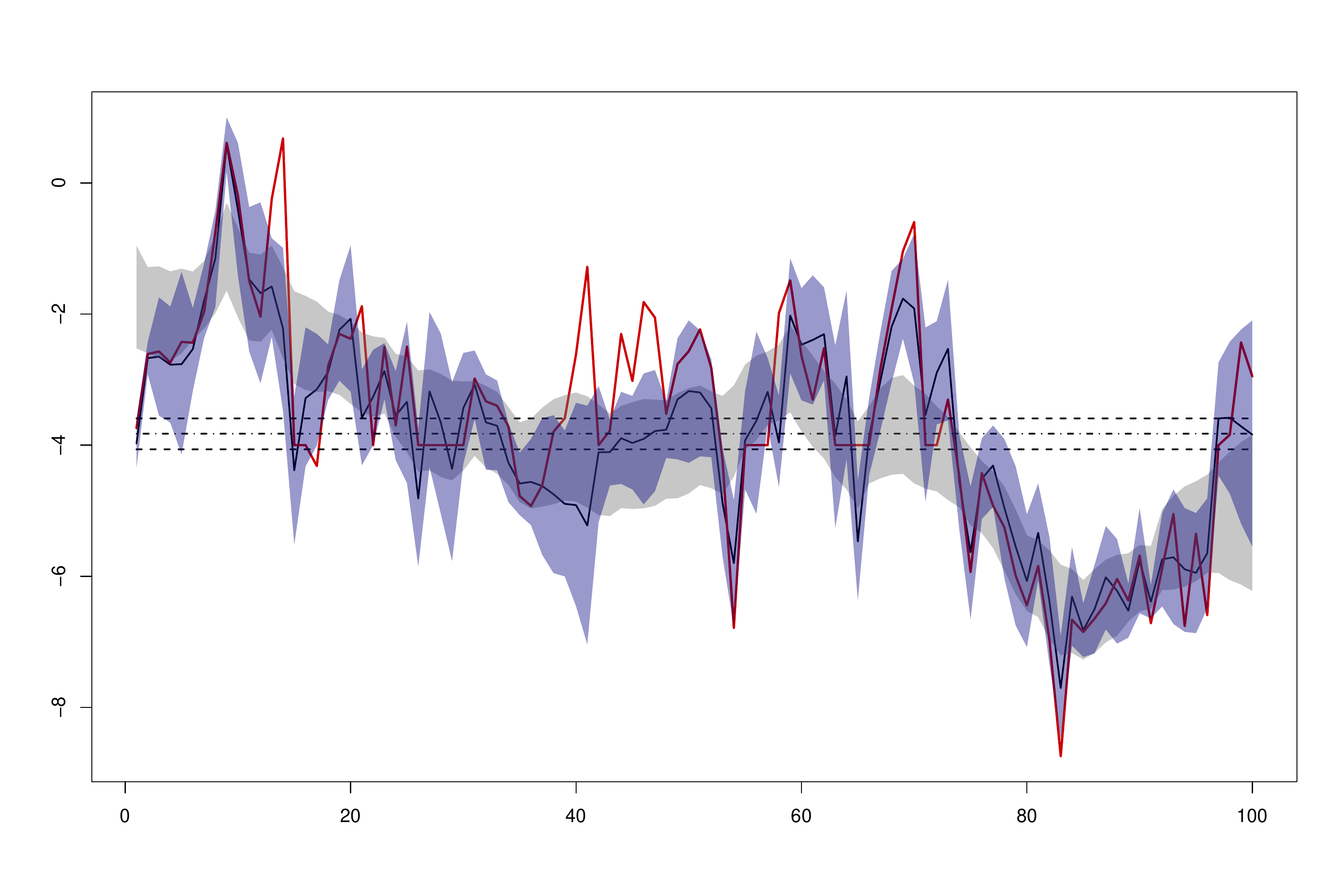}
\end{minipage}
\begin{minipage}[t]{0.19\textwidth}
\centering
\includegraphics[scale=.17]{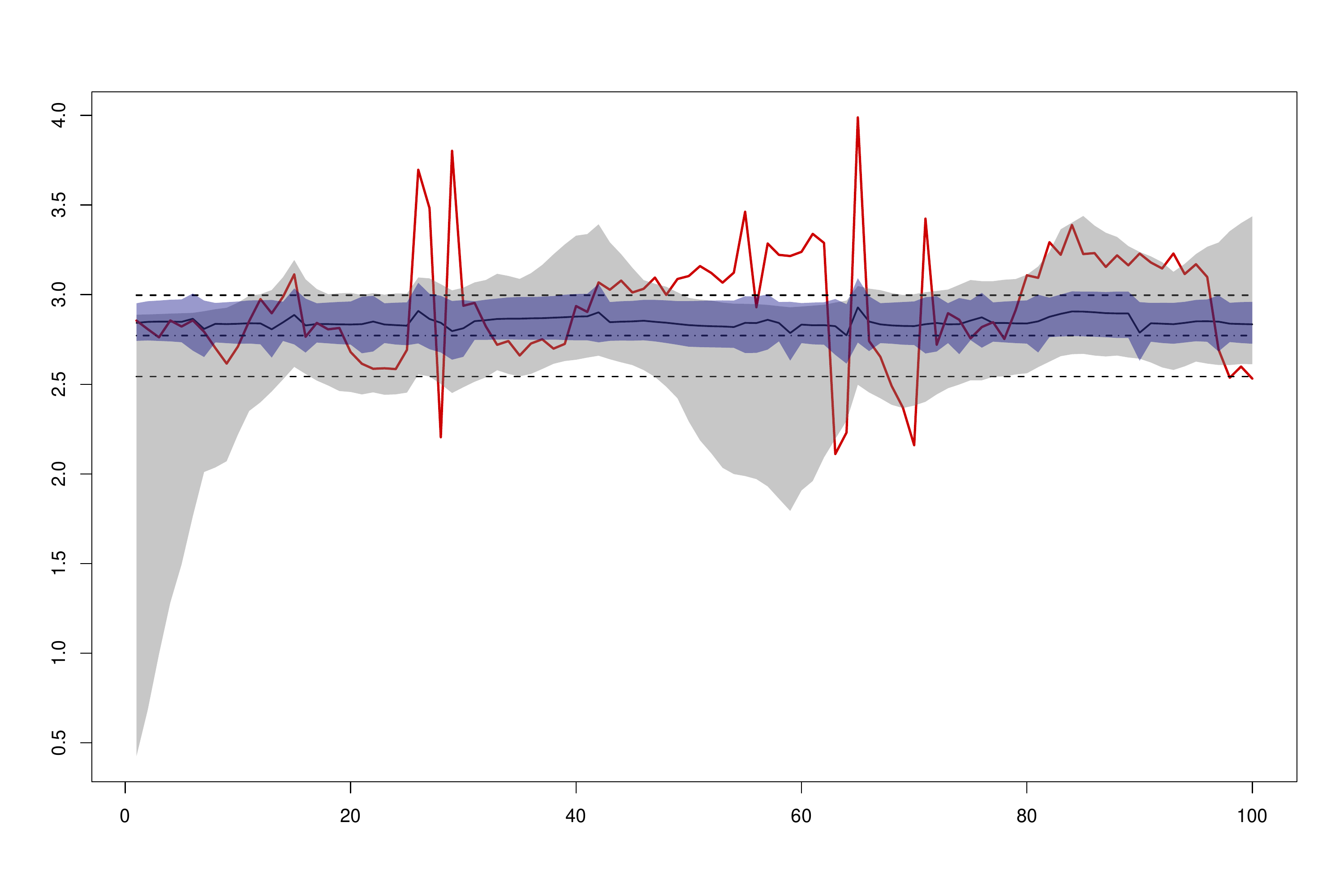}
\end{minipage}
\begin{minipage}[t]{0.19\textwidth}
\centering
\includegraphics[scale=.17]{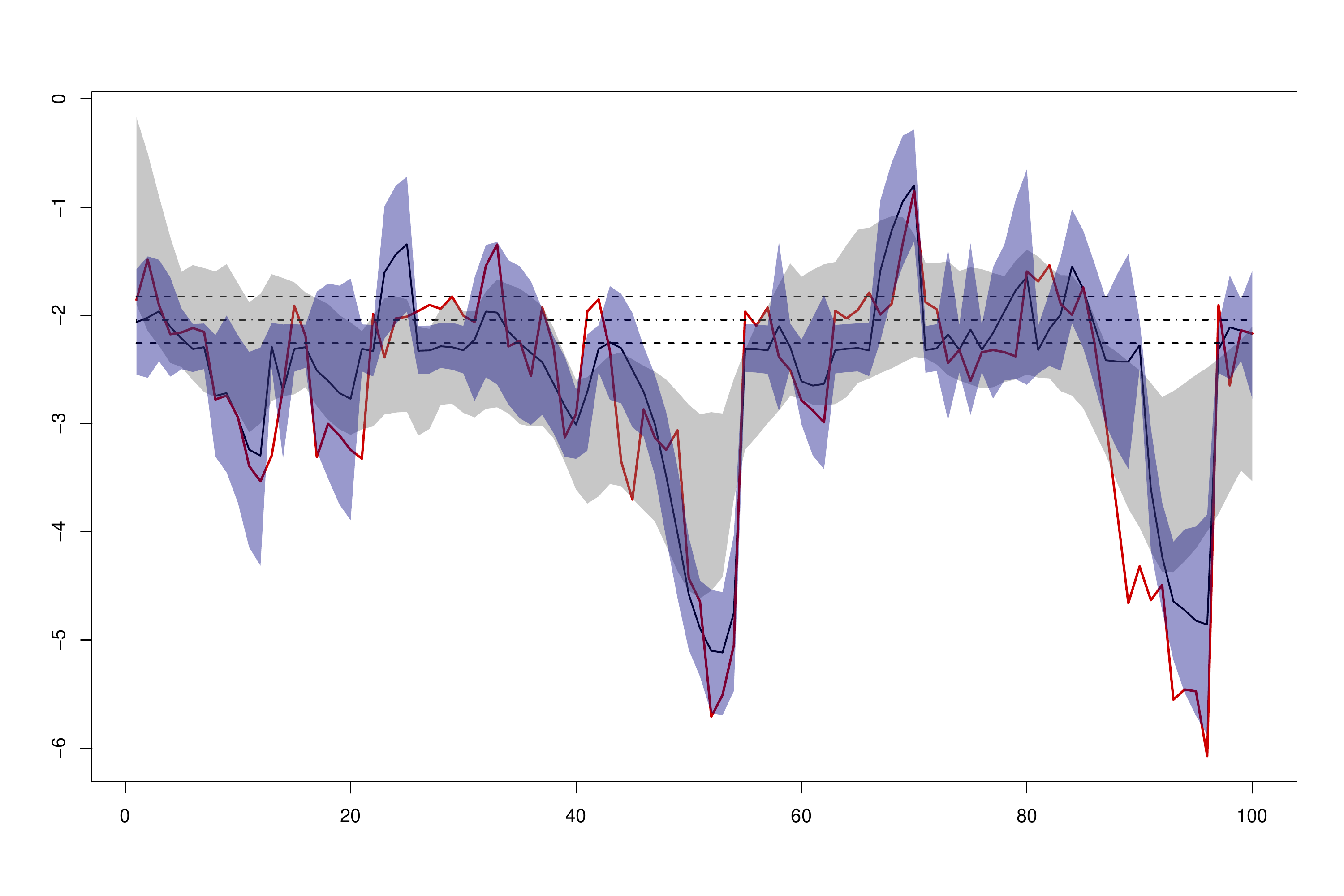}
\end{minipage}
\begin{minipage}[t]{0.19\textwidth}
\centering
\includegraphics[scale=.17]{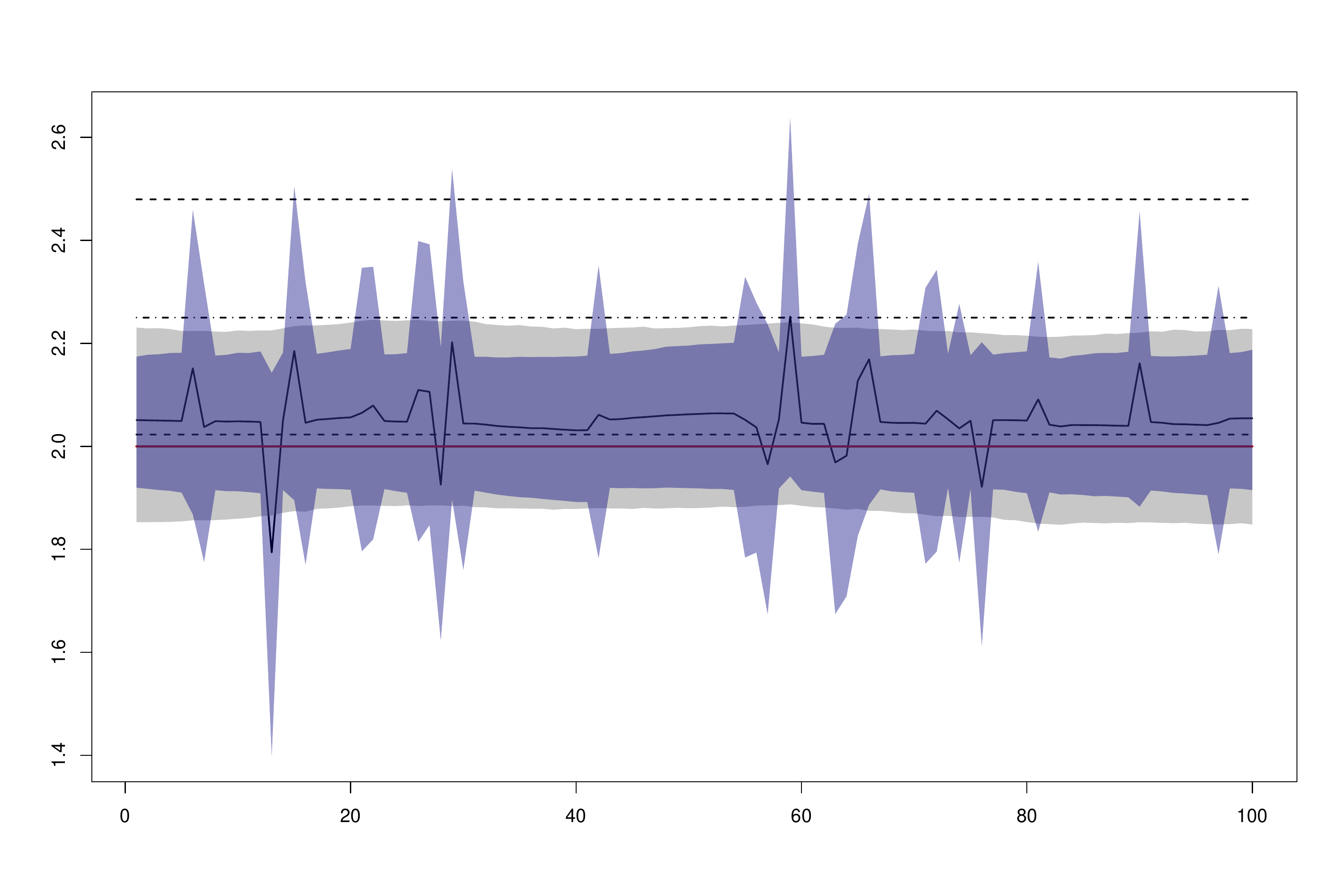}
\end{minipage}
\begin{minipage}[t]{0.19\textwidth}
\centering
\includegraphics[scale=.17]{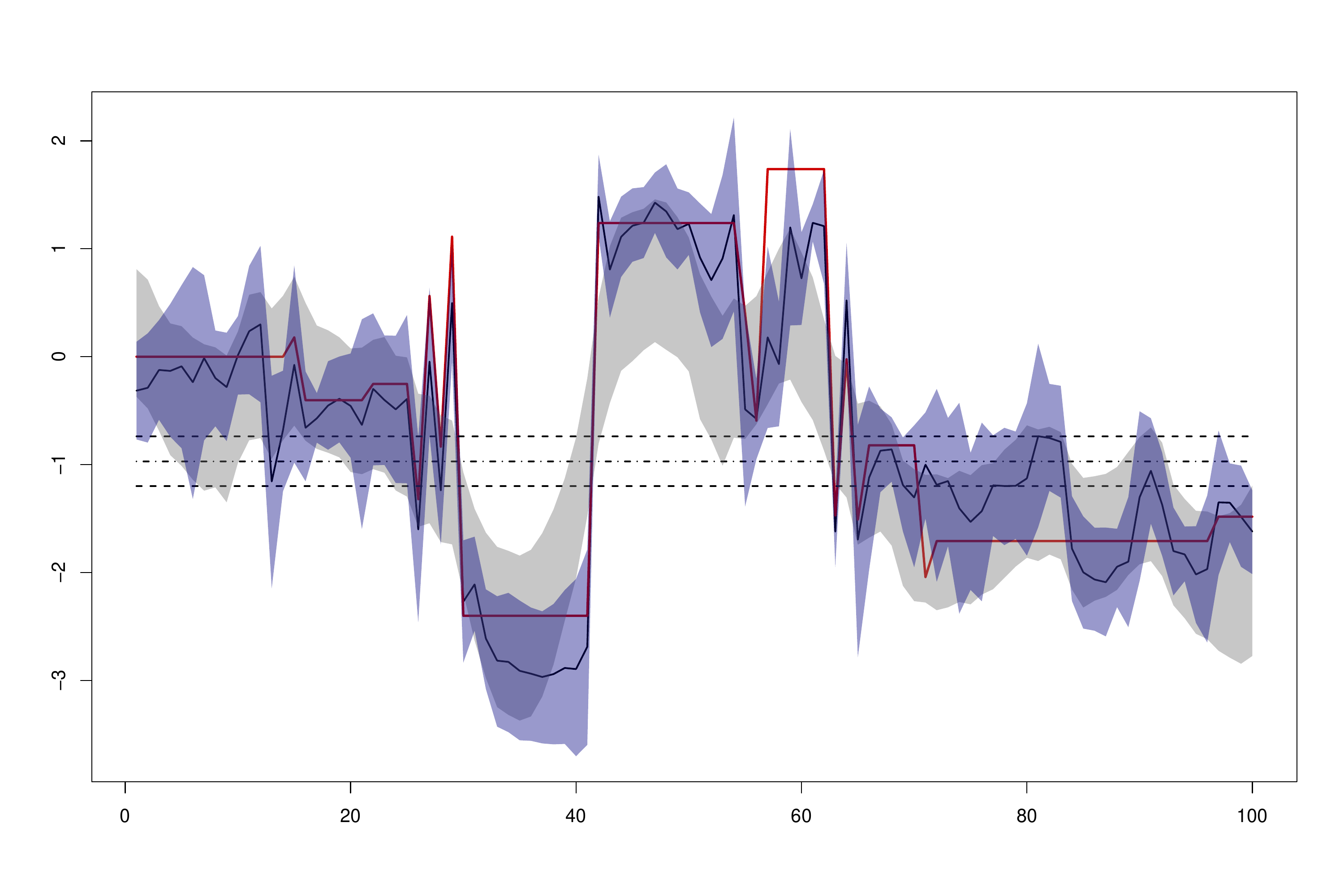}
\end{minipage}
\end{minipage}\vfill

\begin{minipage}[b]{\textwidth}
\centering
(c) \small{\textit{\texttt{TVP-POOL} with flexible state variances (\texttt{FLEX}) and covariate-specific indicators (\texttt{MIX})}:}
\hspace{5pt}
\end{minipage}\vfill

\begin{minipage}[b]{1.6\textwidth}
\begin{minipage}[t]{0.19\textwidth}
\centering
\includegraphics[scale=.17]{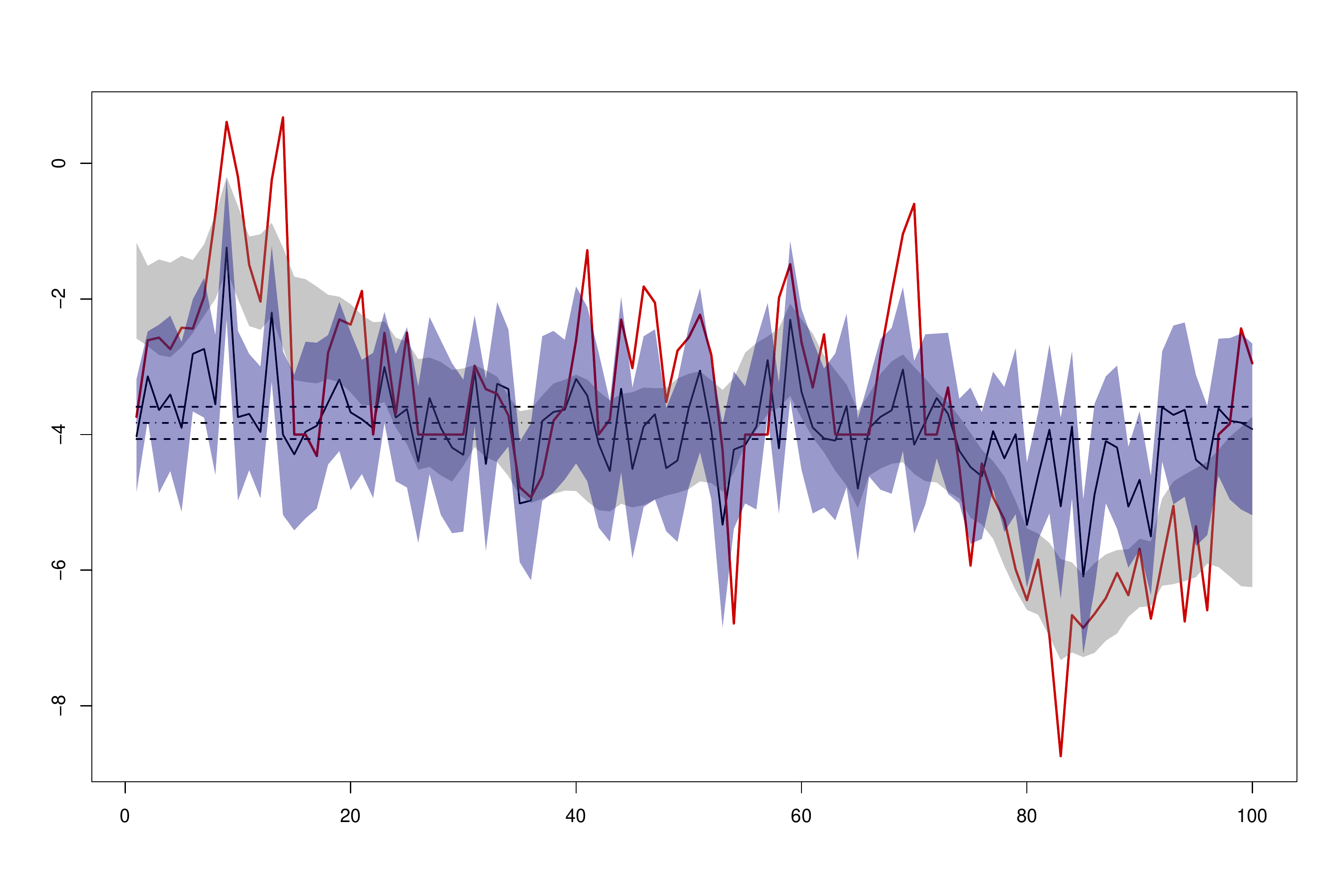}
\end{minipage}
\begin{minipage}[t]{0.19\textwidth}
\centering
\includegraphics[scale=.17]{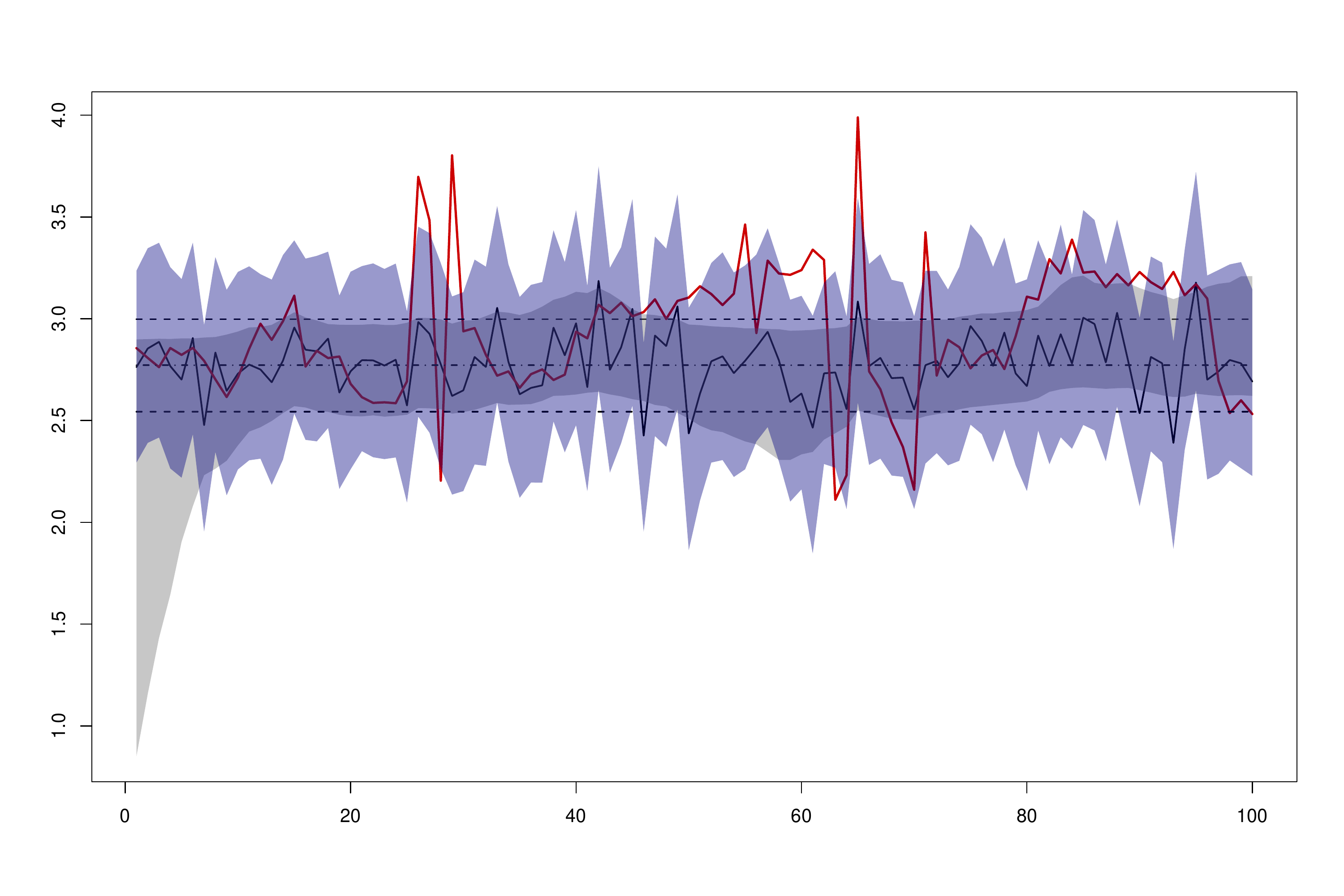}
\end{minipage}
\begin{minipage}[t]{0.19\textwidth}
\centering
\includegraphics[scale=.17]{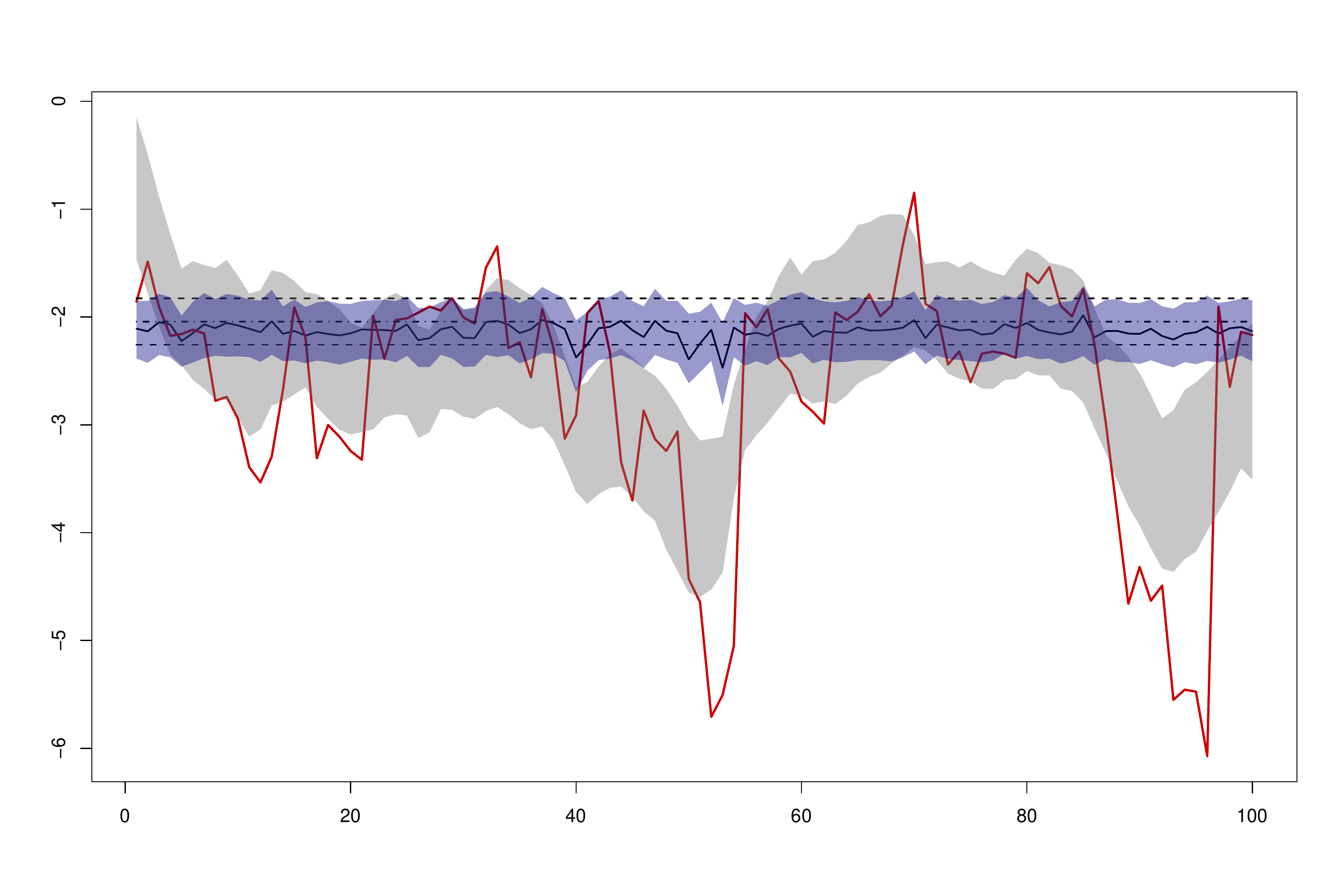}
\end{minipage}
\begin{minipage}[t]{0.19\textwidth}
\centering
\includegraphics[scale=.17]{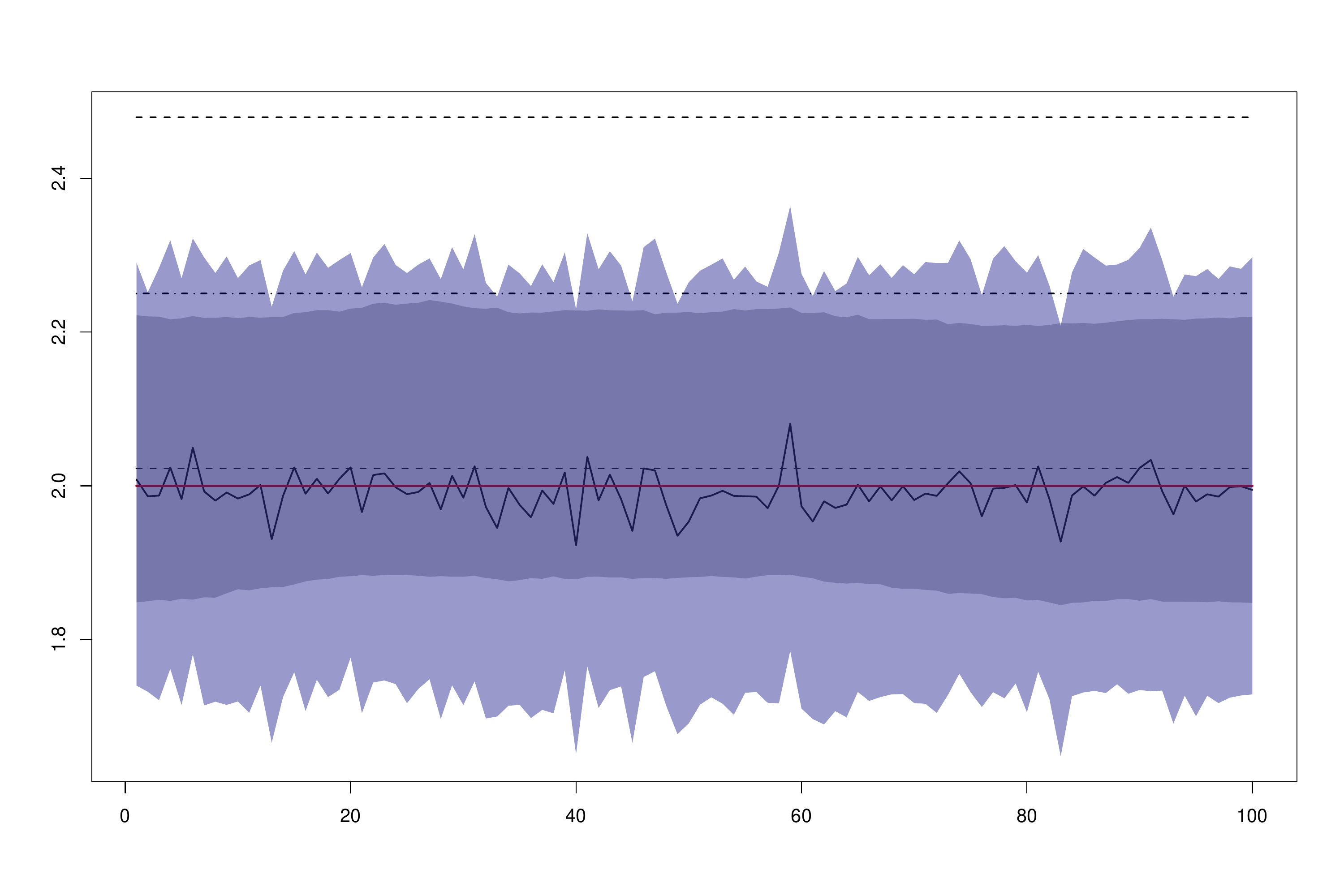}
\end{minipage}
\begin{minipage}[t]{0.19\textwidth}
\centering
\includegraphics[scale=.17]{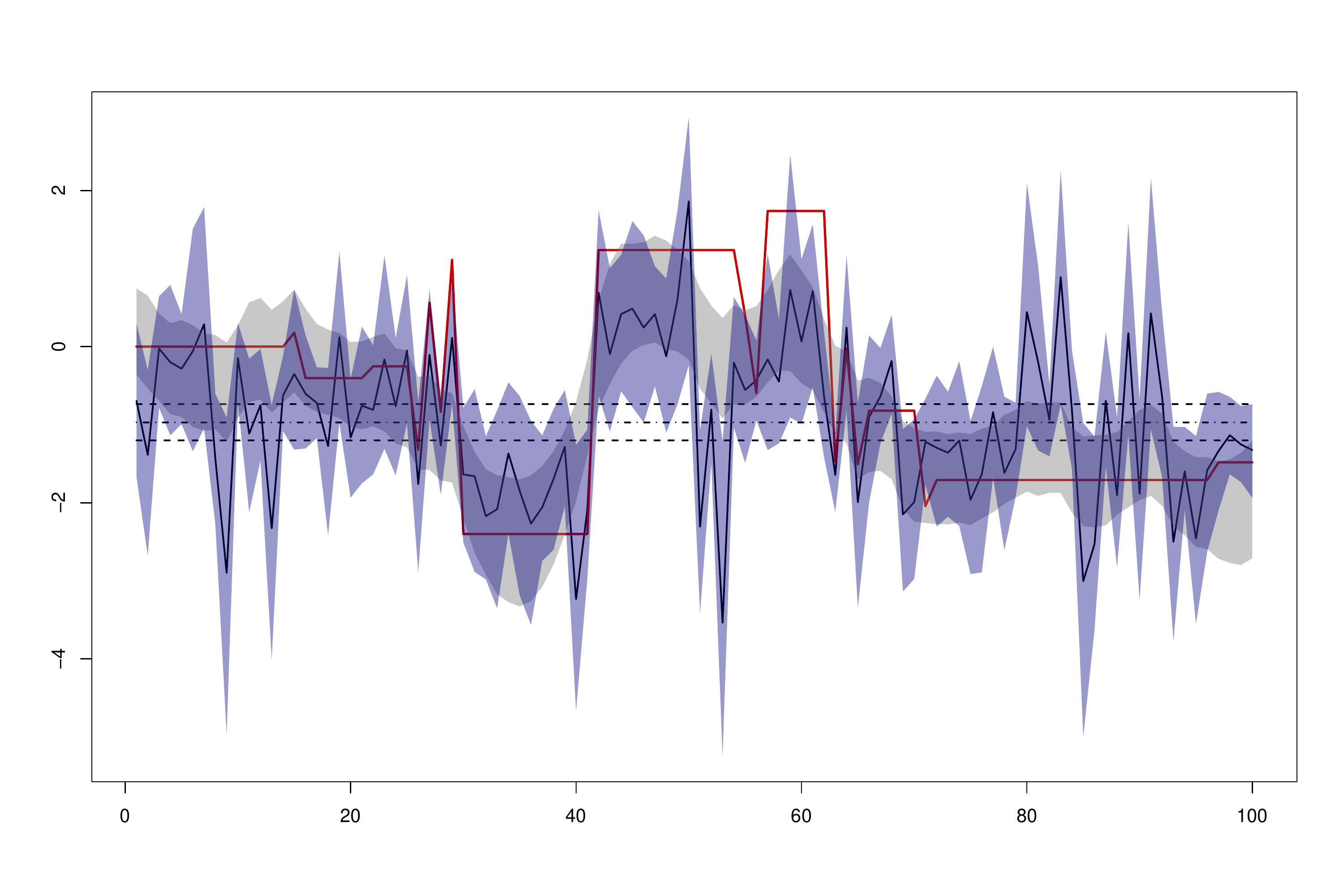}
\end{minipage}
\end{minipage}\vfill

\begin{minipage}[b]{\textwidth}
\centering
(d) \small{\textit{\texttt{TVP-RW} with SSVS-type state variances (\texttt{SSVS}) and covariate-specific indicators (\texttt{MIX})}:}
\hspace{5pt}
\end{minipage}\vfill

\begin{minipage}[b]{1.6\textwidth}
\begin{minipage}[t]{0.19\textwidth}
\centering
\includegraphics[scale=.17]{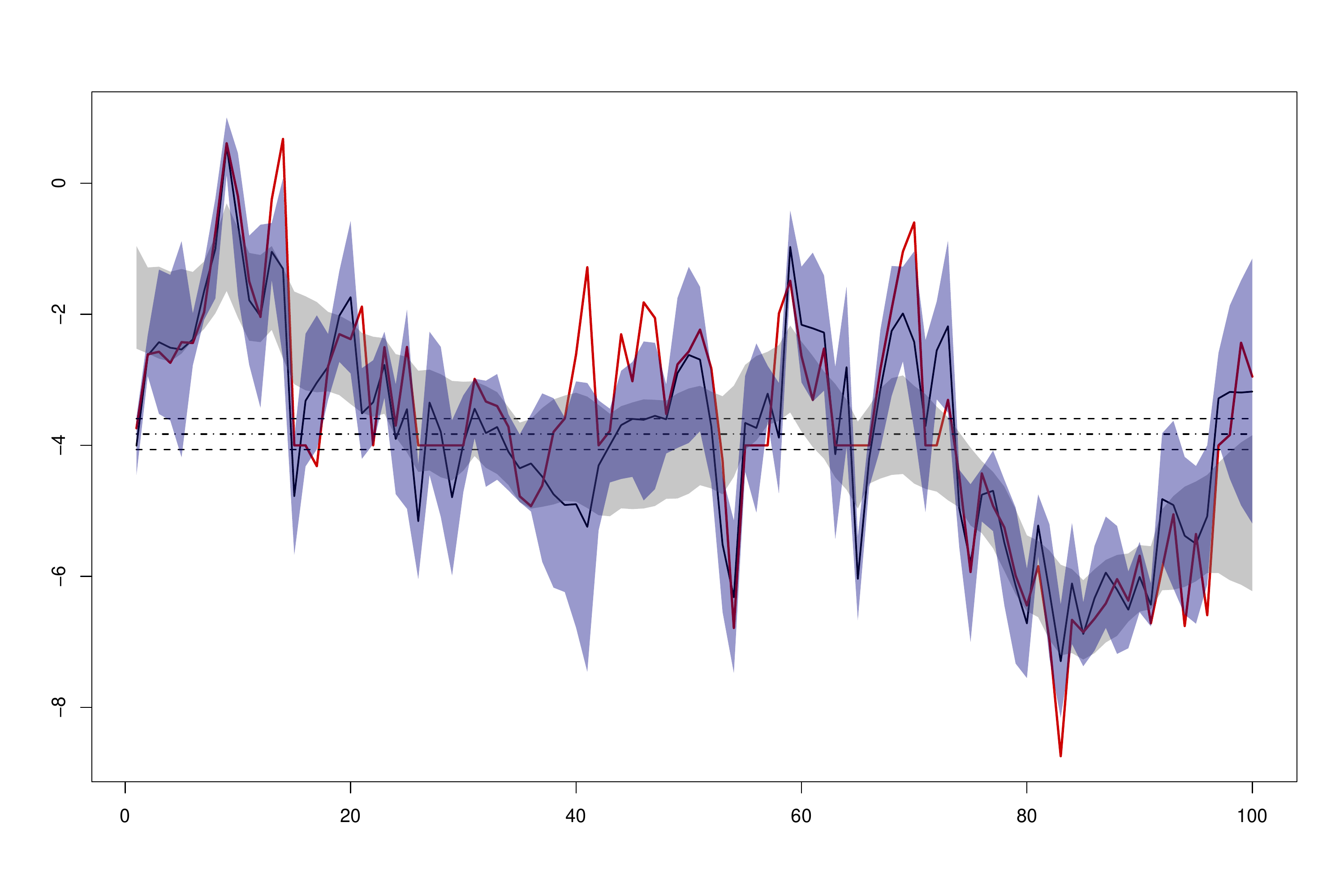}
\end{minipage}
\begin{minipage}[t]{0.19\textwidth}
\centering
\includegraphics[scale=.17]{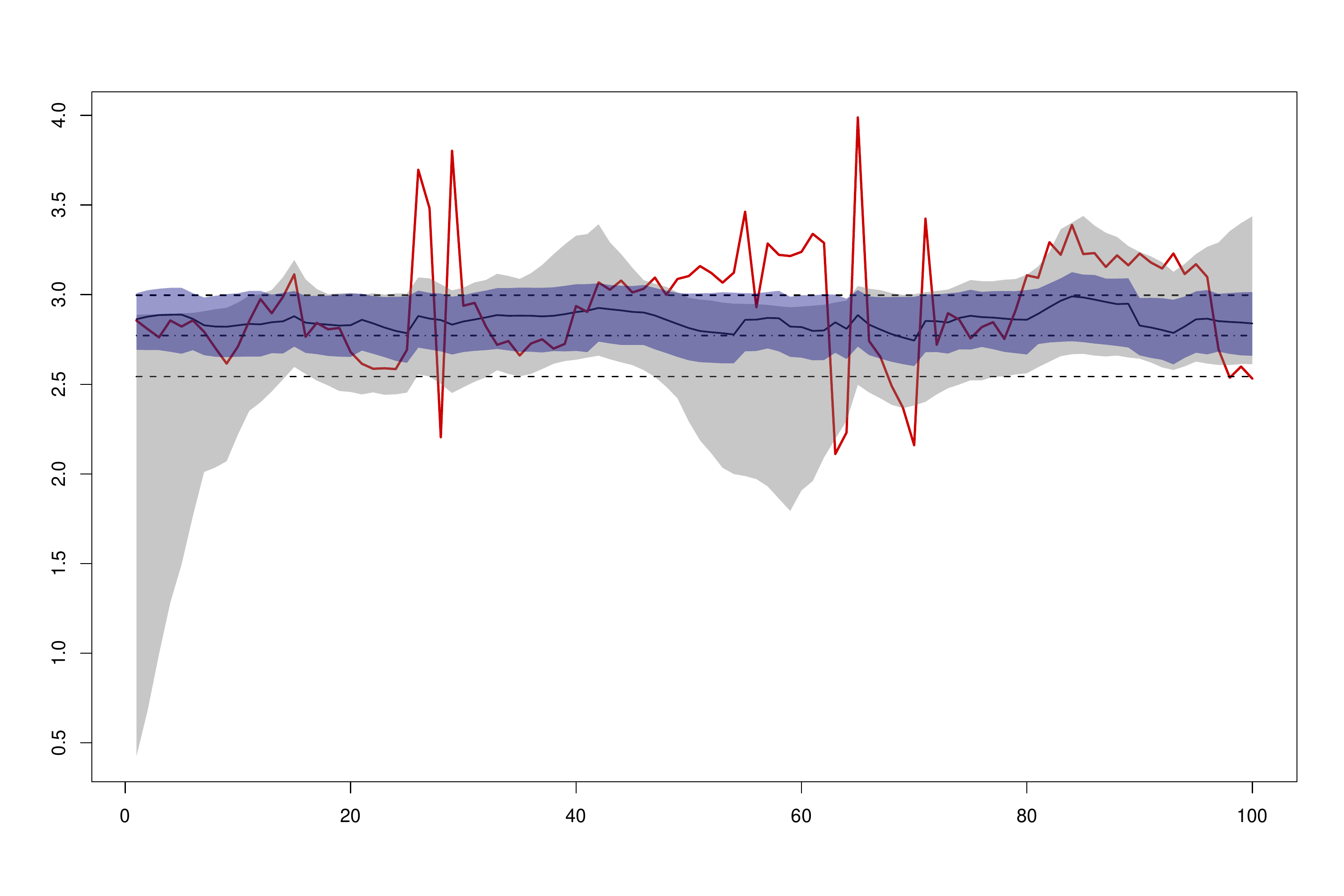}
\end{minipage}
\begin{minipage}[t]{0.19\textwidth}
\centering
\includegraphics[scale=.17]{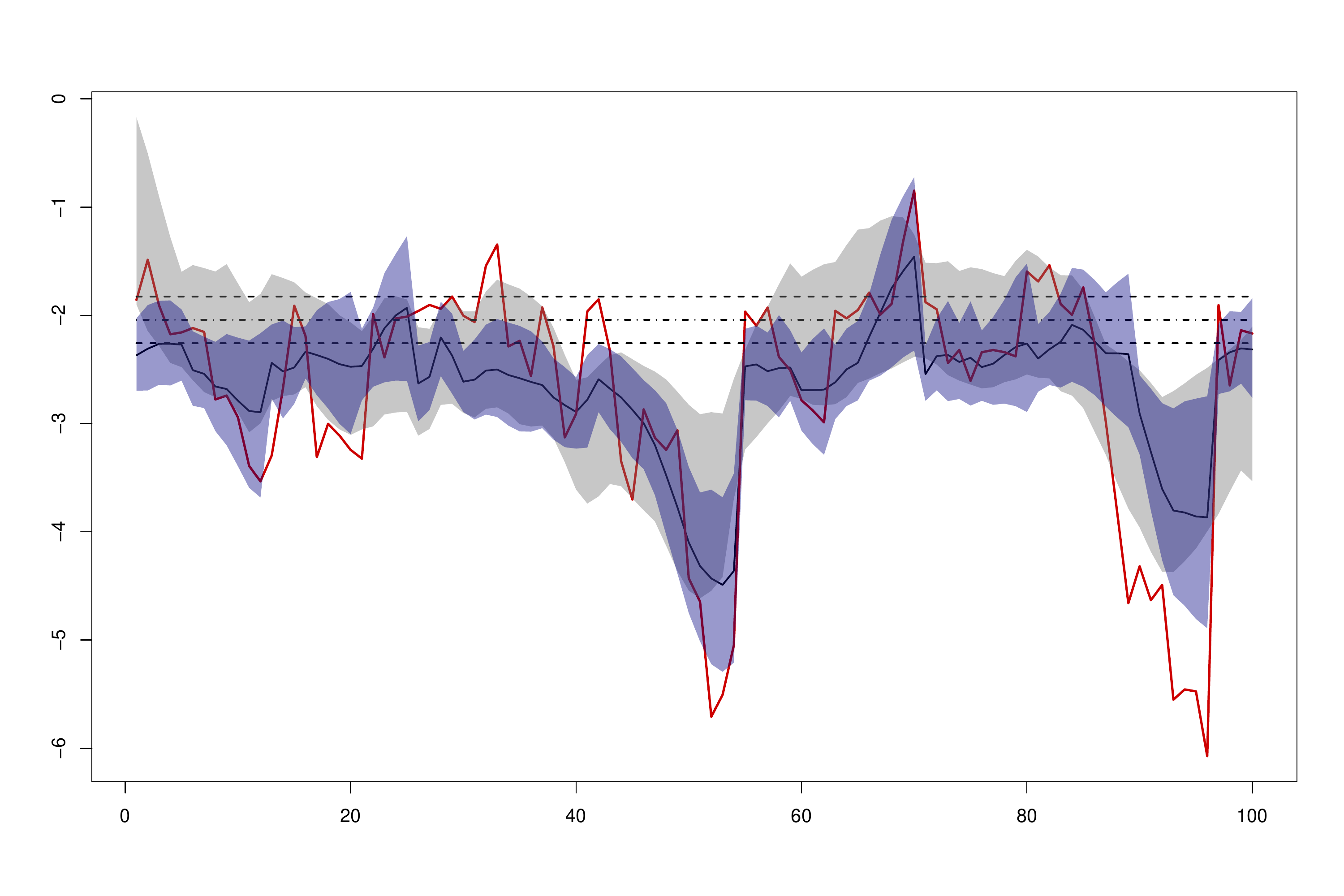}
\end{minipage}
\begin{minipage}[t]{0.19\textwidth}
\centering
\includegraphics[scale=.17]{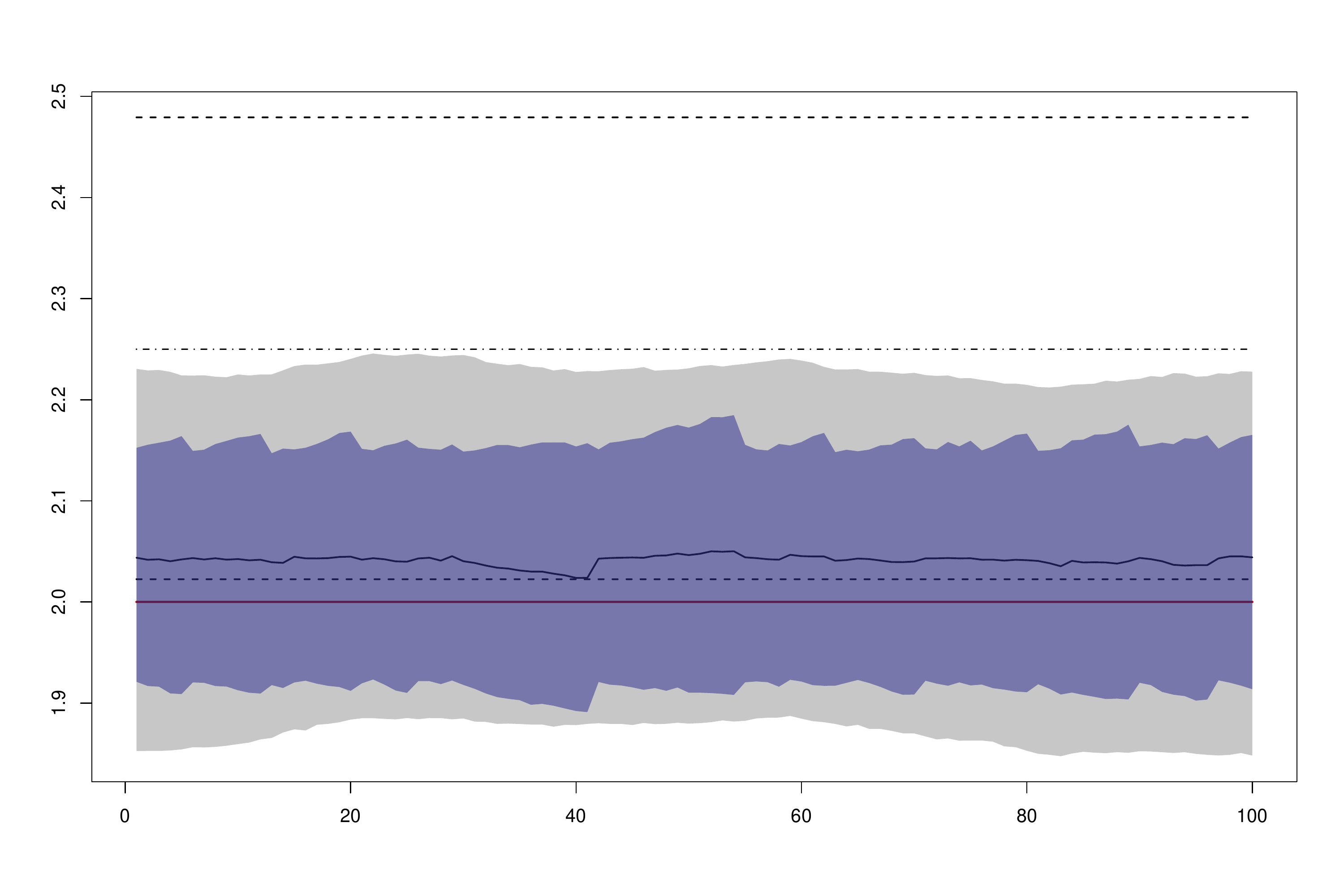}
\end{minipage}
\begin{minipage}[t]{0.19\textwidth}
\centering
\includegraphics[scale=.17]{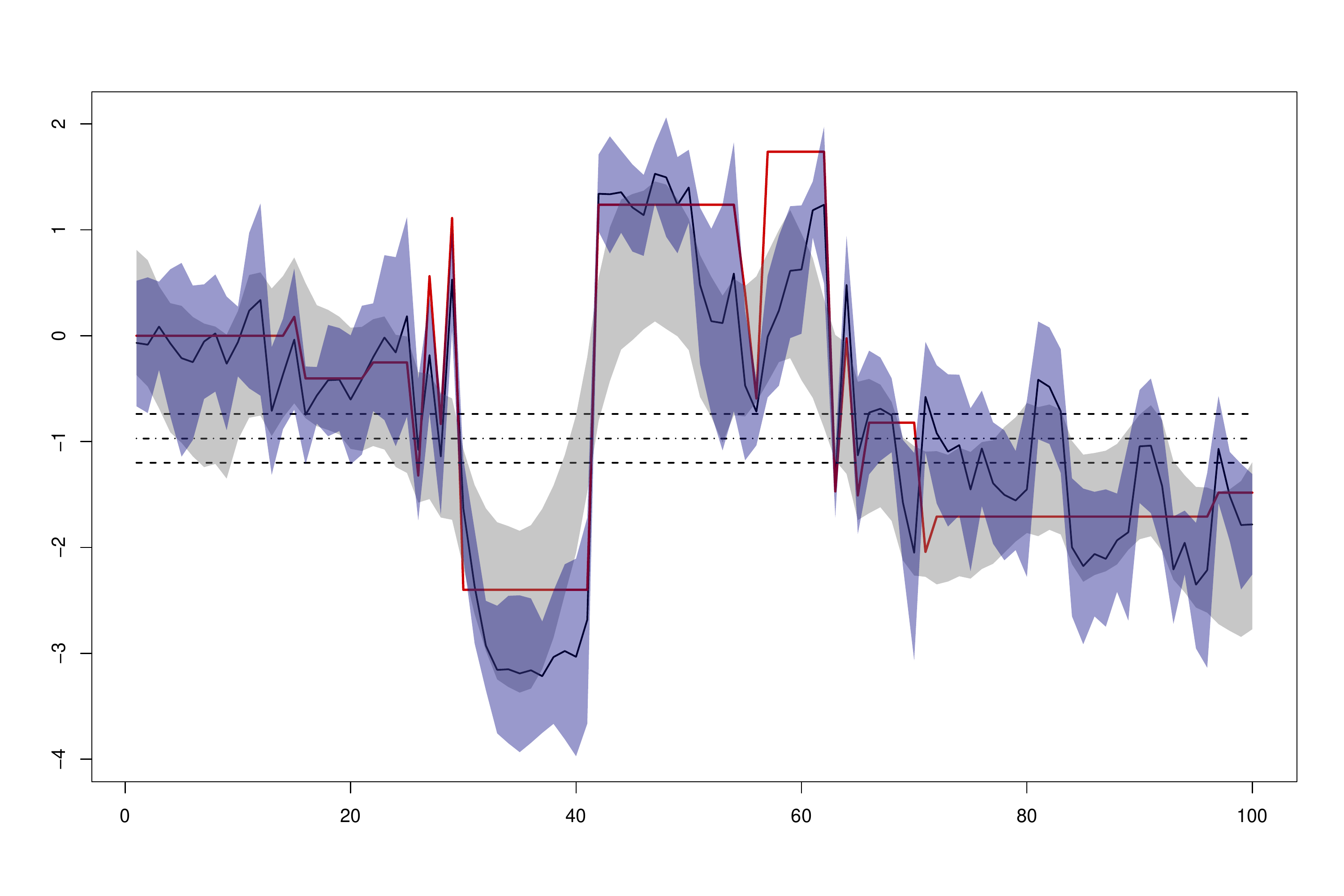}
\end{minipage}
\end{minipage}\vfill
\caption{\small{The blue-shaded areas denote the $68\%$ posterior credible intervals of the proposed methods with the blue solid lines denoting the posterior medians. The gray shaded areas refer to the $68\%$ credible sets of a standard TVP regression with random walk state equation. The black dotted lines indicate the $16^{th}$/$50^{th}$/$84^{th}$ percentiles of a constant coefficient model. Moreover, the red lines denote the true coefficients of $\bm \alpha_t$.} \label{fig:sim}}
\end{figure}
\end{landscape}

Panels (a) to (c) in \autoref{fig:sim} depict the evolution of regression parameters estimated with our proposed methods, while panel (d) shows estimates with a typical mixture innovation model. The red solid lines denote the true coefficients, the blue shaded areas represent the $68\%$ posterior credible interval (with the blue solid lines referring to the posterior median), the gray shaded area represent the respective credible set of a standard TVP model with random walk state equation and the black-dotted lines indicate 
the $16^{th}$/$50^{th}$/$84^{th}$ percentiles of a constant coefficient model. 

The results with artificial data reveal at least three important features. First, all TVP models yield reasonable estimates for constant coefficients, which is most important for forecasting applications. Focussing on the the fourth parameter, considering a more flexible model pays off to produce less biased and more precise estimates, especially when compared to the constant coefficient model. Second, the \texttt{TVP-MIX} specifications are capable in capturing both rapid shifts and smooth adjustments in the regression coefficients. Our methods in panel (a) and (b) tend to quickly adjust when facing high frequency changes, rendering the methods even more flexible when compared to a typical mixture innovation specification in panel (d). Third, the \texttt{TVP-POOL} model in panel (c) tends to detect sudden changes in the parameters quite well, but is less capable in capturing low frequency movements. This feature differs form  a standard random walk evolution assumption on the TVPs. Assuming a standard random walk implies smoothly evolving coefficients makes capturing high frequency changes difficult. Interestingly, the time-varying intercept (the fifth coefficients) tends to soak up movements of other parameters. Models that do not truly detect the large breaks of the third coefficient are particularly prone to this issue (\texttt{TVP-POOL} but also \texttt{TVP-RW} specifications).


\section{Empirical application}\label{sec:real}
Structural analysis and forecasting key macroeconomic indicators is of great relevance for policy makers. In the empirical work, we focus on output growth, inflation, unemployment, and/or the interest rate. Focussing on these variables we investigate the merits of our approach by using the popular quarterly US data described in \cite{mccracken2016fred}. The data set includes $165$ macroeconomic and financial variables and ranges from $1959$:Q$1$ to $2019$:Q$4$.\footnote{In the empirical application we start with $1962$:Q$1$ and use the first observations for transformations.} 

In Subsection \ref{ssec:insmp} we show some stylized in-sample features of our methods for a small-scale model. By including the four target variables in a small-scale VAR (henceforth \texttt{S-VAR}) we present posterior probabilities of the state indicator matrix $\bm S_t$ and estimate the low-frequency relationship between unemployment and inflation (Phillips Curve). For instance, the recent literature highlights the existence of potential non-linear dynamics in both the inflation persistence and the relationship of the Phillips Curve in the US \citep[for a thorough discussion, see][]{cogley2010inflation, BallMazumder, watson2014inflation, coibion2015phillips}.  

Moreover, in Subsection \ref{ssec:outsmp}, this variable set forms the basis for evaluating the predictive performance of our methods in a comprehensive forecast exercise. For the forecasting exercise, we consider two additional information sets. In our largest specification (\texttt{L-VAR}) we pick $20$ macroeconomic indicators, which are commonly considered by the recent literature for forecasting \citep[see, for example,][]{hko2020, pfarrhofer2020forecasts}.\footnote{In Appendix \ref{app:data} we provide further details on the specific variable set, included in the largest specification, and the transformation applied.} In particular, we include financial market indicators that carry important information about the future stance of the economy \citep[see][]{banbura2010large}. Moreover, we consider a factor-augmented VAR (\texttt{FA-VAR}). Here, we augment the target variables with six principal components compromising information of the remaining variables in the data set, effectively leading to VAR with ten endogenous variables.\footnote{The number of principal components is motivated by the specification in \cite{stock2012disentangling}, who also consider six factors.} In such larger scale-models our methods are capable of handling less frequent (but important) parameter instabilities in a genuine way. 

Especially forecasting these important macroeconomic aggregates remains a challenging task, since (at least) two issues arise.  
First, we have to decide on a set of variables, which we want to include in our econometric model. The recent literature on constant parameter VARs highlights that exploiting large information sets yields forecast gains \citep[see, for example,][]{banbura2010large, koop2013forecasting}. Second, it is well documented that important economic indicators feature instabilities in structural parameters and innovation volatilities.\footnote{See, for example \cite{stock2012disentangling}, \cite{ng2013facts} and \cite{aastveit2017have}, which put special emphasis on the recent financial crisis.} In the literature there is strong agreement that SV is important in macroeconomic applications \citep[see][]{clark2011}. There is also strong empirical support for shifting parameters in small-scale models \citep[see][]{giannone2013}. However, there is less consensus for time-varying parameters in larger-scale models. With increasing amount of information overall time-variation in parameters tends to reduce. Recent contributions dealing with large-scale TVP-VARs argue that in smaller models the TVP part controls for an omitted variable bias \citep[see][]{feldkircher2017sophisticated, hkp2020}.\footnote{Since estimating TVP models with typical MCMC methods remains computationally demanding, several studies take this argument as a reason to opt for approximating the TVP part or rely on dimension reduction techniques, yielding fast inference while accepting a certain risk of misspecification \citep[see, inter alia][]{eisenstat2019reducing, korobilis2019high,hhk2020, hkp2020, korobilis2020bayesian}.} 

In the following empirical application, note thate we consider two lags for every model and allow for SV. 
\subsection{In-sample evidence}\label{ssec:insmp}
Before proceeding, we briefly elaborate on a potential identification problem when interpreting the state indicators $\bm S_t$ \citep[see][]{fruhwirth2001markov}. For the \texttt{TVP-MIX} models, identification is ensured by construction (if coefficients indeed feature time variation). Assuming $\bm \phi_t = \bm S_t$ (see \autoref{eq:phi}) automatically imposes inequality constraints on the autoregressive coefficients in the state equation. However, non-identifiability can occur when coefficients are constant. In such a case, elements in $\bm S_t$ are hard to interpret, since a no change evolution is supported by both a random walk and a white noise process. Interpreting $\bm S_t$ for the \texttt{TVP-POOL} specification is an even more challenging task, since in these models $\bm S_t$ solely controls the evolution of state innovations. Here, inference about the state indicator matrix is only useful in combination with inference about the size of state innovation variances $\bar{\bm \Psi}_0$ and $\bar{\bm \Psi}_1$ and with imposing an inequality restriction ex-post (for example, $\bar{\psi}_{i0} < \bar{\psi}_{i1}$).  

\begin{figure}
\begin{minipage}[b]{\textwidth}
\centering
(a) \textit{$\bm S_t = s_t \bm I_K$ with $s_t$ following an MS process:} 
\hspace{5pt}
\end{minipage}\vfill
\begin{minipage}{\textwidth}
\centering
\includegraphics[scale=.3]{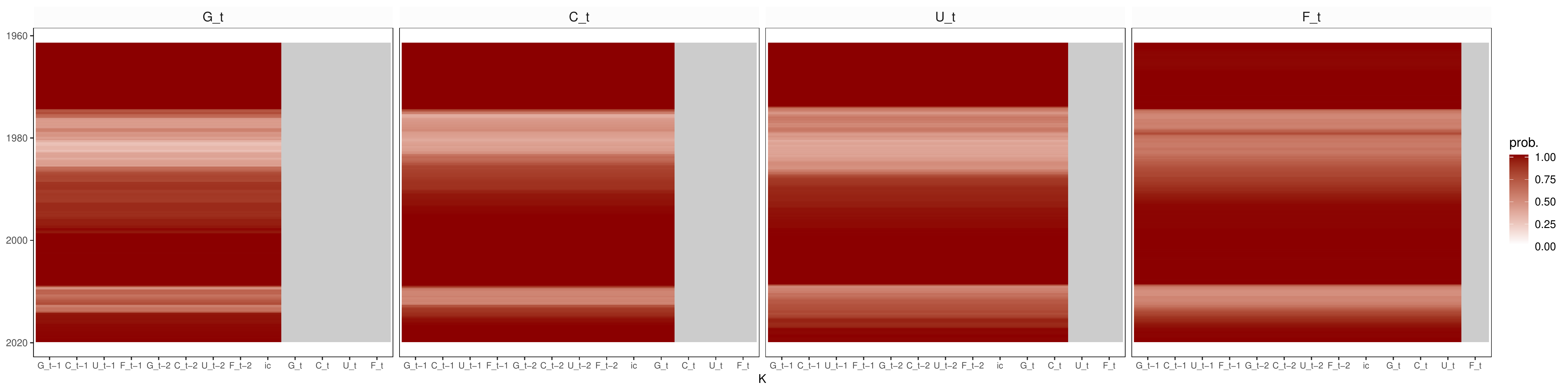}
\end{minipage}
\centering
\begin{minipage}{\textwidth}
\centering
(b) \textit{Elements in $\bm S_t$ follow an independent mixture specification:}
\hspace{5pt}
\end{minipage}\vfill
\begin{minipage}{\textwidth}
\centering
\includegraphics[scale=.3]{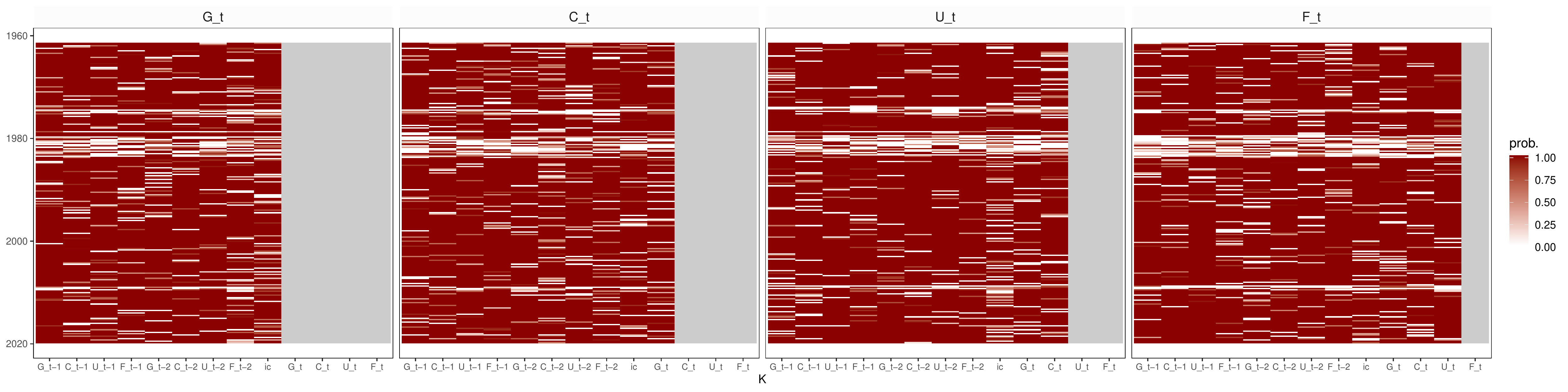}
\end{minipage}
\caption{Posterior distribution of $s_{it}$, for $i = \{1, \dots, K\}$, for two small-scale \texttt{TVP-MIX} models. Here, $G$ denotes output growth (GDPC1), $C$ the inflation (CPIAUCSL), $U$ the unemployment rate (UNRATE), $F$ refers to the interest rate (FEDFUNDS) and $ic$ to an intercept. Moreover, the structural form in \autoref{eq:strVAR} implies that
some parameters are not part of the $i^{th}$ equation (denoted by grey shaded areas), due to the strictly lower triangular structure of $\bm B_{0t}$. \label{fig:St}}
\end{figure}

Therefore, we solely focus on two variants of a \texttt{TVP-MIX} model to illustrate the switching behaviour. 
\autoref{fig:St} depicts the posterior median of the diagonal elements in $\bm S_t$. Panel (a) shows a \texttt{TVP-MIX} model with $\bm S_t = s_t \bm I_K$ and $s_t$ following a first-order Markov process (\texttt{MS}). Panel (b) depicts a specification with  elements in $\bm S_t$ following an independent mixture distribution (\texttt{MIX}). A comparison between both approaches highlights that a joint indicator evidently leads to a different posterior median of $\bm S_t$ than covariate-specific indicators. 
By restricting $\bm S_t = s_t \bm I_K$, all covariates are driven solely by a single indicator that pushes all covariates towards either a random walk or white noise state equation in period $t$. Conversely, with covariate-specific indicators, we see more dispersion across covariates. However, both approaches agree on a white noise state equation in times of turmoil, suggesting a need for abruptly adjusting parameters in these periods. This model feature is in line with the discussion in \cite{primiceri2005}, who suggests that an economically stable period favours more gradual changes (which are more consistent with a random walk state equation) in the coefficients, while shifts in policy rules require quickly adjusting coefficients (which is better captured by using a white noise state equation). 

To further illustrate the proposed methods, we estimate the low-frequency relationship between unemployment and inflation. This low-frequency measure corresponds to a long-run coefficient of distributed-lag regression models \citep{whiteman1984lucas} and disentangles systematic co-movements from short-run fluctuations.\footnote{\cite{sargent2011two} and \cite{kliem2016low} suggest that a TVP-VAR framework, additionally, allows to account for changes in the transmission channels (time-varying coefficients) and changes in the error volatilities (SV). For further details see Appendix \ref{app:spectral}.} 

Panel (a) to (c) in \autoref{fig:lowfreq} depict the obtained low-frequency component with our proposed approaches and  panel (d) shows estimates with a standard TVP model assuming a standard random evolution assumption. Starting with a comparison between the random walk/white noise mixture (\texttt{TVP-MIX}) and a classic random walk TVP model, we observe a similar pattern for both approaches during tranquil periods.
However, during recessions, the approaches significantly differ. Both \texttt{TVP-MIX} models are capable of detecting a major structural break in the low-frequency relationship after the oil crisis in the 1970s and strongly support a long-lasting stagflation period (i.e. positive relationship between unemployment and inflation). While \texttt{TVP-MIX} methods are designed to quickly capture these large abrupt breaks in parameters, a standard random walk state equation translates into a low-frequency component that gradually adapts over time.
However, the \texttt{TVP-MIX} model with covariate-specific indicators \texttt{MIX} is slightly more sensitive with respect to abrupt changes in parameters than the \texttt{TVP-MIX MS} model. 

Panel (c) shows the sparse scale-location mixture (\texttt{TVP-POOL}) approach with covariate-specific indicators (\texttt{MIX}). We observe that this method almost resembles a constant coefficient specification with SV. In the mid 1980s and in the financial crisis movement in the low-frequency relationship is slightly more erratic compared to other periods, but it stays mostly constant and significant.

Overall, considering \texttt{TVP-MIX} methods seem to improve the economic interpretability of the low-frequency component, while a \texttt{TVP-POOL} model aggressively pushes coefficients towards a constant evolution, which could pay off for forecasting. 


\begin{figure}
\begin{minipage}[b]{\textwidth}
\centering
(a) \textit{\texttt{TVP-MIX} with flexible state variances (\texttt{FLEX}) and $\bm S_t = s_t \bm I_K$ (\texttt{MS})}:
\hspace{5pt}
\end{minipage}\hfill
\begin{minipage}{\textwidth}
\centering
\includegraphics[scale=.5]{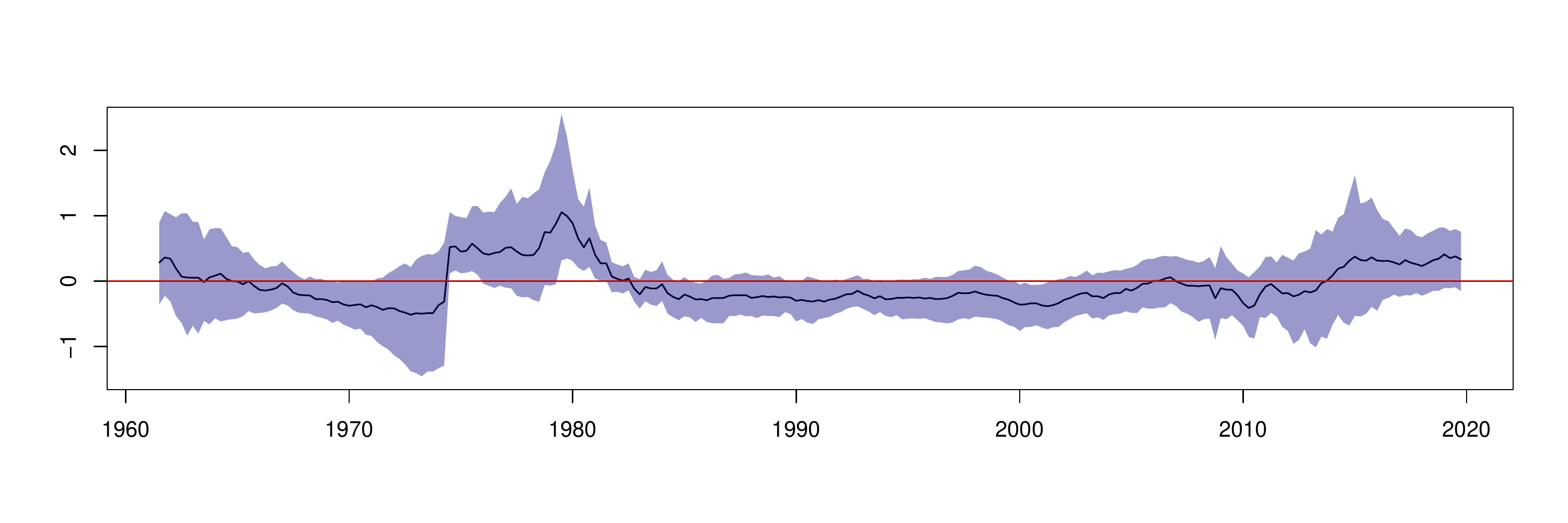}
\end{minipage}
\begin{minipage}[b]{\textwidth}
\centering
(b) \textit{\texttt{TVP-MIX} with flexible state variances (\texttt{FLEX}) and covariate-specific indicators (\texttt{MIX})}:
\hspace{5pt}
\end{minipage}\hfill
\begin{minipage}{\textwidth}
\centering
\includegraphics[scale=.5]{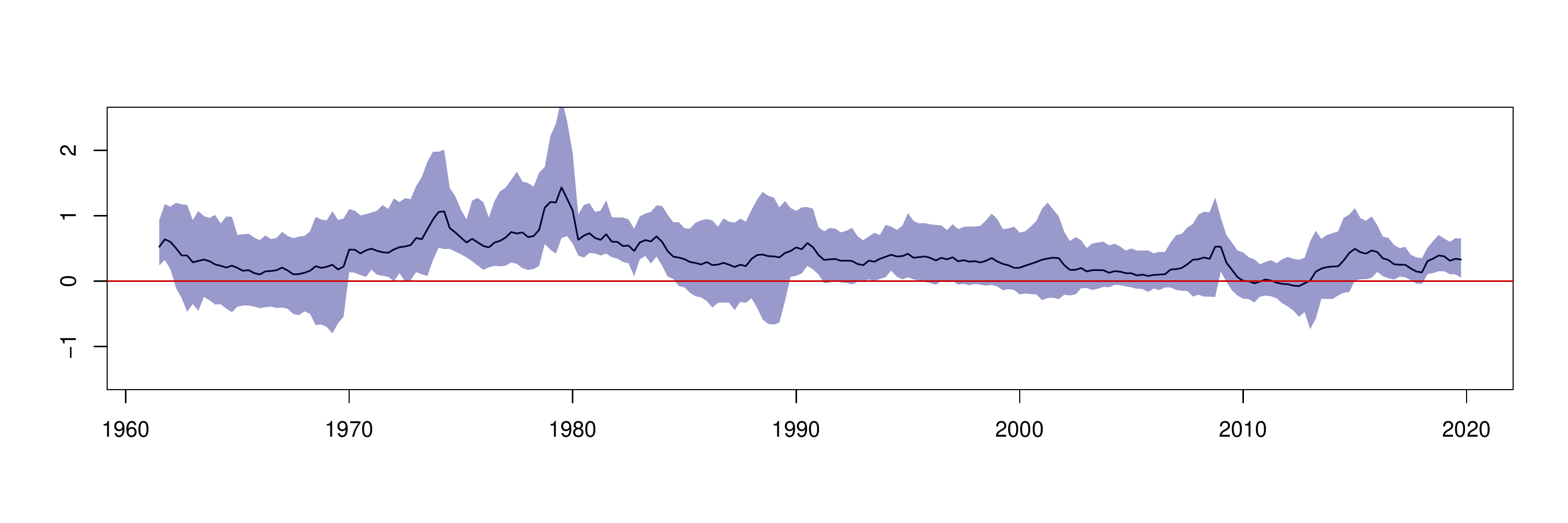}
\end{minipage}
\centering
\begin{minipage}{\textwidth}
\centering
(c) \textit{\texttt{TVP-POOL} with flexible state variances (\texttt{FLEX}) and covariate-specific indicators (\texttt{MIX})}:
\hspace{5pt}
\end{minipage}\hfill
\begin{minipage}{\textwidth}
\centering
\includegraphics[scale=.5]{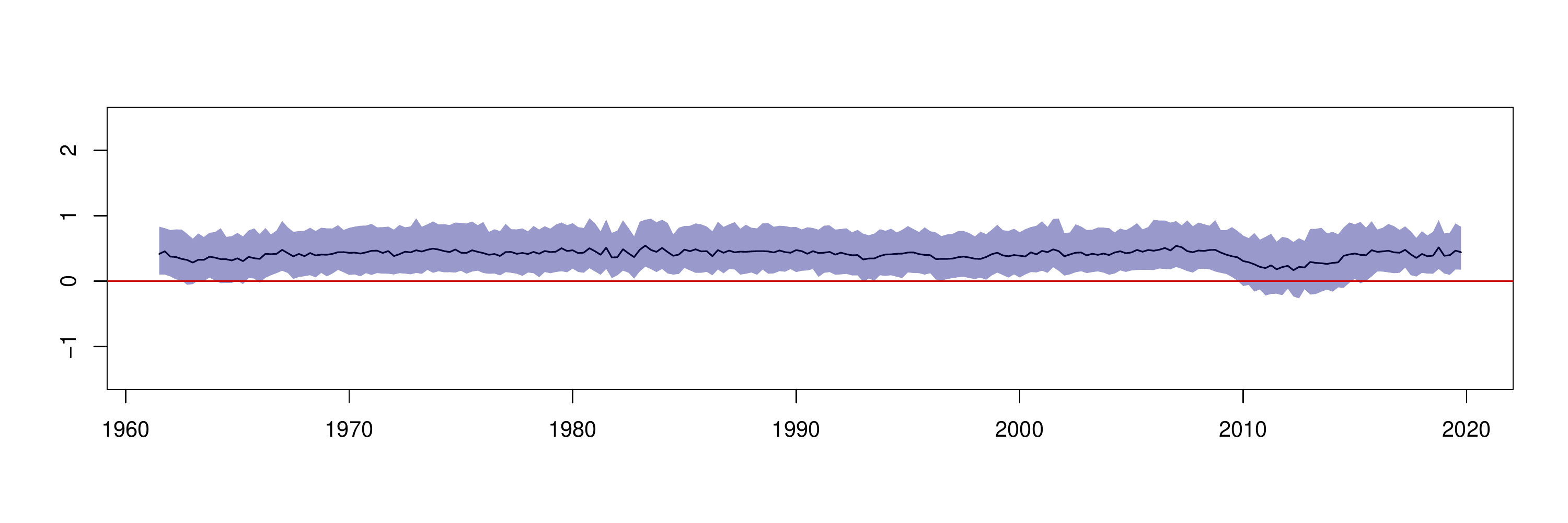}
\end{minipage}
\begin{minipage}{\textwidth}
\centering
(d) \textit{Standard TVP-VAR with random walk state equation:}
\hspace{5pt}
\end{minipage}\hfill
\begin{minipage}{\textwidth}
\centering
\includegraphics[scale=.5]{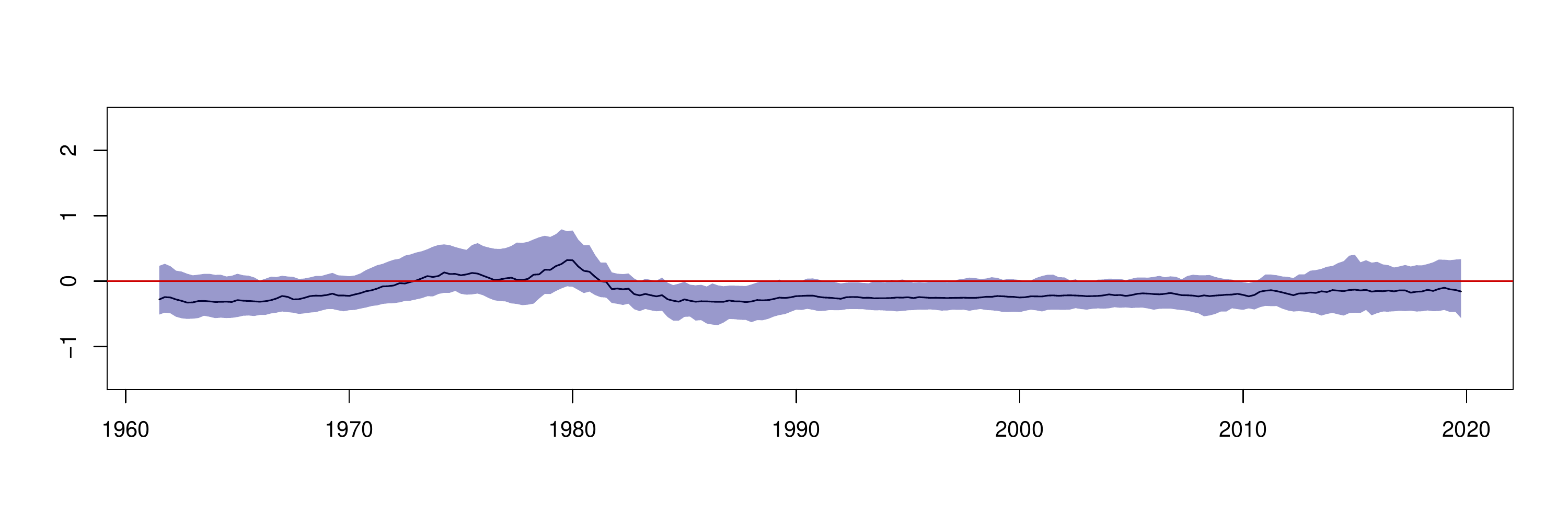}
\end{minipage}
\caption{Low-frequency relationship between the unemployment rate and the inflation. The blue line refers to the posterior median, while the blue-shaded area indicates the $68\%$ posterior credible set. The red line indicates zero. \label{fig:lowfreq}}
\end{figure}

\subsection{Forecasting evidence}\label{ssec:outsmp}
In the forecast exercise we consider a wide range of models varying along the evolution assumption of parameters and the information set considered. 

With respect to the evolution of parameters, it proves convenient to summarize the different specifications (see \autoref{tab:spec}). The models differ along three dimensions: the autoregressive parameters $\bm \phi_t$, the innovation variances $\bm \Psi_t$ and the state indicator matrix $\bm S_t$. First, our main specifications vary between a model that assumes a binary indicator matrix on the autoregressive parameter with $\bm \phi_t = \bm S_t$ (labeled as \texttt{TVP-MIX}) and a model that introduces a hierarchical prior on the TVP-part (\texttt{TVP-POOL}). For the latter we implicitly assume that $\bm \phi_t = \bm 0_{K \times K}$, for all $t$, in \autoref{eq:v2}. Regarding the autoregressive parameter, a natural competing model is a standard random walk assumption with $\phi_t = \bm I_K$, for all $t$ (\texttt{TVP-RW}). Second, the models differ in the treatment of the state innovation variances. The most flexible innovation variance specification does not restrict $\bar{\bm \Psi}_0$ and $\bar{\bm \Psi}_1$ (labeled as \texttt{FLEX}), a second specification assumes $\bar{\bm \Psi}_0 = \kappa \hat{\bm \Psi}_0$ (\texttt{SSVS}), while the most restrictive specification fixes $\bm \Psi_t = \bar{\bm \Psi}$, for all $t$, to a a single variance state (\texttt{SINGLE}). In the empirical exercise, we set $\kappa = 0.1^{6}$ and $\hat{\bm \Psi}_0 = \text{diag}~(\hat{\psi}_1, \dots, \hat{\psi}_K)$ with $\hat{\psi}_i$, for, $i = \{1, \dots, K\}$, denoting ordinary least square (OLS) variances obtained from an AR($p$) model \citep[see][]{huber2019should}. Third, with regards to the state indicator matrix $\bm S_t$, we discriminate between a joint Markov-switching indicator (labeled as \texttt{MS}) and covariate-specific indicators following an independent mixture distribution (\texttt{MIX}). Recall, that for the \texttt{TVP-MIX} models $\bm S_t$ adjusts both the autoregressive parameters and the state innovation variances, while for the \texttt{TVP-POOL} and \texttt{TVP-RW} models $\bm S_t$ only controls the state innovations. In the following, we define \texttt{TVP-MIX}, \texttt{TVP-POOL} and \texttt{TVP-RW} as the \texttt{Class} of the TVP model and the combination of the acronyms for the innovation variances and indicator matrix as the \texttt{Subclass} of the specification. A single model is identified by a combination of all three acronyms. For example, a \texttt{TVP-MIX FLEX MIX} specification denotes a model with a random walk/white noise mixture for the state equation, with unrestricted two-state variances and with the elements in $\bm S_t$ following an independent mixture distribution. 
\begin{table}[!htb]
{\small
\begin{center}
\begin{tabular}{lllll}
\toprule
\multicolumn{1}{l}{\texttt{TVP-MIX}}&\multicolumn{1}{c}{$\bm \phi_t  = \bm S_t$}&\multicolumn{1}{c}{$\bm \Psi_t = $}&\multicolumn{1}{c}{$\bm S_t = $}&\multicolumn{1}{c}{Related to:} \tabularnewline
\midrule
\texttt{FLEX MS} &  & $\bm S_t \bar{\bm \Psi}_1 + (\bm I_K - \bm S_t) \bar{\bm \Psi}_0$ & $s_t \bm I_K$ &  \\
\texttt{FLEX MIX}&  & $\bm S_t \bar{\bm \Psi}_1 + (\bm I_K - \bm S_t) \bar{\bm \Psi}_0$ & $ \text{diag}~(s_{1t}, \dots, s_{Kt})$  &  \\
\texttt{SINGLE}  &  & $\bar{\bm \Psi}$ & &\\
\texttt{SSVS MIX}&  & $\bm S_t \bar{\bm \Psi}_1 + \kappa (\bm I_K - \bm S_t) \hat{\bm \Psi}_0$ & $\text{diag}~(s_{1t}, \dots, s_{Kt})$ &  \cite{chan2012time} \\
\midrule
\multicolumn{1}{l}{\texttt{TVP-POOL}}&\multicolumn{1}{c}{$\bm \phi_t  = \bm 0_{K \times K}$}&\multicolumn{1}{c}{}&\multicolumn{1}{c}{}&\multicolumn{1}{c}{}\tabularnewline
\midrule
\texttt{FLEX MS} &  & $\bm S_t \bar{\bm \Psi}_1 + (\bm I_K - \bm S_t) \bar{\bm \Psi}_0$ & $s_t \bm I_K$ &  \\
\texttt{FLEX MIX}&  & $\bm S_t \bar{\bm \Psi}_1 + (\bm I_K - \bm S_t) \bar{\bm \Psi}_0$ & $ \text{diag}~(s_{1t}, \dots, s_{Kt})$ &  \\
\texttt{SINGLE}  &  & $\bar{\bm \Psi}$ & & \cite{hhko2019} \\
\texttt{SSVS MIX}&  & $\bm S_t \bar{\bm \Psi}_1 + \kappa (\bm I_K - \bm S_t) \hat{\bm \Psi}_0$ & $\text{diag}~(s_{1t}, \dots, s_{Kt})$ &  \\
\midrule
\multicolumn{1}{l}{\texttt{TVP-RW}}&\multicolumn{1}{c}{$\bm \phi_t  = \bm I_K$}&\multicolumn{1}{c}{}&\multicolumn{1}{c}{}&\multicolumn{1}{c}{}\tabularnewline
\midrule
\texttt{FLEX MS} &  & $\bm S_t \bar{\bm \Psi}_1 + (\bm I_K - \bm S_t) \bar{\bm \Psi}_0$ & $s_t \bm I_K$  & \\
\texttt{FLEX MIX}&  & $\bm S_t \bar{\bm \Psi}_1 + (\bm I_K - \bm S_t) \bar{\bm \Psi}_0$ & $ \text{diag}~(s_{1t}, \dots, s_{Kt})$ &  \\
\texttt{SINGLE}  &  & $\bar{\bm \Psi}$ & & Standard \texttt{TVP-RW}\\
\texttt{SSVS MIX}&  & $\bm S_t \bar{\bm \Psi}_1 + \kappa (\bm I_K - \bm S_t) \hat{\bm \Psi}_0$ & $\text{diag}~(s_{1t}, \dots, s_{Kt})$  & e.g. \cite{huber2019should} \\
\bottomrule
\end{tabular}
\caption{Overview of specifications. \label{tab:spec}}\end{center}}
\end{table}

All these TVP models feature a Normal-Gamma \citep{griffin2010inference} on $\hat{\bm \alpha}$.\footnote{Note with a single-state variance ($\bm \Psi_t = \hat{\bm \Psi}$), $\hat{\bm \alpha}$ collapses to a $2K$-dimensional vector \citep[see][]{bitto_sfs2019JoE}.} We compare our methods to two constant parameter models.\footnote{A constant coefficient model can be obtained by either offsetting $\tilde{\bm \alpha} = \bm 0_{\nu \times 1}$ or setting $\bm \{\Psi_t\}_{t = 1}^{T} \approx \bm 0_{K \times K}$.} One variant features a Normal-Gamma (\texttt{const.} \texttt{(NG)}) prior, while the second variant assumes a Minnesota (\texttt{const.} \texttt{(MIN)}) prior. We consider a non-conjugate Minnesota prior, capturing the notion that own lags are more important than lags from other variables \citep{doan1984forecasting, litterman1986forecasting}. 
We estimate this set of models for three information sets (\texttt{FA-VAR}, \texttt{L-VAR} and \texttt{S-VAR}) with each featuring a different number of endogenous variables. Every considered specification features two lags and SV. 

To asses one-quarter-, one-year- and two-year-ahead predictions, we treat observations ranging from $1962$:Q$1$ to $1999$:Q$4$ as an initial sample and the periods from $2000$:Q$1$ to $2019$:Q$4$ as a hold-out sample. The initial sample is then recursively expanded until the penultimate quarter ($2019$:Q$3$) is reached.   
For each forecast comparison, a small-scale Minnesota VAR with constant parameters (\texttt{S-VAR} \texttt{const.} \texttt{(MIN)}) serves as our benchmark. 
In the following, \autoref{tab:best} shows the best performing models for point and density forecasts, being a tractable summary of  \autoref{tab:Point} and \autoref{tab:LPS}. \autoref{tab:Point} depicts root-mean squared error ratios (RMSEs) as point forecast measures and \autoref{tab:LPS} the log predictive Bayes factors (LPBFs) as density forecast metrics. The best performing models within each column are indicated by bold numbers.
In \autoref{tab:CRPS} we provide additional results on continuous rank probability score (CRPS) ratios. This alternative density forecast measure is more robust to outliers than log predictive scores \citep{gneiting2007strictly}. 
With three different measures at three different horizons we obtain a comprehensive picture to evaluate our methods jointly and marginally along the four target variables.

\begin{table}[!htb]
{\tiny
\begin{center}
\scalebox{0.9}{
\begin{tabular}{lllllllllllll}
\toprule
\multicolumn{1}{l}{\bfseries Variable}&\multicolumn{1}{c}{}& \multicolumn{3}{c}{\bfseries 1-quarter-ahead}&\multicolumn{1}{c}{}&\multicolumn{3}{c}{\bfseries 1-year-ahead}&\multicolumn{1}{c}{}&\multicolumn{3}{c}{\bfseries 2-years-ahead} \tabularnewline
\cline{3-5} \cline{7-9} \cline{11-13}
\multicolumn{1}{l}{}&\multicolumn{1}{l}{}&\multicolumn{1}{c}{Size}&\multicolumn{1}{c}{Class}&\multicolumn{1}{c}{Subclass}&\multicolumn{1}{l}{}&\multicolumn{1}{c}{Size}&\multicolumn{1}{c}{Class}&\multicolumn{1}{c}{Subclass}&\multicolumn{1}{l}{}&\multicolumn{1}{c}{Size}&\multicolumn{1}{c}{Class}&\multicolumn{1}{c}{Subclass} \tabularnewline
\midrule
\textbf{Point forecasts}& & & & & & & & & & & & \\
& & & & & & & & & & & & \\
\textbf{RMSE ratios} & & & & & & & & & & & & \\[5 pt]
TOT && \texttt{L-VAR}& \texttt{TVP-POOL} & \texttt{SINGLE} && \texttt{L-VAR}& \texttt{TVP-POOL} & \texttt{SINGLE} && \texttt{FA-VAR}& \texttt{TVP-POOL} & \texttt{SINGLE}  \\[5 pt]
GDPC1 && \texttt{L-VAR}& \texttt{TVP-POOL} & \texttt{FLEX MIX} && \texttt{FA-VAR}& \texttt{TVP-POOL} & \texttt{FLEX MS} && \texttt{FA-VAR}& \texttt{TVP-MIX} &  \texttt{FLEX MIX} \\
\shadeRow CPIAUCSL && \texttt{S-VAR}& \texttt{TVP-MIX} & \texttt{FLEX MIX} && \texttt{S-VAR}& \texttt{TVP-RW} & \texttt{FLEX MIX} && \texttt{L-VAR}& \texttt{TVP-RW}  &  \texttt{SINGLE} \\
UNRATE && \texttt{L-VAR}& \texttt{TVP-POOL} & \texttt{SSVS MIX} && \texttt{L-VAR}& \texttt{TVP-POOL} & \texttt{SINGLE} && \texttt{FA-VAR}& \texttt{TVP-POOL} & \texttt{SSVS MIX} \\
\shadeRow FEDFUNDS && \texttt{FA-VAR}& \texttt{TVP-RW} & \texttt{SSVS MIX} && \texttt{FA-VAR}& \texttt{TVP-RW} & \texttt{SSVS MIX} && \texttt{FA-VAR}& \texttt{TVP-POOL} & \texttt{SINGLE} \\
\midrule
\textbf{Density forecasts}& & & & & & & & & & & & \\
& & & & & & & & & & & & \\
\textbf{LPBFs} & & & & & & & & & & & & \\[5 pt]
TOT && \texttt{L-VAR}& \texttt{TVP-POOL} & \texttt{FLEX MIX} && \texttt{L-VAR}& \texttt{TVP-POOL} & \texttt{SSVS MIX} && \texttt{L-VAR}& \texttt{const (NG.)} &  \\[5 pt]
GDPC1 && \texttt{L-VAR}& \texttt{TVP-POOL} & \texttt{FLEX MS} && \texttt{FA-VAR}& \texttt{TVP-POOL} & \texttt{SSVS MIX} && \texttt{FA-VAR}& \texttt{TVP-POOL} & \texttt{SINGLE}  \\
\shadeRow CPIAUCSL && \texttt{S-VAR}& \texttt{TVP-MIX} & \texttt{SSVS MIX} && \texttt{L-VAR}& \texttt{const. (NG)} & && \texttt{L-VAR}& \texttt{const. (NG)}   & \\
UNRATE && \texttt{L-VAR}& \texttt{TVP-POOL} & \texttt{SSVS MIX} && \texttt{L-VAR}& \texttt{TVP-POOL} & \texttt{SSVS MIX} && \texttt{L-VAR}& \texttt{TVP-MIX}  & \texttt{SSVS MIX} \\
\shadeRow FEDFUNDS && \texttt{L-VAR}& \texttt{TVP-POOL} & \texttt{SSVS MIX} && \texttt{FA-VAR}& \texttt{TVP-RW} & \texttt{FLEX MIX} && \texttt{FA-VAR}& \texttt{const. (Min)}  & \\
\bottomrule
\end{tabular}}
\caption{Overview of the best performing models, indicated by bold numbers in \autoref{tab:Point} and \autoref{tab:LPS}. \label{tab:best}}
\end{center}}
\end{table}

\autoref{tab:best} summarizes the main findings of our forecast exercise. \textit{First}, larger-scale models (\texttt{FA-VAR}, \texttt{L-VAR}) generally outperform the small-scale specifications across horizon-variable combinations, indicating that an increasing amount of information pays off for forecasting \citep[see][]{banbura2010large}. One exception is inflation. For inflation, flexible \texttt{S-VAR}s yield more accurate forecasts than \texttt{FA-VAR}s and \texttt{L-VAR}s for one-quarter- and one-year-ahead point forecasts and one-quarter-ahead density forecasts. Comparing \texttt{FA-VAR}s with \texttt{L-VAR}s, the results are mixed. One pattern worth noting is that \texttt{L-VAR}s tend to outperform \texttt{FA-VAR}s for the one-quarter-ahead horizon while the picture reverses for higher-order forecasts. 
Second, with respect to parameter changes we see that the \texttt{TVP-POOL} specifications forecast particularly well across all horizons and target variables. These models substantially improve upon a wide range of benchmarks. Overall, \autoref{tab:best} shows that all TVP classes that provide accurate point predictions generally also perform well in terms of density forecasts.  

\begin{landscape}\begin{table}[!tbp]
{\tiny
\begin{center}
\scalebox{0.85}{
\begin{tabular}{lllclllllclllllclllll}
\toprule
\multicolumn{1}{l}{\bfseries }&\multicolumn{2}{c}{\bfseries Specification}&\multicolumn{1}{c}{\bfseries }&\multicolumn{5}{c}{\bfseries 1-quarter-ahead}&\multicolumn{1}{c}{\bfseries }&\multicolumn{5}{c}{\bfseries 1-year-ahead}&\multicolumn{1}{c}{\bfseries }&\multicolumn{5}{c}{\bfseries 2-years-ahead}\tabularnewline
\cline{2-3} \cline{5-9} \cline{11-15} \cline{17-21}
\multicolumn{1}{l}{}&\multicolumn{1}{c}{Class}&\multicolumn{1}{c}{Subclass}&\multicolumn{1}{c}{}&\multicolumn{1}{c}{TOT}&\multicolumn{1}{c}{GDPC1}&\multicolumn{1}{c}{CPIAUCSL}&\multicolumn{1}{c}{UNRATE}&\multicolumn{1}{c}{FEDFUNDS}&\multicolumn{1}{c}{}&\multicolumn{1}{c}{TOT}&\multicolumn{1}{c}{GDPC1}&\multicolumn{1}{c}{CPIAUCSL}&\multicolumn{1}{c}{UNRATE}&\multicolumn{1}{c}{FEDFUNDS}&\multicolumn{1}{c}{}&\multicolumn{1}{c}{TOT}&\multicolumn{1}{c}{GDPC1}&\multicolumn{1}{c}{CPIAUCSL}&\multicolumn{1}{c}{UNRATE}&\multicolumn{1}{c}{FEDFUNDS}\tabularnewline
\midrule
{\scshape }&&&&&&&&&&&&&&&&&&&&\tabularnewline
   ~~&   \textbf{FA-VAR}&   &   &   &   &   &   &   &   &   &   &   &   &   &   &   &   &   &   &   \tabularnewline
   ~~&   const. (Min.)&   &   &   0.91*&   0.87&   0.94&   0.79&   1.22&   &   0.92**&   0.84*&   1.03&   0.80&   1.11&   &   0.93*&   1.00&   1.04&   0.86&   0.78***\tabularnewline
   ~~&   const. (NG)&   &   &   0.88**&   0.81*&   0.95&   0.80&   1.14&   &   0.90**&   0.84**&   0.99&   0.81&   1.02&   &   0.92&   1.00&   1.01&   0.87&   0.75**\tabularnewline
   ~~&   \textbf{}&   &   &   &   &   &   &   &   &   &   &   &   &   &   &   &   &   &   &   \tabularnewline
   ~~&   TVP-MIX&   FLEX MIX&   &   0.89**&   0.80*&   0.97&   0.81&   0.91&   &   0.94*&   0.92&   0.98&   0.88&   0.98&   &   1.01&   \textbf{0.86}&   1.10*&   1.08&   0.95\tabularnewline
   ~~&   &   FLEX MS&   &   0.91**&   0.83&   0.98&   0.77&   0.89&   &   0.91&   0.89&   1.00&   0.80&   0.83*&   &   0.94&   0.96&   1.00&   0.92&   0.89\tabularnewline
   ~~&   &   SINGLE&   &   0.92*&   0.82&   1.01&   0.79&   0.85**&   &   0.91*&   0.87&   1.00&   0.81&   0.86*&   &   0.97&   0.93&   1.12&   0.94&   0.87\tabularnewline
   ~~&   &   SSVS MIX&   &   0.89**&   0.83&   0.96&   0.79&   0.95&   &   0.92*&   0.89&   0.98&   0.83&   0.98&   &   0.95&   0.93&   1.09&   0.88&   0.92\tabularnewline
   ~~&   \textbf{}&   &   &   &   &   &   &   &   &   &   &   &   &   &   &   &   &   &   &   \tabularnewline
   ~~&   TVP-POOL&   FLEX MIX&   &   0.88**&   0.79*&   0.95&   0.77&   1.06&   &   0.88**&   0.80**&   0.99&   0.77&   0.96&   &   0.89&   0.95&   1.01&   0.82&   0.73*\tabularnewline
   ~~&   &   FLEX MS&   &   0.88***&   0.79*&   0.96&   0.76&   1.05&   &   0.87**&   \textbf{0.80**}&   0.98&   0.77&   0.97&   &   0.89&   0.96*&   1.00&   0.82&   0.73*\tabularnewline
   ~~&   &   SINGLE&   &   0.87***&   0.79*&   0.95&   0.77&   1.05&   &   0.87**&   0.81**&   0.98&   0.77&   0.95&   &   \textbf{0.88}&   0.94**&   1.00&   0.82&   \textbf{0.73*}\tabularnewline
   ~~&   &   SSVS MIX&   &   0.89**&   0.81*&   0.96&   0.76&   1.06&   &   0.88**&   0.81**&   1.00&   0.77&   0.98&   &   0.89&   0.98&   1.01&   \textbf{0.82}&   0.73*\tabularnewline
   ~~&   \textbf{}&   &   &   &   &   &   &   &   &   &   &   &   &   &   &   &   &   &   &   \tabularnewline
   ~~&   TVP-RW&   FLEX MIX&   &   0.89**&   0.83&   0.95&   0.77&   0.89**&   &   0.88*&   0.86&   0.97&   0.76&   0.81**&   &   0.93&   1.00&   1.00&   0.88&   0.79\tabularnewline
   ~~&   &   FLEX MS&   &   0.89**&   0.80&   0.96&   0.80&   0.91&   &   0.90*&   0.88&   0.99&   0.79&   0.87&   &   0.95&   1.00&   1.00&   0.93&   0.83\tabularnewline
   ~~&   &   SINGLE&   &   0.91*&   0.83&   0.99&   0.81&   0.92&   &   1.04&   0.99&   1.07&   1.02&   1.23&   &   1.20&   0.90&   1.24&   1.30*&   1.36\tabularnewline
   ~~&   &   SSVS MIX&   &   0.92*&   0.84&   0.99&   0.79&   \textbf{0.83**}&   &   0.90*&   0.90&   0.96&   0.83&   \textbf{0.80*}&   &   0.96&   0.99&   1.02&   0.94&   0.83\tabularnewline
\midrule
{\scshape }&&&&&&&&&&&&&&&&&&&&\tabularnewline
   ~~&   \textbf{L-VAR}&   &   &   &   &   &   &   &   &   &   &   &   &   &   &   &   &   &   &   \tabularnewline
   ~~&   const. (Min.)&   &   &   1.02&   0.99&   1.05&   0.72&   1.52&   &   0.92**&   0.93&   0.95*&   0.79&   1.00&   &   0.97**&   1.02&   0.98&   0.95&   0.88**\tabularnewline
   ~~&   const. (NG)&   &   &   0.91**&   0.85&   0.96&   0.71&   1.10&   &   0.88***&   0.89*&   0.92**&   0.74&   0.98&   &   0.92**&   1.00&   0.97&   0.89&   0.79**\tabularnewline
   ~~&   \textbf{}&   &   &   &   &   &   &   &   &   &   &   &   &   &   &   &   &   &   &   \tabularnewline
   ~~&   TVP-MIX&   FLEX MIX&   &   1.04&   0.90&   1.14&   0.71&   2.19&   &   0.91*&   0.82&   1.02&   0.81&   1.09&   &   0.92&   0.93&   0.99&   0.89&   0.90\tabularnewline
   ~~&   &   FLEX MS&   &   1.06&   0.92&   1.17&   0.72&   1.51&   &   0.90*&   0.86&   0.95&   0.79&   1.08&   &   1.22&   1.56&   1.30&   0.90&   1.13\tabularnewline
   ~~&   &   SINGLE&   &   0.92*&   0.86&   0.98&   0.74&   0.90**&   &   0.87*&   0.82*&   0.94&   0.83&   0.87*&   &   0.92&   0.94&   0.97&   0.91&   0.86\tabularnewline
   ~~&   &   SSVS MIX&   &   0.96&   0.98&   0.94&   0.72&   1.20&   &   0.90&   0.86&   0.98&   0.78&   1.00&   &   0.95&   0.96&   1.06&   0.91&   0.89\tabularnewline
   ~~&   \textbf{}&   &   &   &   &   &   &   &   &   &   &   &   &   &   &   &   &   &   &   \tabularnewline
   ~~&   TVP-POOL&   FLEX MIX&   &   0.88***&   \textbf{0.79**}&   0.96&   0.68&   0.86***&   &   0.87**&   0.87&   0.94&   0.74&   0.89**&   &   0.91&   0.99&   1.01&   0.84&   0.79**\tabularnewline
   ~~&   &   FLEX MS&   &   0.88***&   0.82*&   0.94&   0.67&   0.87***&   &   0.86**&   0.85*&   0.94&   0.72&   0.90**&   &   0.90&   0.99&   0.99&   0.83&   0.77**\tabularnewline
   ~~&   &   SINGLE&   &   \textbf{0.87***}&   0.79**&   0.95&   0.68*&   0.88***&   &   \textbf{0.86**}&   0.86*&   0.93&   \textbf{0.71}&   0.88**&   &   0.91&   0.99&   0.99&   0.84&   0.78**\tabularnewline
   ~~&   &   SSVS MIX&   &   0.88***&   0.81*&   0.95&   \textbf{0.67*}&   0.89***&   &   0.86**&   0.85*&   0.94&   0.72&   0.88**&   &   0.90&   0.98&   1.01&   0.83&   0.78**\tabularnewline
   ~~&   \textbf{}&   &   &   &   &   &   &   &   &   &   &   &   &   &   &   &   &   &   &   \tabularnewline
   ~~&   TVP-RW&   FLEX MIX&   &   0.96&   0.97&   0.97&   0.71&   1.09&   &   0.88*&   0.85&   0.94&   0.80&   0.93*&   &   0.95&   1.00&   0.97&   0.94&   0.87\tabularnewline
   ~~&   &   FLEX MS&   &   0.98&   0.97&   0.99&   0.73&   0.93**&   &   0.86*&   0.81*&   0.92&   0.84&   0.96&   &   0.96&   0.97&   0.96&   0.96&   0.93\tabularnewline
   ~~&   &   SINGLE&   &   1.24&   1.09&   1.37&   0.75&   1.75&   &   0.93&   0.88&   0.97***&   0.84&   1.23&   &   0.94&   0.91&   \textbf{0.93}&   0.96&   0.95\tabularnewline
   ~~&   &   SSVS MIX&   &   0.94&   0.93&   0.95&   0.73&   1.17&   &   0.88*&   0.85&   0.93&   0.82&   0.97&   &   0.96&   0.95&   0.98&   0.95&   0.95\tabularnewline
\midrule
{\scshape }&&&&&&&&&&&&&&&&&&&&\tabularnewline
   ~~&   \textbf{S-VAR}&   &   &   &   &   &   &   &   &   &   &   &   &   &   &   &   &   &   &   \tabularnewline
\shadeBench   ~~&   const. (Min.)&   &   &   0.60&   0.83&   0.85&   0.15&   0.11&   &   0.76&   1.00&   0.87&   0.62&   0.41&   &   0.91&   0.93&   0.85&   1.10&   0.73\tabularnewline
   ~~&   const. (NG)&   &   &   0.99**&   0.99*&   0.99&   1.00&   1.01&   &   0.96**&   0.94*&   0.98&   0.98*&   0.98&   &   0.97&   0.99&   0.98&   0.97**&   0.95\tabularnewline
   ~~&   \textbf{}&   &   &   &   &   &   &   &   &   &   &   &   &   &   &   &   &   &   &   \tabularnewline
   ~~&   TVP-MIX&   FLEX MIX&   &   0.93**&   0.94*&   \textbf{0.92}&   0.92&   0.91***&   &   0.89&   0.84&   0.94&   0.92&   0.93&   &   0.93&   0.89&   1.01&   0.93&   0.92\tabularnewline
   ~~&   &   FLEX MS&   &   0.96&   0.96&   0.96&   0.91&   0.87&   &   0.93&   0.92&   0.95&   0.91&   0.96&   &   0.96&   0.97&   1.01&   0.92&   0.95\tabularnewline
   ~~&   &   SINGLE&   &   0.97&   0.97&   0.96&   0.97&   0.85*&   &   0.93&   0.91&   0.95&   0.95&   0.97&   &   0.95&   0.96&   0.99&   0.94&   0.93\tabularnewline
   ~~&   &   SSVS MIX&   &   0.95**&   0.96&   0.94**&   0.98&   0.94***&   &   0.91*&   0.86&   0.94&   0.93&   0.92&   &   0.95&   0.94&   1.00&   0.94&   0.92\tabularnewline
   ~~&   \textbf{}&   &   &   &   &   &   &   &   &   &   &   &   &   &   &   &   &   &   &   \tabularnewline
   ~~&   TVP-POOL&   FLEX MIX&   &   0.98***&   0.97**&   0.99&   0.95&   0.98&   &   0.95**&   0.93*&   0.98&   0.94&   0.94&   &   0.95&   0.99&   0.96&   0.93&   0.90\tabularnewline
   ~~&   &   FLEX MS&   &   0.98**&   0.98*&   0.98&   0.94**&   0.98*&   &   0.95**&   0.94*&   0.98&   0.94&   0.94&   &   0.95&   0.99&   0.97&   0.93&   0.90*\tabularnewline
   ~~&   &   SINGLE&   &   0.98**&   0.98*&   0.99&   0.94&   0.98&   &   0.95**&   0.93*&   0.99&   0.94&   0.94&   &   0.95&   0.99&   0.97&   0.93&   0.90\tabularnewline
   ~~&   &   SSVS MIX&   &   0.98**&   0.98*&   0.99&   0.95&   0.97*&   &   0.95**&   0.93*&   0.98*&   0.95&   0.94&   &   0.94&   0.98&   0.97&   0.92&   0.91\tabularnewline
   ~~&   \textbf{}&   &   &   &   &   &   &   &   &   &   &   &   &   &   &   &   &   &   &   \tabularnewline
   ~~&   TVP-RW&   FLEX MIX&   &   0.96**&   0.96&   0.95&   0.97&   0.92**&   &   0.93*&   0.94&   \textbf{0.91}&   0.95&   0.94&   &   0.96&   0.98&   0.96&   0.95&   0.94\tabularnewline
   ~~&   &   FLEX MS&   &   0.98&   0.99&   0.97&   0.98&   0.88*&   &   0.93&   0.91&   0.93&   0.96&   0.98&   &   0.95&   0.94&   0.99&   0.95&   0.94\tabularnewline
   ~~&   &   SINGLE&   &   0.96&   0.92&   1.00&   0.92&   0.86***&   &   0.89&   0.82&   0.95&   0.94&   0.90&   &   0.93&   0.88&   1.00&   0.96&   0.85\tabularnewline
   ~~&   &   SSVS MIX&   &   0.98&   1.00&   0.96&   0.98&   0.89*&   &   0.94&   0.92&   0.93&   0.99&   0.97&   &   0.97&   0.96&   0.97&   0.97&   0.95\tabularnewline
\bottomrule
\end{tabular}}
\caption{Point forecast performance (RMSE ratios) relative to the benchmark (\texttt{const (Min.)}). The red shaded row denotes the benchmark (and its RMSE values). Asterisks indicate statistical significance for each model relative to \texttt{const (Min.)} at the $1$ ($^{***}$), $5$ ($^{**}$) and $10$ ($^{*}$) percent significance levels. \label{tab:Point}}\end{center}}
\end{table}\end{landscape}

\begin{landscape}\begin{table}[!tbp]
{\tiny
\begin{center}
\scalebox{0.85}{
\begin{tabular}{lllclllllclllllclllll}
\toprule
\multicolumn{1}{l}{\bfseries }&\multicolumn{2}{c}{\bfseries Specification}&\multicolumn{1}{c}{\bfseries }&\multicolumn{5}{c}{\bfseries 1-quarter-ahead}&\multicolumn{1}{c}{\bfseries }&\multicolumn{5}{c}{\bfseries 1-year-ahead}&\multicolumn{1}{c}{\bfseries }&\multicolumn{5}{c}{\bfseries 2-years-ahead}\tabularnewline
\cline{2-3} \cline{5-9} \cline{11-15} \cline{17-21}
\multicolumn{1}{l}{}&\multicolumn{1}{c}{Class}&\multicolumn{1}{c}{Subclass}&\multicolumn{1}{c}{}&\multicolumn{1}{c}{TOT}&\multicolumn{1}{c}{GDPC1}&\multicolumn{1}{c}{CPIAUCSL}&\multicolumn{1}{c}{UNRATE}&\multicolumn{1}{c}{FEDFUNDS}&\multicolumn{1}{c}{}&\multicolumn{1}{c}{TOT}&\multicolumn{1}{c}{GDPC1}&\multicolumn{1}{c}{CPIAUCSL}&\multicolumn{1}{c}{UNRATE}&\multicolumn{1}{c}{FEDFUNDS}&\multicolumn{1}{c}{}&\multicolumn{1}{c}{TOT}&\multicolumn{1}{c}{GDPC1}&\multicolumn{1}{c}{CPIAUCSL}&\multicolumn{1}{c}{UNRATE}&\multicolumn{1}{c}{FEDFUNDS}\tabularnewline
\midrule
{\scshape }&&&&&&&&&&&&&&&&&&&&\tabularnewline
   ~~&   \textbf{FA-VAR}&   &   &   &   &   &   &   &   &   &   &   &   &   &   &   &   &   &   &   \tabularnewline
   ~~&   const. (Min.)&   &   &    11.35&    12.01**&     2.23&    8.56&   -4.19&   &     14.59&     8.30**&     2.02&    14.03&     8.69&   &     37.98**&     2.12&     5.63&     12.49&   \textbf{30.86***}\tabularnewline
   ~~&   const. (NG)&   &   &    18.88&    13.48***&     4.19&    8.66&    0.46&   &     25.90***&     9.01***&     2.34&    14.21&    10.93&   &     40.82***&     3.08&     7.01&     12.22&    28.66***\tabularnewline
   ~~&   \textbf{}&   &   &   &   &   &   &   &   &   &   &   &   &   &   &   &   &   &   &   \tabularnewline
   ~~&   TVP-MIX&   FLEX MIX&   &    23.23&    12.71**&     2.78&   11.15&   12.95***&   &     39.12**&     7.60&     2.33&    26.49&    16.00&   &     50.99&     3.80&     3.97&     23.44&    13.09\tabularnewline
   ~~&   &   FLEX MS&   &    29.37&    11.74**&     5.32&   13.84&    9.59*&   &     49.54**&     7.49&     2.32&    27.51&    17.64&   &     41.96&     0.10&     3.65&     22.67&    13.22\tabularnewline
   ~~&   &   SINGLE&   &    24.87&    11.07**&     2.24&   10.12&   17.01***&   &     52.21**&     6.25&     0.25&    25.16&    24.39&   &     55.19&     0.70&     3.17&     22.70&    18.79\tabularnewline
   ~~&   &   SSVS MIX&   &    25.93&    13.23**&     2.59&   10.60&    6.38*&   &     40.29*&     7.92*&     1.90&    23.53&    10.64&   &     40.57&     3.38&     1.82&     25.83&     7.45\tabularnewline
   ~~&   \textbf{}&   &   &   &   &   &   &   &   &   &   &   &   &   &   &   &   &   &   &   \tabularnewline
   ~~&   TVP-POOL&   FLEX MIX&   &    26.23&    13.49***&     3.44&   10.51&    7.28***&   &     39.69***&    10.63**&     2.06&    19.71&    15.86&   &     53.01**&     3.61&     6.18&     22.98&    28.67*\tabularnewline
   ~~&   &   FLEX MS&   &    28.06&    14.49***&     3.98*&   10.24&    7.62***&   &     34.75**&    10.85***&     2.71&    17.29&    14.71&   &     52.55**&     4.07&     6.27&     21.13&    28.34*\tabularnewline
   ~~&   &   SINGLE&   &    25.63&    14.42***&     2.98&   10.61&    6.20***&   &     34.67**&    10.54***&     2.21&    18.32&    16.06&   &     46.04**&   \textbf{4.36}&     6.22&     19.52&    26.47\tabularnewline
   ~~&   &   SSVS MIX&   &    25.54&    14.08***&     3.17*&   10.06&    6.84***&   &     35.28**&   \textbf{11.02***}&     1.59&    18.01&    15.29&   &     50.57**&     4.04&     6.74&     23.07&    28.09\tabularnewline
   ~~&   \textbf{}&   &   &   &   &   &   &   &   &   &   &   &   &   &   &   &   &   &   &   \tabularnewline
   ~~&   TVP-RW&   FLEX MIX&   &    37.26&    11.40**&     7.35**&   12.30&   17.50***&   &     63.22**&     7.80&     1.93&    29.72&   \textbf{27.66}&   &     69.42*&    -1.76&     5.94&     27.99&    29.14\tabularnewline
   ~~&   &   FLEX MS&   &    26.44&    11.88***&     5.01&   11.44&   11.10**&   &     53.18**&     7.23&     0.70&    24.70&    20.65&   &     38.45&    -0.47&     4.42&     21.35&    18.47\tabularnewline
   ~~&   &   SINGLE&   &    20.52&    11.81**&    -2.21&   11.01&   13.10***&   &     46.68**&     4.46&     2.24&    19.20&    17.68&   &     45.86&    -1.09&    -0.24&     13.83&    16.93\tabularnewline
   ~~&   &   SSVS MIX&   &    38.78&    11.12**&     6.46&    9.69&   16.94***&   &     47.25***&     6.07&     0.96&    21.05&    27.08&   &     47.06*&    -1.64&     5.28&     17.38&    25.21\tabularnewline
\midrule
{\scshape }&&&&&&&&&&&&&&&&&&&&\tabularnewline
   ~~&   \textbf{L-VAR}&   &   &   &   &   &   &   &   &   &   &   &   &   &   &   &   &   &   &   \tabularnewline
   ~~&   const. (Min.)&   &   &    15.05&    13.32*&    -0.29&   12.15&   -3.68***&   &     56.73**&     2.78&     3.22***&    32.41&    11.78***&   &     61.59***&    -3.86&     3.18&     15.52&    18.28**\tabularnewline
   ~~&   const. (NG)&   &   &    28.70&    15.97***&     1.06&   14.87&    2.34&   &     70.27***&     7.40**&   \textbf{6.76***}&    36.48&    16.42*&   &   \textbf{73.50***}&     0.33&   \textbf{8.80*}&     17.46**&    21.68***\tabularnewline
   ~~&   \textbf{}&   &   &   &   &   &   &   &   &   &   &   &   &   &   &   &   &   &   &   \tabularnewline
   ~~&   TVP-MIX&   FLEX MIX&   &    21.50&    13.39**&    -3.42&   14.97&    6.06&   &     57.99*&     7.63&    -0.49&    36.44&     6.38&   &     50.44&    -3.40&    -0.52&     27.14&    -1.18\tabularnewline
   ~~&   &   FLEX MS&   &    13.65*&    10.32*&    -0.35&   13.39&    7.88&   &     46.02&     6.19&    -0.16&    32.55&     7.15&   &     44.48&    -6.69&    -0.73&     28.69&     0.75\tabularnewline
   ~~&   &   SINGLE&   &    28.03&    12.75**&     1.04&   13.65&   12.62&   &     60.71*&     8.02&     2.15&    32.82&    12.23&   &     60.21&    -3.28&     1.74&     28.66&     7.41\tabularnewline
   ~~&   &   SSVS MIX&   &    16.40*&    11.52**&     1.00&   12.90&   11.67&   &     55.25*&     8.09&    -0.77&    34.05&     9.97&   &     41.64&    -0.87&     1.23&   \textbf{31.10}&     1.53\tabularnewline
   ~~&   \textbf{}&   &   &   &   &   &   &   &   &   &   &   &   &   &   &   &   &   &   &   \tabularnewline
   ~~&   TVP-POOL&   FLEX MIX&   &   \textbf{54.32}&    16.63***&     0.86&   16.78&   22.62***&   &     71.86***&     8.08*&     3.09&    35.86&    24.24&   &     63.16***&     0.27&     2.73&     22.08&    21.52**\tabularnewline
   ~~&   &   FLEX MS&   &    50.88&   \textbf{17.28***}&     1.72&   16.95&   21.29***&   &     70.43**&     8.02**&     2.27&    36.55&    24.52&   &     65.07***&    -0.06&     2.43&     20.10&    22.50**\tabularnewline
   ~~&   &   SINGLE&   &    51.15&    16.22***&     1.44&   16.75&   22.24***&   &     72.35***&     8.14*&     2.47&    36.30&    24.31&   &     60.97***&    -0.50&     2.74&     24.76&    20.51\tabularnewline
   ~~&   &   SSVS MIX&   &    49.90&    16.39***&    -0.50&   \textbf{17.15}&   \textbf{23.32***}&   &   \textbf{73.08**}&     6.80*&     2.06&   \textbf{36.87}&    25.34&   &     63.73***&    -1.29&     2.69&     19.82&    22.33*\tabularnewline
   ~~&   \textbf{}&   &   &   &   &   &   &   &   &   &   &   &   &   &   &   &   &   &   &   \tabularnewline
   ~~&   TVP-RW&   FLEX MIX&   &    29.38&    10.54**&     1.22&   11.09&   12.63&   &     54.24&     3.78&     1.28&    27.88&    17.50&   &     59.54&    -7.53&    -0.86&     28.15&    15.32\tabularnewline
   ~~&   &   FLEX MS&   &    15.46&     9.99**&    -8.43&   11.68&   13.28&   &     55.15*&     5.85&     1.40&    28.01&    13.82&   &     41.09&    -5.25&    -0.64&     25.89&     6.06\tabularnewline
   ~~&   &   SINGLE&   &    12.39&    12.04**&    -1.38&   12.34&    7.23&   &     41.96&     4.35&     0.38&    28.51&    -2.50&   &     28.63&    -9.08*&     0.76&     20.03&    -6.91\tabularnewline
   ~~&   &   SSVS MIX&   &    18.47&     9.93*&     0.81&   11.50&   13.63&   &     55.11&     4.67&     2.12&    27.41&    13.44&   &     46.83&    -3.83&     1.76&     23.01&     6.32\tabularnewline
\midrule
{\scshape }&&&&&&&&&&&&&&&&&&&&\tabularnewline
   ~~&   \textbf{S-VAR}&   &   &   &   &   &   &   &   &   &   &   &   &   &   &   &   &   &   &   \tabularnewline
\shadeBench   ~~&   const. (Min.)&   &   &   -22.11&   -82.64&   -80.74&   45.02&   86.64&   &   -256.48&   -97.26&   -83.70&   -69.04&   -29.94&   &   -383.84&   -92.27&   -89.84&   -125.53&   -87.85\tabularnewline
   ~~&   const. (NG)&   &   &     4.55&     1.97*&     0.71&    0.62*&    1.39***&   &      4.97&     2.64*&     1.95*&     0.03&     2.30&   &      9.15&     0.58&     3.07&      2.99&     1.37\tabularnewline
   ~~&   \textbf{}&   &   &   &   &   &   &   &   &   &   &   &   &   &   &   &   &   &   &   \tabularnewline
   ~~&   TVP-MIX&   FLEX MIX&   &    24.41&     3.97*&     5.52&    2.90&   12.18***&   &     15.60&     9.64**&     3.98&     8.50&     5.03&   &     -3.76&     3.00&     3.74&     -0.02&    -7.71\tabularnewline
   ~~&   &   FLEX MS&   &    18.00&     2.66&     3.57&    6.42&    5.86**&   &      9.48&     5.25*&    -0.44&    12.27&    -0.22&   &      0.77&    -1.60&     0.83&     10.35&   -11.26\tabularnewline
   ~~&   &   SINGLE&   &     7.18**&     2.45&    -1.98&    0.85&   13.26***&   &     12.51&     5.42&     3.33&     6.38&     7.12*&   &     -0.80&    -2.74&     1.19&      2.14&    -5.03\tabularnewline
   ~~&   &   SSVS MIX&   &    23.21&     3.44*&   \textbf{8.67**}&    2.67&    7.84***&   &     22.03&     6.18&     1.77&    10.22&     5.15&   &     15.68&     1.32&     6.23&     14.46&    -3.63\tabularnewline
   ~~&   \textbf{}&   &   &   &   &   &   &   &   &   &   &   &   &   &   &   &   &   &   &   \tabularnewline
   ~~&   TVP-POOL&   FLEX MIX&   &    16.39&     2.47***&     2.63**&    1.29&    6.18***&   &      9.34&     3.10*&     1.70&     3.87*&     5.15&   &     14.86&     0.07&     3.60*&      6.37&     0.56\tabularnewline
   ~~&   &   FLEX MS&   &    14.35&     2.30**&     1.26*&    1.66&    7.91***&   &     11.74&     3.65***&     2.04**&     7.55*&     6.02&   &     14.99&    -0.16&     4.04**&      8.28&     1.86\tabularnewline
   ~~&   &   SINGLE&   &    16.46&     2.68***&     2.61*&    1.84&    6.99***&   &      5.61&     3.52**&     1.73&     3.43&     5.59&   &     14.77&     0.21&     3.13&     10.10&    -0.22\tabularnewline
   ~~&   &   SSVS MIX&   &    18.75&     2.40**&     2.84*&    1.84&    7.03***&   &     11.67&     3.66**&     2.70***&     4.44**&     5.74&   &     12.63&     0.51&     3.68**&      7.64&     0.59\tabularnewline
   ~~&   \textbf{}&   &   &   &   &   &   &   &   &   &   &   &   &   &   &   &   &   &   &   \tabularnewline
   ~~&   TVP-RW&   FLEX MIX&   &    22.15&     2.35&     5.20***&    2.14&   11.56***&   &     31.13*&     4.67&     6.61*&    10.96&     9.42&   &     23.66&    -2.07&     7.42&     11.93&     2.23\tabularnewline
   ~~&   &   FLEX MS&   &    17.02&     1.76&     2.62&    1.98&    8.41**&   &     10.37&     4.80&     1.87&     8.30&     2.86&   &      7.89&    -2.64&     4.15&      3.79&    -7.05\tabularnewline
   ~~&   &   SINGLE&   &    13.68&     6.02**&    -3.98&    1.79&   12.17***&   &      6.72&     8.60&     1.15&     1.63&    13.13&   &    -11.49&     2.34&     0.87&     -7.78&     3.09\tabularnewline
   ~~&   &   SSVS MIX&   &    20.25&     1.26&     5.76*&    0.31&   12.10***&   &     28.93*&     5.73&     5.24&     6.34&     9.14&   &     16.00&    -2.42&     6.40&      4.74&    -2.84\tabularnewline
\bottomrule
\end{tabular}}
\caption{Density forecast performance (LPBFs) relative to the benchmark (\texttt{const (Min.)}). The red shaded row denotes the benchmark (and its LPS values). Asterisks indicate statistical significance for each model relative to \texttt{const (Min.)} at the $1$ ($^{***}$), $5$ ($^{**}$) and $10$ ($^{*}$) percent significance levels.
 \label{tab:LPS}}\end{center}}
\end{table}\end{landscape}

When examining \autoref{tab:Point} and \autoref{tab:LPS} in greater detail, note that a large number of models shown in the tables outperform the Minnesota benchmark in terms of RMSEs (indicated by ratios below one) and in terms of LPBFs (indicated by values above zero). However, the benchmark is a tough competitor when predicting inflation and for higher-order point forecasts. In particular, the result that inflation is hard to predict is a well-known fact in the literature \citep{stock2007has}. Commonly, it is found here that more sophisticated models with large information sets are outperformed by rather small-sized and simple models \cite[see, for example,][]{giannone2015prior}. 

When focussing on the differences occuring through the varying treatment of parameter evolutions, our proposed methods, the \texttt{TVP-POOL} and \texttt{TVP-MIX} specifications, show that their good performance is mainly driven by improved forecast accuracy for output growth and unemployment. In terms of the innovation variance assumption for these specifications, we observe that additional flexibility tends to improve density forecasts performance and yields accurate point forecasts. For the \texttt{TVP-POOL} models this higher degree of flexibility generally pays off across variables and model sizes. For flexible \texttt{TVP-MIX} specifications, forecast ability tends to improve for \texttt{S-VAR}s and \texttt{FA-VARs} and is competitive for the \texttt{L-VAR}s. Especially the \texttt{TVP-MIX SSVS MIX} and \texttt{TVP-MIX FLEX MIX} models using a small information set yield quite accurate inflation forecasts, being the best performing models for the one-quarter-ahead horizon. Across variables, a notable exceptions is the interest rate for \texttt{TVP-MIX} models. Here, a \texttt{TVP-MIX SINGLE} specification is superior to models assuming a mixture on innovation volatilities. 

When assessing random walk state equation (\texttt{TVP-RW}) specifications across the information sets, two things are worth noting. First, a standard TVP model with random  walk assumption \texttt{TVP-RW SINGLE} is only competitive for one-year- and two-year-ahead forecasts and otherwise forecasts poorly. Second, more flexible \texttt{TVP-RW} variants produce quite accurate forecasts for \texttt{FA-VAR}s. Constant parameter models with a Normal-Gamma prior show reasonable forecasts for \texttt{L-VAR}s (especially for inflation), but lack flexibility in smaller-scale models. This observation is in line with the fact that in larger-scale model, time-variation in coefficients vanishes \citep[see][]{hkp2020}. However, few parameter instabilities might still be present since, apart from some exceptions, our methods provide improvements when compared to constant coefficient models. 
 

\begin{figure}
\begin{minipage}{\textwidth}
\centering
(a) \texttt{FA-VAR}
\hspace{5pt}
\end{minipage}
\begin{minipage}{\textwidth}
\centering
\includegraphics[scale=.54]{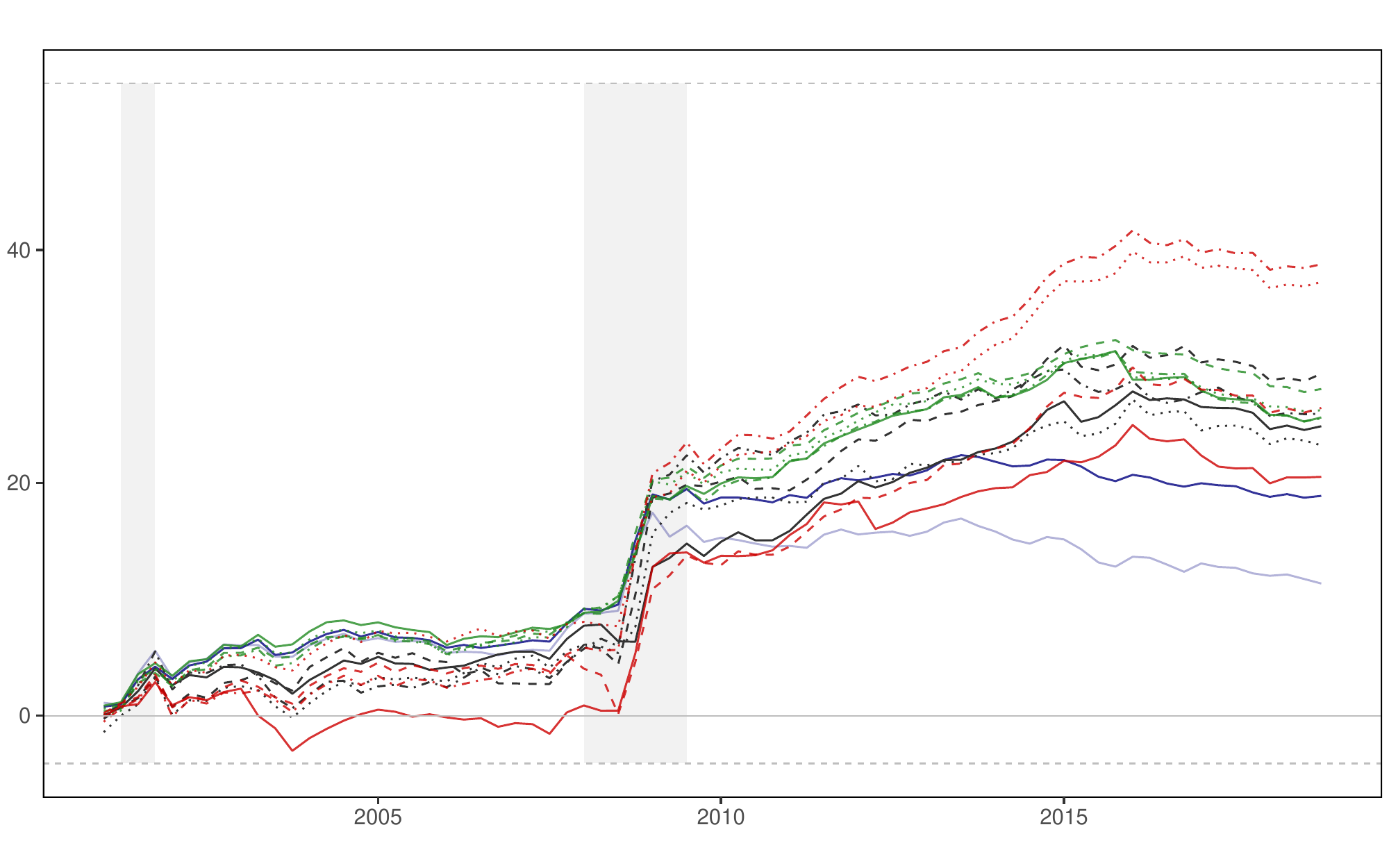}
\end{minipage}
\begin{minipage}{\textwidth}
\centering
(b) \texttt{L-VAR} 
\hspace{5pt}
\end{minipage}
\begin{minipage}{\textwidth}
\centering
\includegraphics[scale=.54]{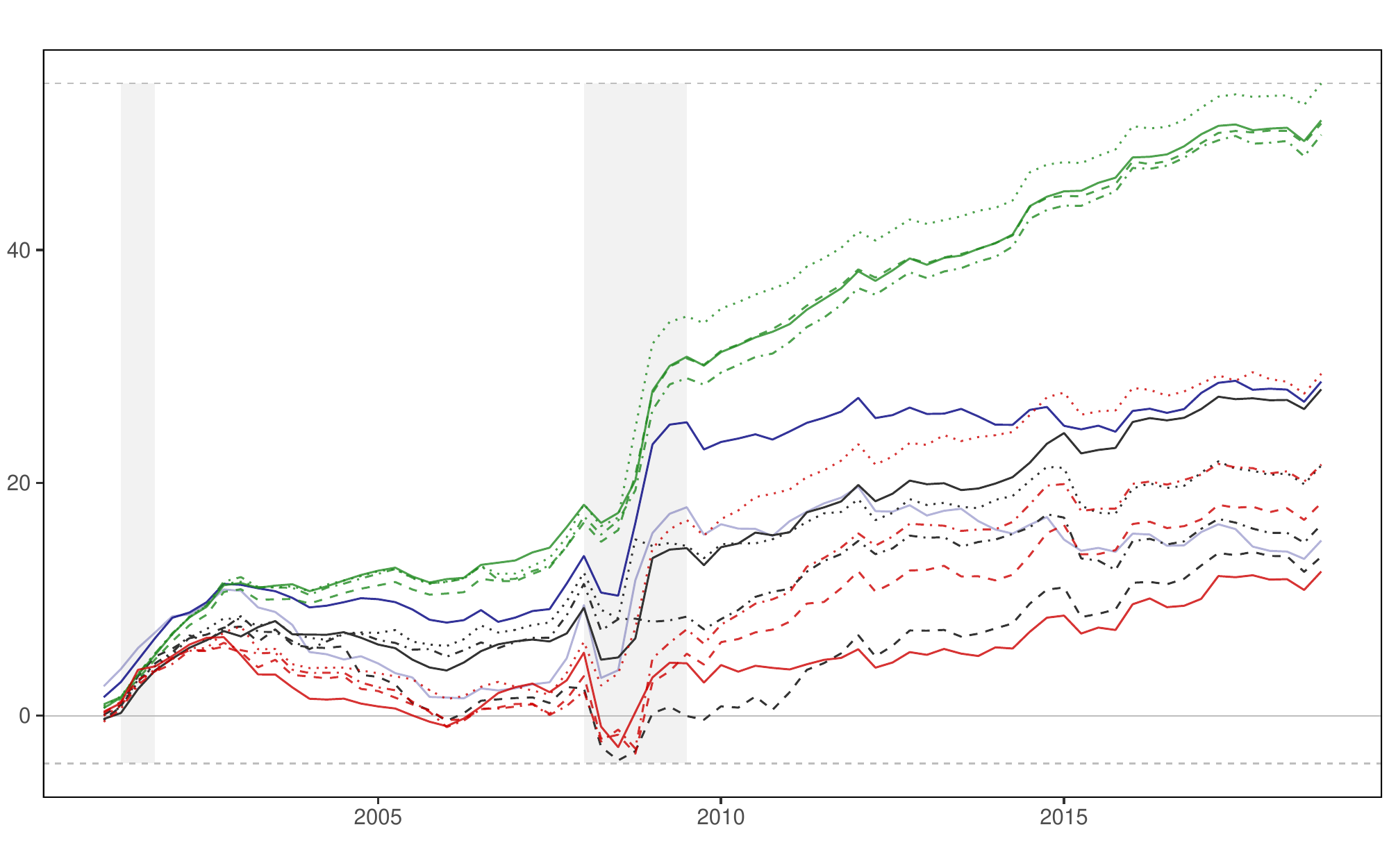}
\end{minipage}
\begin{minipage}{\textwidth}
\centering
(c) \texttt{S-VAR}
\hspace{5pt}
\end{minipage}
\begin{minipage}{\textwidth}
\centering
\includegraphics[scale=.54]{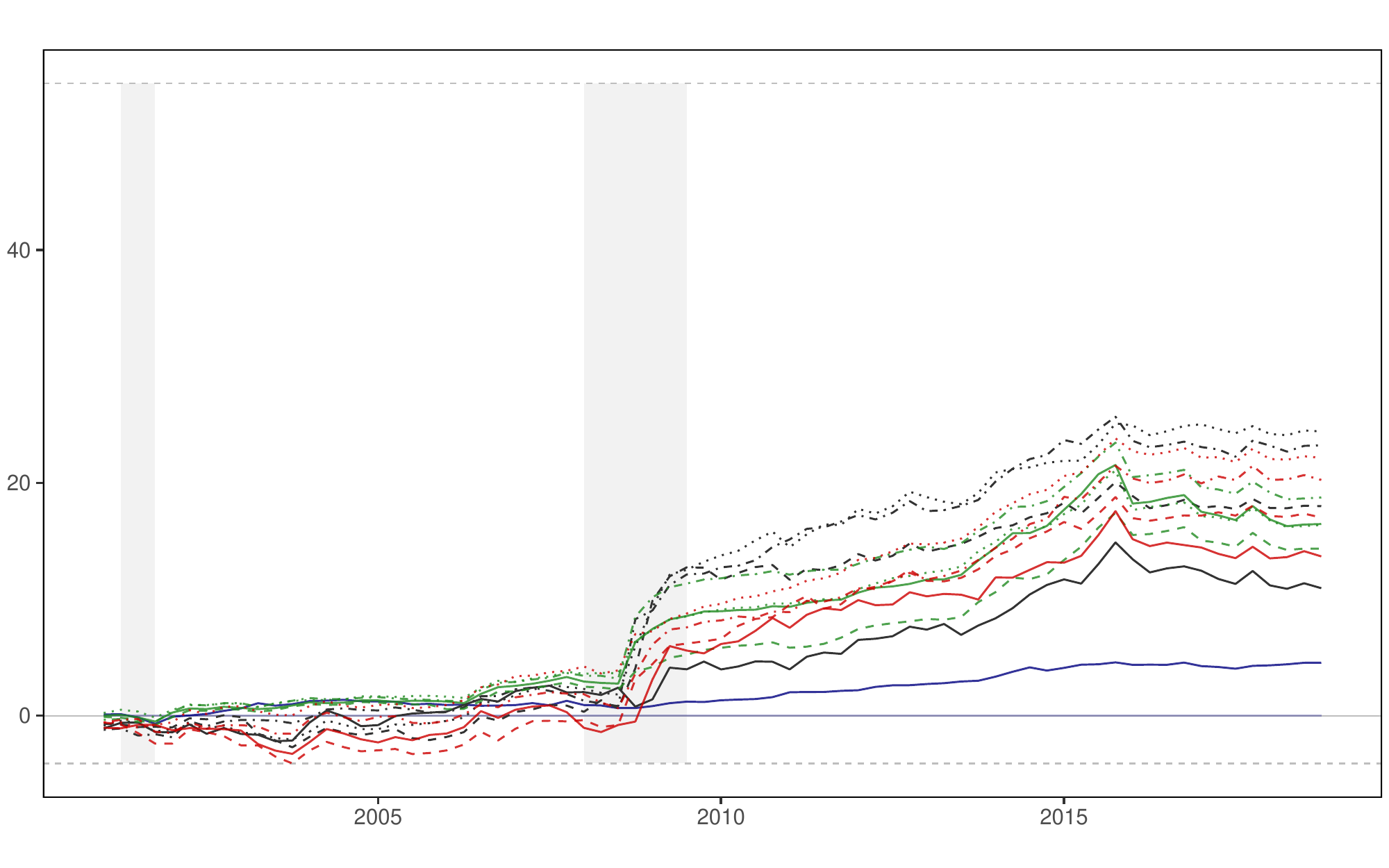}
\end{minipage}
\begin{minipage}{\textwidth}
\vspace{-30pt}
\centering
\includegraphics[scale=.4]{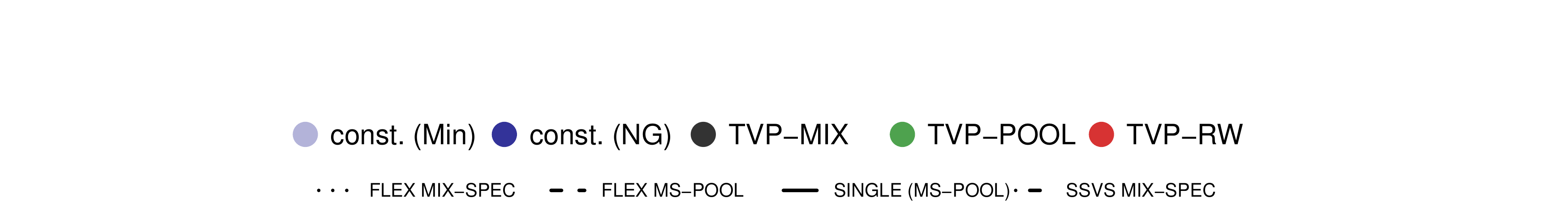}
\end{minipage}\vfill
\caption{Evolution of one-quarter-ahead total cumulative LPBFs relative to the benchmark. The gray dashed
lines refer to the maximum/minimum Bayes factor over the full hold-out sample. The light gray shaded areas indicate the NBER recessions in the US.\label{fig:lps1}}
\end{figure}

To illustrate the forecast performance over time, \autoref{fig:lps1} depicts the evolution of cumulated joint LPBFs relative to our benchmark. 
Overall we find that for all four target variable jointly our proposed methods never forecast poorly. Both \texttt{TVP-MIX} (black lines) and \texttt{TVP-POOL} (green lines) methods outperform a standard TVP model with random walk state equation (\texttt{TVP-RW SINGLE} denoted by the red solid line) across information sets (one exception is the \texttt{TVP-MIX SINGLE} model for the \texttt{S-VAR}). Moreover, allowing for occasional parameter changes during and in the aftermath of financial crisis tends to increase predictive ability.  

A few points are worth discussing in greater detail. First, at the beginning of the hold-out sample is characterized by the early 2000s recession. Although this was a quite short crisis, it already leads to a quite diverse model performance across information sets. During this episode and its consecutive three years, \texttt{L-VAR}s strictly dominate the other two information sets (\texttt{FA-VAR}s and \texttt{S-VAR}s). This implies, that for any TVP evolution assumption, the large-scale model outperforms its smaller-scale counterparts. Moreover, during the financial crisis, we observe a substantial increase in LPBFs for a wide range of \texttt{FA-VAR}s  and \texttt{L-VAR}s, while for \texttt{S-VAR}s we see similar improvements solely for some TVP-VARs. This feature might indicate that TVPs are capable of mitigating a potential omitted variable bias \citep{hkp2020}.

Second, within each information set, performance across parameter evolution assumptions is mixed. Evidently, performance of the four \texttt{L-VAR}s featuring a \texttt{TVP-POOL} specification stands out (depicted by green lines). In more tranquil periods they show constant improvements and yield substantial predictive gains during the financial crisis. Especially in the aftermath of the Great Recession, the LPBFs steadily increase compared to other large TVP-VARs. This episode also includes a time characterized by a (sluggish) recovery of the US economy after the financial crisis and the interest rate hitting the zero lower bound. With monetary policy shifting towards unconventional measures it might not only pay off to include financial market variables, such as longer-term yields, but also to allow for occasional changes in transmission channels in these variables. Moreover, it is worth noting that the four \texttt{TVP-POOL} variants tend to perform almost identical until 2010, while afterwards slight performance differences become evident. \cite{hhko2019} have made a similar observation, when varying the  hyperparameters on their conjugate prior of the state equation. 

Third, \texttt{TVP-MIX} methods generally forecast well for \texttt{S-VARs} and \texttt{FA-VARs}, while for \texttt{L-VARs} only the \texttt{TVP-MIX SINGLE} specification yields substantial gains. In particular for \texttt{FA-VARs} and \texttt{S-VAR}s, a flexible variance modelling (\texttt{TVP-MIX FLEX MIX} and \texttt{TVP-MIX SSVS MIX}) generally pays off. For \texttt{L-VAR}s these two models also yield reasonable forecast accuracy, while the \texttt{TVP-MIX MS} forfeits forecast accuracy. Thus, for a large-scale model, the assumption that a joint indicator governs the evolution of large number of coefficients might be less appropriate. Moreover, comparing \texttt{TVP-MIX} with \texttt{TVP-RW} specifications reveals that the random walk/white noise mixture yields gains in tranquil periods for larger-scale models (\texttt{FA-VAR}s and \texttt{L-VAR}s) and does particularly well in recessions for the small information set (\texttt{S-VAR}). Especially for small-scale VARs the \texttt{TVP-MIX} variants, featuring a mixture distribution on the state innovation volatilities, greatly improve predictive performance relative to \texttt{TVP-RW} models during the financial crisis. 
Moreover, it is worth noting that the \texttt{TVP-RW SINGLE} model forecasts poorly for larger-scale models (\texttt{FA-VAR}s and \texttt{L-VAR}s) during tranquil periods previous to the financial crisis, while performance slightly recovers in the middle of the Great Recession. A plausible explanation for this pattern might be that spurious movements in coefficients lead to overfitting, widening the predictive density of the \texttt{TVP-RW SINGLE} model. This feature is harming predictive accuracy in stable times, while it is to some extent helpful in times of turmoil (periods characterized by large outliers).
In contrast to the \texttt{TVP-RW SINGLE} the other three flexible \texttt{TVP-RW} variants forecast particularly well, suggesting that flexible state innovation volatilities greatly increase precision for TVP coefficients. In particular for \texttt{FA-VAR}s these models show improved forecast accuracy after the Great Recession.  

\section{Closing remarks}\label{sec:conclusions}
It is empirically well documented that macroeconomic time series feature instabilities in the parameters and innovation volatilities. In the literature there is strong agreement that stochastic volatility is important, while, especially in larger-scale models, there is less consensus for time-varying coefficients. With increasing amount of information overall time-variation in parameters tends to reduce, but might be still present at few points in time for some parameters.
Detecting such occasional changes is challenging and requires highly flexible modeling techniques.   
To achieve such flexibility we introduce mixture priors on the time-varying part of the parameters.  
By additionally using hierarchical shrinkage priors on dynamic state variances, these methods are capable in imposing dynamic sparsity, as well as capturing a wide range of parameter changes. 
In a simulation study we show that our methods detect both sudden and gradual changes in parameters. 
In an empirical exercise we find that some coefficients tend to change abruptly in times of turmoil. Moreover, all proposed approaches forecast well. Even for large VARs flexible mixture priors improve forecast accuracy upon a wide range of benchmarks, suggesting that capturing these infrequent instabilities pays off.
\end{spacing}
\newpage
\small{\setstretch{0.95}
\addcontentsline{toc}{section}{References}
\bibliographystyle{custom.bst}
\bibliography{tvpvar.bib}

\clearpage

\normalsize\setstretch{1.5}
\begin{appendices}\crefalias{section}{appsec}
\setcounter{equation}{0}

\renewcommand\theequation{A.\arabic{equation}}
\section{Technical appendix}\label{app:tech}
\subsection{Stochastic volatility specification}\label{app:sv}
A stochastic volatility specifications assumes that $h_t = \log(\sigma^2_t)$ follows an AR($1$)-process:
\begin{equation}
h_t = \mu_h + \phi_h (h_{t-1}- \mu_h) + \vartheta_t, \quad \vartheta_t \sim \mathcal{N}(0, \psi_h). \label{eq:sv}
\end{equation}
Following \cite{kastner2014ancillarity}, we assume Gaussian priors on the initial state $h_0 \sim \mathcal{N}\left(\mu_h, \frac{\psi_h}{1-\phi_h^2}\right)$ and the unconditional mean $\mu_h \sim \mathcal{N}(0, 100)$, a Beta prior on the autoregressive parameter $\frac{\psi_h+1}{2}\sim \mathcal{B}(25, 1.5)$ and a Gamma prior on the state variance $\psi_h \sim \mathcal{G}(1/2, 1/2)$. This quite informative prior on $\psi_h$ pushes the specification towards a random walk. 

\subsection{The Normal-Gamma prior \citep{griffin2010inference}}\label{app:ng}
Similar to \cite{bitto_sfs2019JoE}, we introduce class-specific global shrinkage parameters, differentiating between the constant part of the coefficients (labeled $\lambda_a$) and regime-switching variances (labeled $\lambda_{\psi_0}$ and  $\lambda_{\psi_1}$, respectively). In the following, specify $\tau_j|\lambda_j \sim \mathcal{G}(\varrho_j, \varrho_j \lambda_j/2)$ and $\lambda_j \sim \mathcal{G}(\zeta, \zeta)$ with $\lambda_j = \lambda_k$ and $\varrho_j = \varrho_k$ if $j \in P_k$ for $k = \{a, \psi_0, \psi_1\}$. $P_k$ denotes a classifier (i.e. defines the set of coefficients belonging to the $k^{th}$ group). In the following, 
\begin{equation*}
\begin{aligned}
P_{a} =&  \left\{j: \hat{\alpha}_j \in \bm \alpha_0\right\}, \quad P_{\psi_0} =&  \left \{j: \hat{\alpha}_j \in \left  \{\sqrt{\bar{\psi}_{i0}}\right\}_{i = 1}^{K} \right\}, \quad \text{and} \quad P_{\psi_1} =& \left \{j: \hat{\alpha}_j \in \left \{\sqrt{\bar{\psi}_{i1}}\right\}_{i = 1}^{K} \right \}.
\end{aligned}
\end{equation*}
Moreover, we learn the hyperparameter $\varrho_k$ in a fully Bayesian fashion and specify $\zeta = 0.01$.

\subsection{Detailed MCMC algorithm}\label{app:post}
In this section, we provide details on each sampling step of the MCMC algorithm and on the full conditional posterior distributions. After defining appropriate starting values, we iterate through the following steps $20,000$ times and discard the first $10,000$ draws as burn-in:

\begin{enumerate}
\item The sampling steps (and conditional posteriors) for $\hat{\bm \alpha}_t$, $\lambda_k$, $\tau_j$, for $k = \{a, \psi_0, \psi_1\}$ and $j = 1, \dots, 3K$ and $\varrho_k$ are of standard form \citep{griffin2010inference}: 
\begin{enumerate}
\item Draw $\hat{\bm\alpha}$ from a multivariate Gaussian distribution:
\begin{equation*}
\hat{\bm \alpha}|\bm y, \hat{\bm X}, \bm \Sigma, \{\tau_j\}_{j = 1}^{3K} \sim \mathcal{N}\left(\hat{\bm \alpha}_1, \hat{\bm V}_1 \right) \label{eq: post_ahat}.
\end{equation*}
Here, $\hat{\bm X}$ is a $T \times 3K$-dimensional matrix with $\hat{\bm x}_t'$ on the $t^{th}$ position and:
\begin{align*}
\hat{\bm V}_{1}^{-1} =& \left((\bm \Sigma^{-1} \hat{\bm X})'(\bm \Sigma^{-1}\hat{\bm X}) + \text{diag}~(\tau_1^{-1}, \dots, \tau_{3K}^{-1})\right),\\
\hat{\bm \alpha}_1 =& \hat{\bm V}_{1} \left((\bm \Sigma^{-1} \hat{\bm X})'(\bm \Sigma^{-1} \bm y) \right).
\end{align*}

\item Sample the local shrinkage scalings $\{\tau_j\}_{j=1}^{3K}$ from a generalized inverse Gaussian (GIG) distribution \citep{griffin2010inference}:\footnotetext{The $\text{GIG}(a,b,c)$ is parameterized as $p(x)\propto x^{a -1} \exp\{-(bx + c/x)/2\}$.}
\begin{equation*}
\tau_j|\hat{\alpha}_j, \lambda_j, \varrho_j \sim \text{GIG}\left(\varrho_j - \frac{1}{2}, \varrho_j \lambda_j, \hat{\alpha}_j^2 \right),  \text{ for} \hspace{2pt} j = \{1, \dots, 3K\}. \label{eq: post_tau}
\end{equation*}
Here, $\lambda_j = \lambda_k$ and $\varrho_j = \varrho_k$ if $j \in P_k$ with $k = \{a, \psi_0, \psi_1\}$. 

\item Sample the associated global shrinkage parameter $\lambda_k$, for $k = \{a, \psi_0, \psi_1\}$, from a Gamma distribution distribution: 
\begin{equation*}
\lambda_k|\{\tau_j\}_{j \in P_k}, \varrho_k \sim \mathcal{G} \left(\zeta + \varrho_k p_k, \zeta + \frac{\varrho_k}{2} \sum_{j \in P_k}\tau_j \right)\label{eq: post_lambda}
\end{equation*}
with $p_k$ denoting the cardinality of the set $P_k$ (see Appendix \ref{app:ng}).

\item The hyperparameter $\varrho_k$, for $k = \{a, \psi_0, \psi_1\}$, are updated with a random walk Metropolis Hastings (MH) step. We refer to \cite{bitto_sfs2019JoE} for details. 

\end{enumerate}
\item Draw the normalized latent states $\tilde{\bm \alpha}$ from a $\nu$-dimensional Gaussian distribution by exploiting the static representation (see Section \ref{sec:postTVP}). 

\item Draw time-varying volatilities $\bm \Sigma$ using the R package \texttt{stochvol} \citep{kastner2016dealing}. 

\item Update binary indicators in $\bm S_t$, depending on its law of motion. We recast state equation back in the centered parameterization and evaluate the following regime-switching specification: 
\begin{equation}
\bm \alpha_t = 
\begin{cases}
\bm \alpha_0 + \bm \gamma_t +  \bar{\bm \phi}_0(\bm \alpha_{t-1} - \bm \alpha_0) + \bm \varsigma_t, \quad  \bm \varsigma_t \sim \mathcal{N}(\bm 0, \bar{\bm \Psi}_0) & \text{if} \quad  s_t = 0, \\
\bm \alpha_0 + \bm \gamma_t +  \bar{\bm \phi}_1(\bm \alpha_{t-1} - \bm \alpha_0) + \bm \varsigma_t, \quad  \bm \varsigma_t \sim \mathcal{N}(\bm 0, \bar{\bm \Psi}_1) & \text{if} \quad s_t = 1,
\end{cases}\label{eq:ms}
\end{equation}
with $\bar{\bm \phi}_0 = \bm 0_{K \times K}$, $\bar{\bm \phi}_1 = \bm I_K$ and $\bm \gamma_t = \bm 0_{K \times 1}$ for the \texttt{TVP-MIX} model. For the \texttt{TVP-POOL} model we set $\bar{\bm \phi}_0 = \bar{\bm \phi}_1 = \bm 0_{K \times K}$.  

\begin{itemize}
\item $s_t$ follows a first-order MS process (\texttt{MS}):
\begin{enumerate}
\item Conditional on the other parameters in \autoref{eq:ms}, we follow \cite{kim1999has} and sample $\{s_t\}_{t = 1}^{T}$ using standard algorithms.
\item Conditional on $\{s_t\}_{t = 1}^{T}$ we update transition probabilities by sampling 
$p_{00}  \sim \mathcal{B}(T_{00} + c_{00}, T_{01} + c_{10})$ and $p_{11} \sim \mathcal{B}(T_{11} + c_{01}, T_{10} + c_{11})$ both from a Beta distribution with $T_{kl}$, denoting the number of transitions from the $k^{th}$ to the $l^{th}$ regime. 
\end{enumerate}
\item Covariate-specific indicators with $\{s_{it}\}_{i = 1}^{K}$ independent over time (\texttt{MIX}): 
\begin{enumerate}
\item Conditional on the other parameters in \autoref{eq:ms} we evaluate both regimes in \autoref{eq:ms} and sample $s_{it}$ for each period and covariate independently from a Bernoulli distribution. 
\item  Conditional on $\{s_{it}\}_{t = 1}^{T}$ we are able to update the success probability for each covariate by sampling from a Beta distribution $p_i \sim \mathcal{B}(T_{i,0} +  c_{i,0}, T_{i,1} + c_{i,1})$, for $i = \{1, \dots, K\}$, with $T_{i,k}$ denoting the number of periods in the $k^{th}$ regime.
\end{enumerate}
\end{itemize}

\item For the specification with a hierarchical prior on $\tilde{\bm \gamma}_t$ and $\bm \phi_1 = \bm \phi_0 = \bm 0_{K \times K}$, we need five additional sampling steps (details can be found in \cite{malsiner2016model} and \cite{hhko2019}):
\begin{enumerate}  
\item Draw mixture weights $\bm \omega$ from a Dirichlet distribution:
\begin{equation*}
\bm \omega|\bm \theta, \xi \sim \mathcal{D}\text{ir}(\xi_1, \dots, \xi_N), \label{eq: post_weights}
\end{equation*}
with $\bm \theta = (\theta_1, \dots, \theta_T)'$ and $\xi_n = \xi + T_n$, where $T_{n}$ denotes the number of periods assigned to group $n$. 

\item Update the hyperparemter $\xi$ with random walk Metropolis-Hastings step \citep[for details, see][]{malsiner2016model}. 

\item Sample group indicators $\theta_t \in \{1, \dots, N\}$ for each $\tilde{\bm \alpha}_t$ from a Multinomial distribution: 
\begin{equation*}
\text{P}(\theta_t = n|\omega_n, \tilde{\bm \mu}_n) \propto \omega_n f_{\mathcal{N}}(\tilde{\bm \alpha}_{t}|\tilde{\bm \mu}_n, \bm I_K), \quad \text{for} \hspace{5pt} n = \{1, \dots, N\}.
\label{eq: post_theta}
\end{equation*}

\item The full conditional posterior of $\tilde{\bm \mu} = \text{vec}(\tilde{\bm \mu}_1, \dots, \tilde{\bm \mu}_N)$ follows a multivariate Gaussian distribution:
\begin{equation*}
\tilde{\bm \mu}|\bm \Lambda_0, \bm \theta \sim \mathcal{N}(\bm c_1, \bm \Lambda_{1}), \label{eq: post_mu}
\end{equation*}
with $\bm \theta = (\theta_1, \dots, \theta_T)'$ and:
\begin{equation*}
\begin{aligned}
\bm \Lambda_{1} &= \left(\bm I_K \otimes \bm \Xi' \bm \Xi + \bm I_N \otimes \bm \Lambda_0^{-1} \right)^{-1},\\
\bm c_1 &= \bm \Lambda_{1} \left(\text{vec}(\bm \Xi' \tilde{\bm A}) \right).
\end{aligned}
\end{equation*}
Here, $\bm \Xi$ denotes a $T \times N$ matrix with $(t,n)^{th}$ element given by $\mathcal{I}(\theta_t=n)$, where $\mathcal{I}(\bullet)$ refers to the indicator function and $\tilde{\bm A}$ collects $\tilde{\bm \alpha}$ in a $T \times K$ matrix. 

\item Sample shrinkage parameters $\{l_j\}_{j=1}^K$  from a GIG distribution: 
\begin{equation*}
l_j|\bm R, \{\tilde{\bm \mu}_n\}_{n = 1}^{N}\sim \text{GIG}\left( e_0 - \frac{N}{2}, 2 e_1, \frac{\sum_{n = 1}^{N} \tilde{\mu}_{jn}^2}{r_j} \right).\label{eq: post_ell}
\end{equation*} 
with $\tilde{\mu}_{jn}$, for $n = \{1, \dots, N\}$, denoting the $j^{th}$ element of $\tilde{\bm \mu}_n$.  
\end{enumerate}
\end{enumerate}

\subsection{The spectral decomposition}\label{app:spectral}
To obtain a time-varying low-frequency measure between two endogenous variable, we follow 
\cite{sargent2011two} and \cite{kliem2016low}. 
We therfore recast a TVP-VAR model in its companion form:
\begin{equation*}
\begin{aligned}
\bm Y_t =& \bm J \bm Z_t \\
\bm Z_t =& \bm F_t \bm Z_{t-1} + \bm E_t, \quad \bm E_t \sim \mathcal{N}(\bm 0, \bm \Upsilon_t)
\end{aligned}
\end{equation*}
In the following, the spectral density of $\bm Y_t$ at the very low frequency $\rho = 0$ is given by: 
\begin{equation*}
\bm \Pi_t(\rho = 0) = \bm J \left(\bm I_{mp+1} - \bm F_t)\bm \Upsilon_t (\bm I_{mp+1} - \bm F'_t)^{-1} \right) \bm J'. 
\end{equation*}
For $\rho = 0$ the low-frequency relationship $\pi_{ij,t}$ between two variables $(Y_{it},Y_{jt}) \in \bm Y_t$ can be derived with: 
\begin{equation*}
\pi_{ij,t} = \frac{\Pi_{ij,t}(\rho = 0)}{\Pi_{jj,t}(\rho = 0)}
\end{equation*} 
with $\Pi_{ij,t}$ denoting the $(i,j)^{th}$ element in $\bm \Pi_t$.

\newpage
\renewcommand\theequation{B.\arabic{equation}}
\section{Additional forecasting results}

\begin{figure}[hbt!]
\begin{minipage}{0.49 \textwidth}
\centering
i. \textit{One-year-ahead }
\hspace{5pt}
\end{minipage}
\begin{minipage}{0.49 \textwidth}
\centering
ii. \textit{Two-years-ahead} 
\hspace{5pt}
\end{minipage}\hfill
\begin{minipage}{\textwidth}
\centering
(a) \texttt{FA-VAR}
\hspace{5pt}
\end{minipage}
\begin{minipage}{0.49\textwidth}
\centering
\includegraphics[scale=.4]{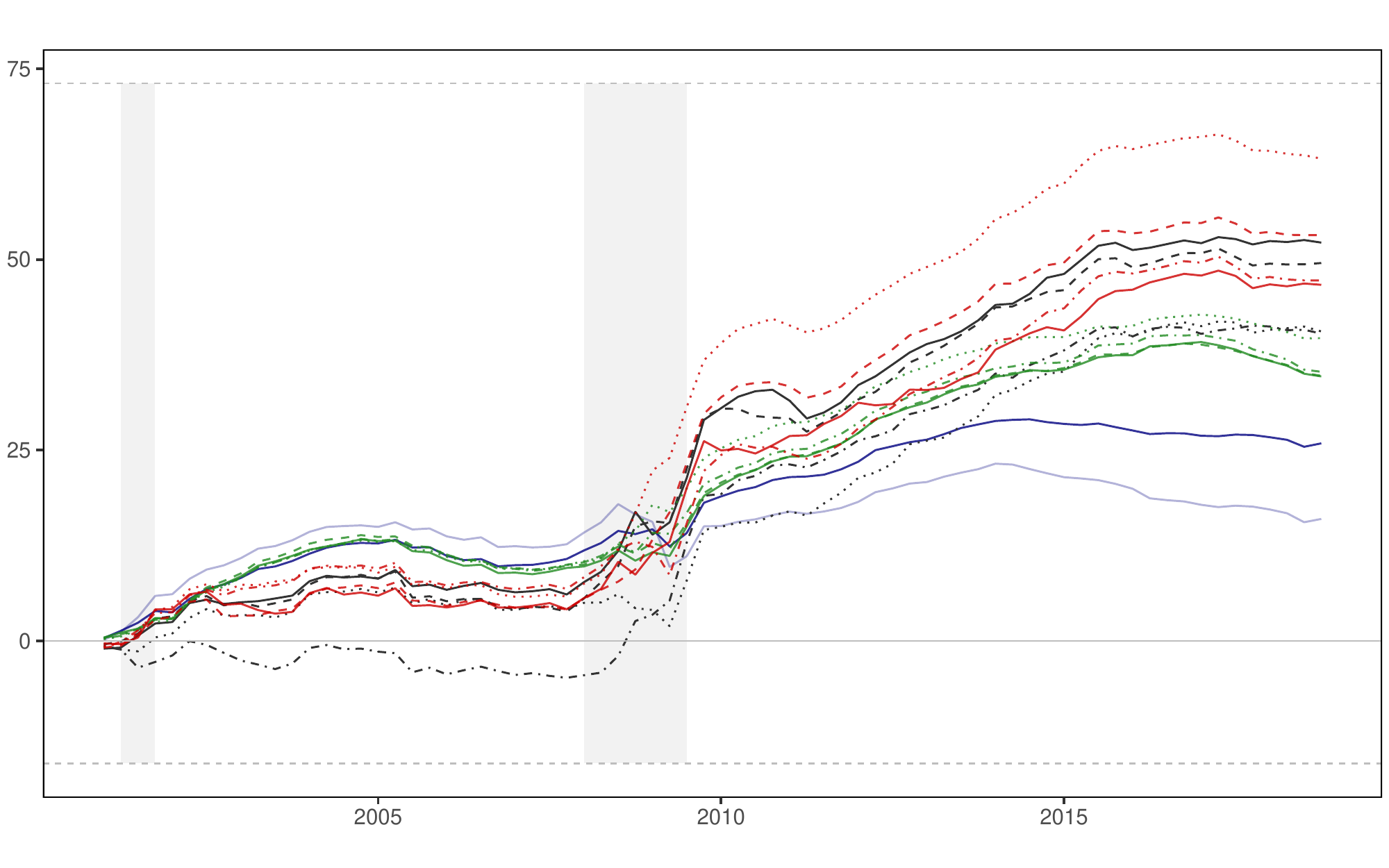}
\end{minipage}
\begin{minipage}{0.49\textwidth}
\centering
\includegraphics[scale=.4]{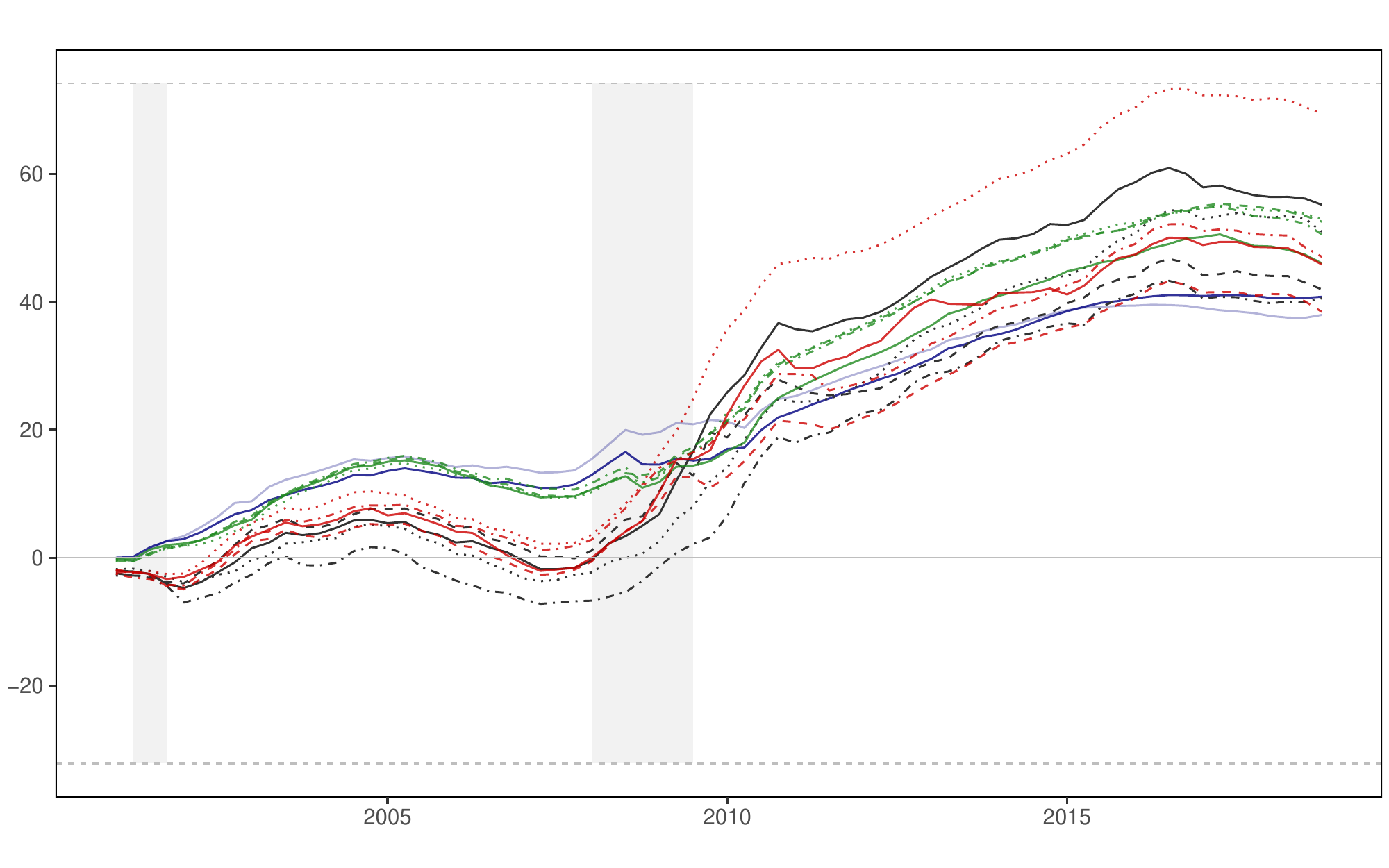}
\end{minipage}\hfill

\begin{minipage}{\textwidth}
\centering
(b) \texttt{L-VAR} 
\hspace{5pt}
\end{minipage}
\begin{minipage}{0.49\textwidth}
\centering
\includegraphics[scale=.4]{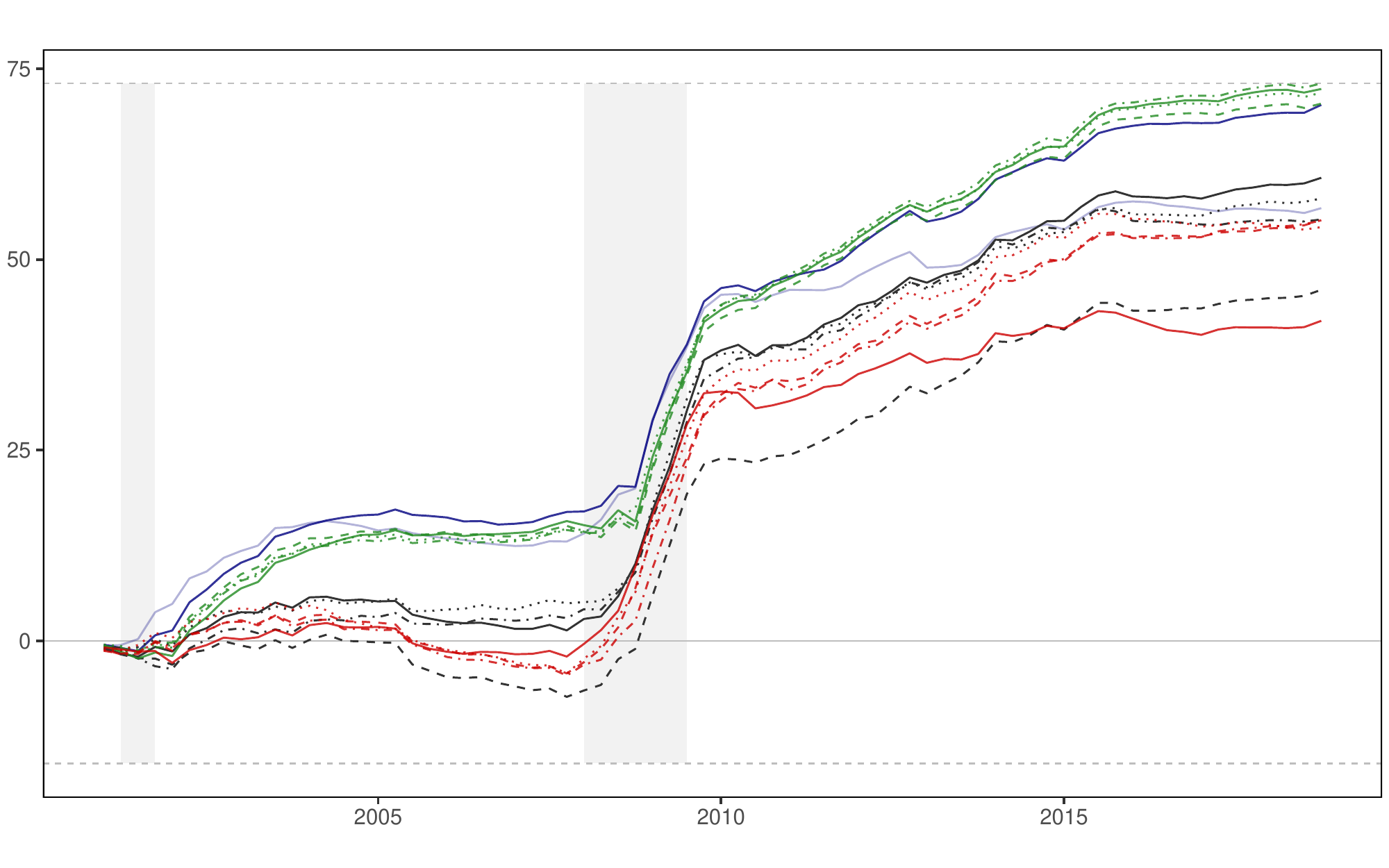}
\end{minipage}
\begin{minipage}{0.49\textwidth}
\centering
\includegraphics[scale=.4]{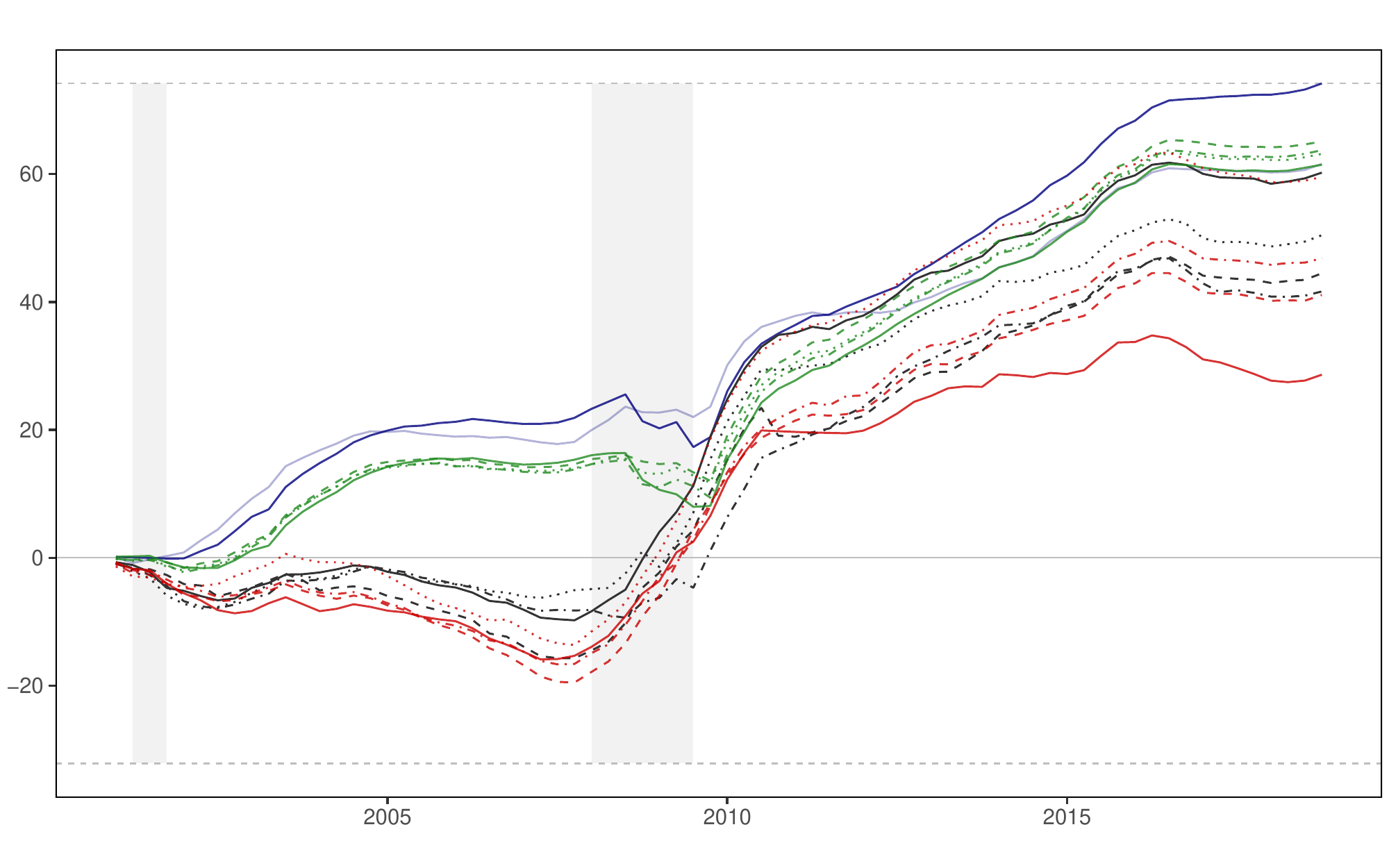}
\end{minipage}\hfill

\begin{minipage}{\textwidth}
\centering
(c) \texttt{S-VAR}
\hspace{5pt}
\end{minipage}
\begin{minipage}{0.49\textwidth}
\centering
\includegraphics[scale=.4]{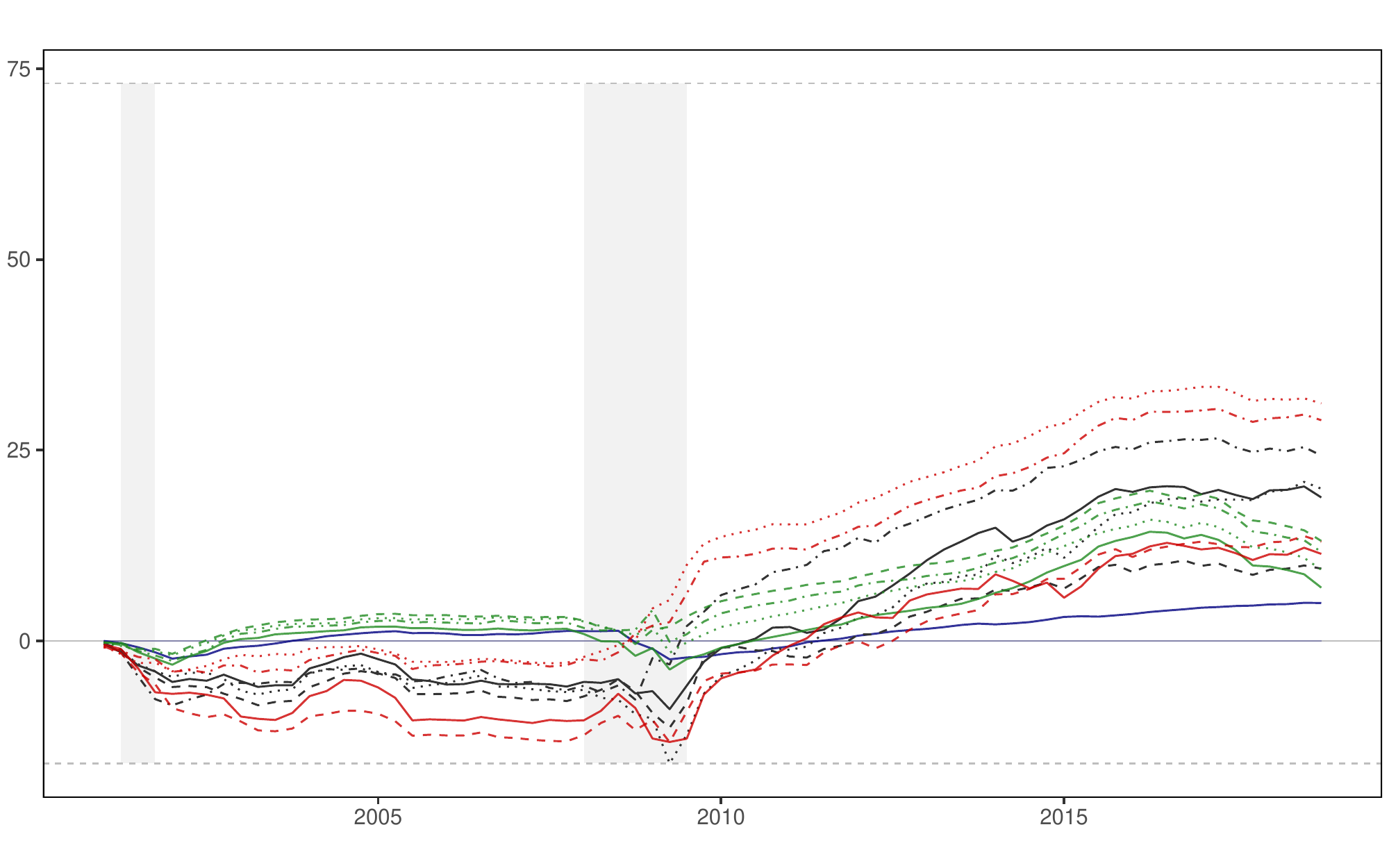}
\end{minipage}
\begin{minipage}{0.49\textwidth}
\centering
\includegraphics[scale=.4]{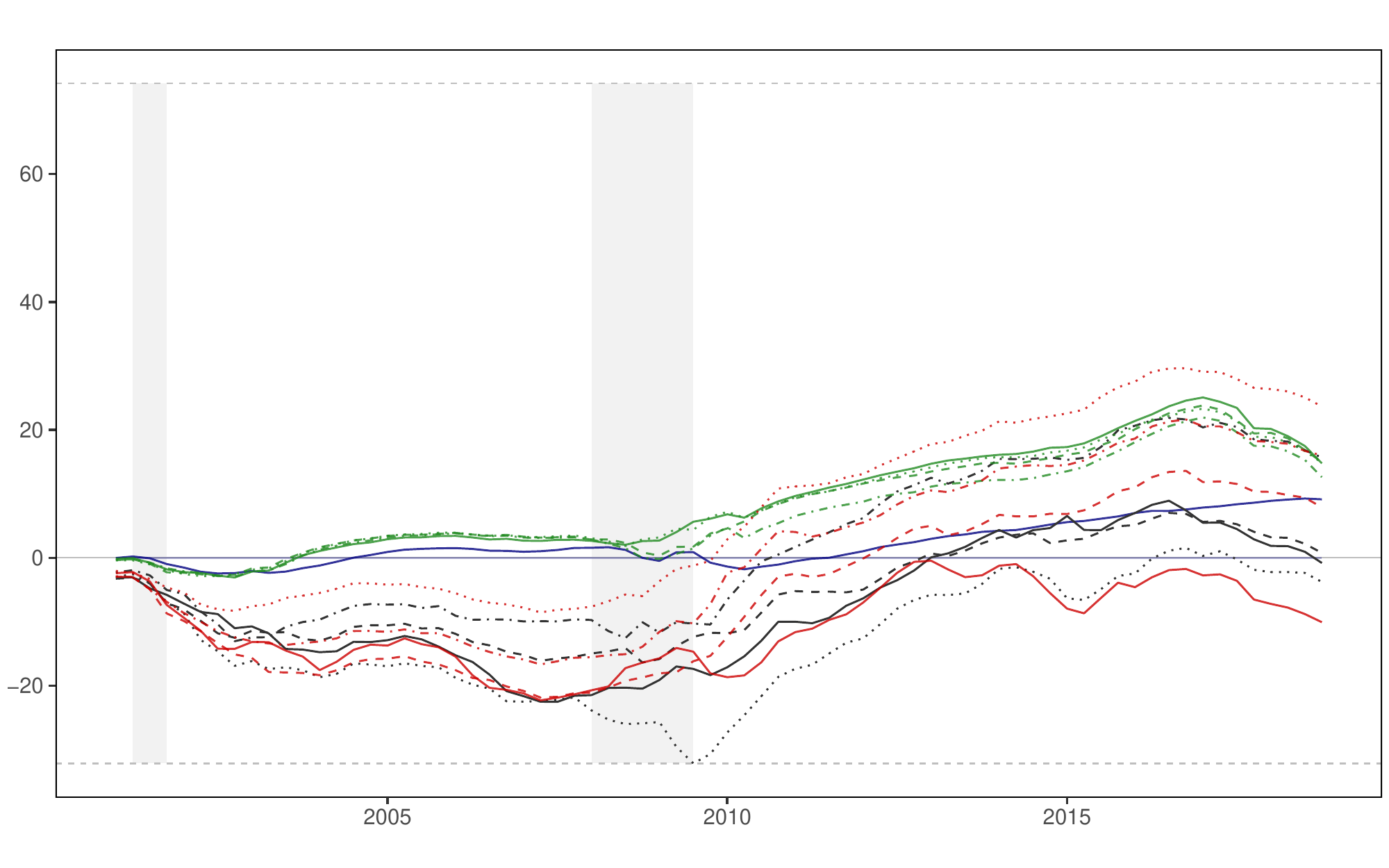}
\end{minipage}\hfill

\begin{minipage}{\textwidth}
\vspace{-30pt}
\centering
\includegraphics[scale=.4]{legend.pdf}
\end{minipage}
\caption{Evolution of one- and two-year-ahead total cumulative LPBFs relative to the benchmark. The gray dashed
lines refer to the maximum/minimum Bayes factor over the full hold-out sample. The light gray shaded areas indicate the NBER recessions in the US.}
\end{figure}

\begin{landscape}\begin{table}[!tbp]
{\tiny
\begin{center}
\scalebox{0.85}{
\begin{tabular}{lllclllllclllllclllll}
\toprule
\multicolumn{1}{l}{\bfseries }&\multicolumn{2}{c}{\bfseries Specification}&\multicolumn{1}{c}{\bfseries }&\multicolumn{5}{c}{\bfseries 1-quarter-ahead}&\multicolumn{1}{c}{\bfseries }&\multicolumn{5}{c}{\bfseries 1-year-ahead}&\multicolumn{1}{c}{\bfseries }&\multicolumn{5}{c}{\bfseries 2-years-ahead}\tabularnewline
\cline{2-3} \cline{5-9} \cline{11-15} \cline{17-21}
\multicolumn{1}{l}{}&\multicolumn{1}{c}{Class}&\multicolumn{1}{c}{Subclass}&\multicolumn{1}{c}{}&\multicolumn{1}{c}{TOT}&\multicolumn{1}{c}{GDPC1}&\multicolumn{1}{c}{CPIAUCSL}&\multicolumn{1}{c}{UNRATE}&\multicolumn{1}{c}{FEDFUNDS}&\multicolumn{1}{c}{}&\multicolumn{1}{c}{TOT}&\multicolumn{1}{c}{GDPC1}&\multicolumn{1}{c}{CPIAUCSL}&\multicolumn{1}{c}{UNRATE}&\multicolumn{1}{c}{FEDFUNDS}&\multicolumn{1}{c}{}&\multicolumn{1}{c}{TOT}&\multicolumn{1}{c}{GDPC1}&\multicolumn{1}{c}{CPIAUCSL}&\multicolumn{1}{c}{UNRATE}&\multicolumn{1}{c}{FEDFUNDS}\tabularnewline
\midrule
{\scshape }&&&&&&&&&&&&&&&&&&&&\tabularnewline
   ~~&   \textbf{FA-VAR}&   &   &   &   &   &   &   &   &   &   &   &   &   &   &   &   &   &   &   \tabularnewline
   ~~&   const. (Min.)&   &   &   0.92&   0.85&   0.97&   0.89&   1.10&   &   0.90**&   0.85**&   1.02&   0.81&   0.94&   &   0.88**&   0.98&   0.97&   0.87&   0.71***\tabularnewline
   ~~&   const. (NG)&   &   &   0.90**&   0.82**&   0.97&   0.89&   1.04&   &   0.89**&   0.84**&   1.00&   0.82&   0.89&   &   0.88**&   0.97&   0.97&   0.87&   0.72**\tabularnewline
   ~~&   \textbf{}&   &   &   &   &   &   &   &   &   &   &   &   &   &   &   &   &   &   &   \tabularnewline
   ~~&   TVP-MIX&   FLEX MIX&   &   0.89***&   0.82**&   0.96&   0.89&   0.85*&   &   0.91**&   0.89&   0.98&   0.86&   0.88&   &   0.97&   0.95&   1.03&   0.99&   0.89\tabularnewline
   ~~&   &   FLEX MS&   &   0.90**&   0.84*&   0.97&   0.85&   0.88&   &   0.91*&   0.89*&   1.02&   0.82&   0.85&   &   0.95&   1.00&   1.00&   0.94&   0.87\tabularnewline
   ~~&   &   SINGLE&   &   0.90**&   0.84*&   0.98&   0.88&   0.80**&   &   0.90**&   0.88**&   1.01&   0.83&   0.79**&   &   0.96&   0.98&   1.04&   0.96&   0.82\tabularnewline
   ~~&   &   SSVS MIX&   &   0.89**&   0.83*&   0.97&   0.88&   0.92&   &   0.91**&   0.88**&   1.01&   0.83&   0.88&   &   0.95&   0.97&   1.05&   0.90&   0.89\tabularnewline
   ~~&   \textbf{}&   &   &   &   &   &   &   &   &   &   &   &   &   &   &   &   &   &   &   \tabularnewline
   ~~&   TVP-POOL&   FLEX MIX&   &   0.88**&   0.80**&   0.97&   0.86&   0.95&   &   0.87**&   0.81**&   1.00&   0.79&   0.85&   &   0.86*&   0.94*&   0.97&   0.82&   0.70**\tabularnewline
   ~~&   &   FLEX MS&   &   0.88***&   0.80**&   0.96&   0.86&   0.95&   &   0.87**&   \textbf{0.81***}&   0.99&   0.79&   0.86&   &   0.85*&   0.94*&   0.95&   0.82&   0.71**\tabularnewline
   ~~&   &   SINGLE&   &   0.88***&   0.80**&   0.95*&   0.86&   0.95&   &   0.87**&   0.81***&   1.00&   0.79&   0.85&   &   \textbf{0.85*}&   0.93**&   0.96&   \textbf{0.81}&   0.70**\tabularnewline
   ~~&   &   SSVS MIX&   &   0.89**&   0.82**&   0.97&   0.86&   0.95&   &   0.87**&   0.81**&   0.99&   0.79&   0.87&   &   0.86*&   0.95*&   0.96&   0.82&   \textbf{0.70**}\tabularnewline
   ~~&   \textbf{}&   &   &   &   &   &   &   &   &   &   &   &   &   &   &   &   &   &   &   \tabularnewline
   ~~&   TVP-RW&   FLEX MIX&   &   0.89**&   0.85*&   0.95&   0.85&   0.82**&   &   0.87*&   0.87**&   0.99&   0.78&   \textbf{0.75**}&   &   0.92&   1.04&   0.97&   0.90&   0.74\tabularnewline
   ~~&   &   FLEX MS&   &   0.89**&   0.83**&   0.96&   0.88&   0.85&   &   0.90*&   0.88*&   1.01&   0.82&   0.81**&   &   0.95&   1.03&   0.99&   0.95&   0.81\tabularnewline
   ~~&   &   SINGLE&   &   0.91**&   0.84*&   1.00&   0.89&   0.87&   &   0.97&   0.94&   1.02&   0.95&   0.98&   &   1.06&   1.00&   1.09&   1.16*&   0.98\tabularnewline
   ~~&   &   SSVS MIX&   &   0.90**&   0.85*&   0.97&   0.88&   0.79**&   &   0.89*&   0.90&   0.98&   0.85&   0.75**&   &   0.95&   1.03&   0.98&   0.97&   0.78\tabularnewline
\midrule
{\scshape }&&&&&&&&&&&&&&&&&&&&\tabularnewline
   ~~&   \textbf{L-VAR}&   &   &   &   &   &   &   &   &   &   &   &   &   &   &   &   &   &   &   \tabularnewline
   ~~&   const. (Min.)&   &   &   0.94&   0.87&   1.03&   0.83&   1.10&   &   0.90**&   0.95*&   0.95***&   0.79&   0.87&   &   0.94**&   1.03&   0.95&   0.94&   0.83***\tabularnewline
   ~~&   const. (NG)&   &   &   0.89**&   0.82***&   0.97&   0.81&   0.99&   &   0.85***&   0.89**&   \textbf{0.90***}&   0.74&   0.84&   &   0.88**&   0.99&   \textbf{0.91}&   0.87&   0.76**\tabularnewline
   ~~&   \textbf{}&   &   &   &   &   &   &   &   &   &   &   &   &   &   &   &   &   &   &   \tabularnewline
   ~~&   TVP-MIX&   FLEX MIX&   &   0.97&   0.87&   1.08&   0.81&   1.17&   &   0.92*&   0.89&   1.04&   0.79&   0.95&   &   1.00&   1.06&   1.09&   0.91&   0.95\tabularnewline
   ~~&   &   FLEX MS&   &   0.97&   0.88&   1.07&   0.84&   1.10&   &   0.96&   0.95&   1.06&   0.81&   1.00&   &   1.44&   1.75&   1.79&   1.05&   1.24\tabularnewline
   ~~&   &   SINGLE&   &   0.92&   0.85**&   1.02&   0.83&   0.88&   &   0.90*&   0.88*&   1.00&   0.81&   0.86*&   &   0.96&   1.04&   1.02&   0.91&   0.88\tabularnewline
   ~~&   &   SSVS MIX&   &   0.91&   0.88&   0.97&   0.82&   0.92&   &   0.90*&   0.89&   1.01&   0.78&   0.88&   &   0.97&   1.03&   1.05&   0.90&   0.91\tabularnewline
   ~~&   \textbf{}&   &   &   &   &   &   &   &   &   &   &   &   &   &   &   &   &   &   &   \tabularnewline
   ~~&   TVP-POOL&   FLEX MIX&   &   0.87***&   \textbf{0.80***}&   0.97&   0.78*&   \textbf{0.79***}&   &   0.85**&   0.88*&   0.95&   0.73&   0.77**&   &   0.89&   1.00&   0.98&   0.83&   0.75**\tabularnewline
   ~~&   &   FLEX MS&   &   \textbf{0.86***}&   0.80***&   0.95&   0.77*&   0.83*&   &   0.85**&   0.87**&   0.94&   \textbf{0.72}&   0.78**&   &   0.88&   0.99&   0.98&   0.82&   0.74**\tabularnewline
   ~~&   &   SINGLE&   &   0.87***&   0.80***&   0.96&   0.77*&   0.85&   &   \textbf{0.84**}&   0.86**&   0.94&   0.72&   0.77***&   &   0.88*&   0.99&   0.97&   0.82&   0.75**\tabularnewline
   ~~&   &   SSVS MIX&   &   0.87***&   0.81***&   0.96&   \textbf{0.77*}&   0.85&   &   0.84**&   0.87**&   0.94&   0.72&   0.76**&   &   0.88&   1.00&   0.97&   0.82&   0.74**\tabularnewline
   ~~&   \textbf{}&   &   &   &   &   &   &   &   &   &   &   &   &   &   &   &   &   &   &   \tabularnewline
   ~~&   TVP-RW&   FLEX MIX&   &   0.93&   0.89&   0.98&   0.83&   0.93&   &   0.90*&   0.91&   0.97&   0.82&   0.85*&   &   0.97&   1.09&   1.02&   0.93&   0.86\tabularnewline
   ~~&   &   FLEX MS&   &   0.93&   0.89&   1.00&   0.83&   0.89&   &   0.92*&   0.90&   0.99&   0.84&   0.92&   &   0.99&   1.07&   1.02&   0.95&   0.94\tabularnewline
   ~~&   &   SINGLE&   &   1.03&   0.92&   1.17&   0.85&   1.14&   &   0.97&   0.93&   1.03&   0.85&   1.10&   &   1.05&   1.12&   1.04&   1.01&   1.04\tabularnewline
   ~~&   &   SSVS MIX&   &   0.92&   0.89&   0.97&   0.84&   0.95&   &   0.91*&   0.90&   0.97&   0.84&   0.92&   &   0.99&   1.04&   1.00&   0.96&   0.94\tabularnewline
\midrule
{\scshape }&&&&&&&&&&&&&&&&&&&&\tabularnewline
   ~~&   \textbf{S-VAR}&   &   &   &   &   &   &   &   &   &   &   &   &   &   &   &   &   &   &   \tabularnewline
\shadeBench   ~~&   const. (Min.)&   &   &   0.24&   0.44&   0.41&   0.07&   0.05&   &   0.38&   0.54&   0.44&   0.31&   0.22&   &   0.49&   0.48&   0.46&   0.58&   0.44\tabularnewline
   ~~&   const. (NG)&   &   &   0.98**&   0.98*&   0.99&   1.00&   1.00&   &   0.96**&   0.94**&   0.97**&   0.96**&   0.96&   &   0.97&   0.99&   0.96&   0.96**&   0.95\tabularnewline
   ~~&   \textbf{}&   &   &   &   &   &   &   &   &   &   &   &   &   &   &   &   &   &   &   \tabularnewline
   ~~&   TVP-MIX&   FLEX MIX&   &   0.92***&   0.92**&   \textbf{0.91*}&   0.96&   0.88***&   &   0.90&   0.84*&   0.94&   0.93&   0.90&   &   0.95&   0.93&   0.97&   0.98&   0.91\tabularnewline
   ~~&   &   FLEX MS&   &   0.95*&   0.96&   0.96&   0.94&   0.92*&   &   0.95&   0.92*&   0.98&   0.93&   0.98&   &   0.98&   1.01&   1.00&   0.95&   0.97\tabularnewline
   ~~&   &   SINGLE&   &   0.96*&   0.96&   0.96&   0.99&   0.85**&   &   0.93&   0.90*&   0.95&   0.97&   0.93&   &   0.98&   1.00&   0.99&   0.99&   0.92\tabularnewline
   ~~&   &   SSVS MIX&   &   0.94***&   0.95&   0.92**&   0.99&   0.90***&   &   0.91*&   0.87*&   0.95&   0.93&   0.90&   &   0.96&   0.98&   0.98&   0.96&   0.91\tabularnewline
   ~~&   \textbf{}&   &   &   &   &   &   &   &   &   &   &   &   &   &   &   &   &   &   &   \tabularnewline
   ~~&   TVP-POOL&   FLEX MIX&   &   0.97***&   0.97**&   0.98&   0.98&   0.95***&   &   0.95**&   0.94**&   0.98*&   0.94&   0.92*&   &   0.94&   0.99&   0.95&   0.92&   0.91\tabularnewline
   ~~&   &   FLEX MS&   &   0.98***&   0.97**&   0.99&   0.97*&   0.95**&   &   0.94**&   0.94**&   0.98**&   0.93&   0.92*&   &   0.94&   1.00&   0.96&   0.92&   0.90\tabularnewline
   ~~&   &   SINGLE&   &   0.97***&   0.97**&   0.98&   0.97&   0.95**&   &   0.95**&   0.93**&   0.99&   0.94&   0.92&   &   0.94&   0.99&   0.95&   0.92&   0.90\tabularnewline
   ~~&   &   SSVS MIX&   &   0.97***&   0.97*&   0.98&   0.97*&   0.95***&   &   0.95**&   0.93**&   0.98*&   0.94&   0.93&   &   0.94&   0.99&   0.95&   0.92&   0.91\tabularnewline
   ~~&   \textbf{}&   &   &   &   &   &   &   &   &   &   &   &   &   &   &   &   &   &   &   \tabularnewline
   ~~&   TVP-RW&   FLEX MIX&   &   0.96***&   0.96&   0.95&   1.00&   0.90***&   &   0.93*&   0.93&   0.92*&   0.97&   0.91&   &   0.96&   1.01&   0.93&   0.98&   0.92\tabularnewline
   ~~&   &   FLEX MS&   &   0.97&   0.98&   0.97&   0.99&   0.89**&   &   0.94&   0.91*&   0.95&   0.98&   0.96&   &   2.26&   3.86&   2.30&   1.62&   1.31\tabularnewline
   ~~&   &   SINGLE&   &   0.95*&   0.91*&   0.99&   0.97&   0.84***&   &   0.91&   0.82*&   0.96&   1.00&   0.88&   &   0.96&   \textbf{0.92}&   0.99&   1.07&   0.84\tabularnewline
   ~~&   &   SSVS MIX&   &   0.97&   0.99&   0.95&   1.01&   0.88**&   &   0.94&   0.91&   0.93&   1.01&   0.93&   &   0.98&   1.01&   0.95&   1.03&   0.93\tabularnewline
\bottomrule
\end{tabular}}
\caption{Density forecast performance (CRPS ratios) relative to the benchmark (\texttt{const (Min.)}). The red shaded row denotes the benchmark (and its CRPS values). Asterisks indicate statistical significance for each model relative to \texttt{const (Min.)} at the $1$ ($^{***}$), $5$ ($^{**}$) and $10$ ($^{*}$) percent significance levels. \label{tab:CRPS}}\end{center}}
\end{table}\end{landscape}

\newpage
\renewcommand\theequation{C.\arabic{equation}}
\section{Data}\label{app:data}
In this section we provide further details on the variable used for the large-scale VAR (\texttt{L-VAR}). \autoref{tab:data-descr} lists the exact description and provides further information on the transformation of the indicators. The gray shaded rows denote our target variables.
\begin{table}[!htb]
{\scriptsize
\begin{center}
\scalebox{0.95}{
\begin{tabular}{lllc}
\toprule
\multicolumn{1}{l}{\ }&\multicolumn{1}{l}{\ FRED.Mnemonic}&\multicolumn{1}{l}{\ Description}&\multicolumn{1}{l}{\ Transformation}\tabularnewline
\midrule
\shadeRow &GDPC1&Real Gross Domestic Product&$5$\tabularnewline
&PCECC96&Real Personal Consumption Expenditures&$5$\tabularnewline
&FPIx&Real private fixed investment &$5$\tabularnewline
&GCEC1&Real Government Consumption Expenditures and Gross Investment&$5$\tabularnewline
&INDPRO&IP:Total index Industrial Production Index (Index 2012=100)&$5$\tabularnewline
&CE16OV&Civilian Employment (Thousands of Persons)&$5$\tabularnewline
\shadeRow &UNRATE&Civilian Unemployment Rate (Percent)&$1$\tabularnewline
&CES0600000007&Average Weekly Hours of Production and Nonsupervisory Employees:  Goods-Producing&$1$\tabularnewline
&HOUST&Housing Starts: Total: New Privately Owned Housing Units Started&$5$\tabularnewline
&PERMIT&New Private Housing Units Authorized by Building Permits&$5$\tabularnewline
&PCECTPI&Personal Consumption Expenditures: Chain-type Price Index &$5$\tabularnewline
&GDPCTPI&Gross Domestic Product: Chain-type Price Index&$5$\tabularnewline
\shadeRow &CPIAUCSL&Consumer Price Index for All Urban Consumers:  All Items&$5$\tabularnewline
&CES0600000008&Average Hourly Earnings of Production and Nonsupervisory Employees&$5$\tabularnewline
\shadeRow &FEDFUNDS&Effective Federal Funds Rate (Percent)&$1$\tabularnewline
&GS1&1-Year Treasury Constant Maturity Rate (Percent)&$1$\tabularnewline
&GS10&10-Year Treasury Constant Maturity Rate (Percent)&$1$\tabularnewline
&TOTRESNS&Total Reserves of Depository Institutions &$5$\tabularnewline
&NONBORRES&Reserves Of Depository Institutions, Nonborrowed&$7$\tabularnewline
&S.P.500&S \& P's Common Stock Price Index:  Composite&$5$\tabularnewline
\bottomrule
\end{tabular}}
\caption{\small{Data for the US is obtained from the FRED data base of the Federal Reserve of St. Louis. The column \texttt{Transformation} shows the transformation applied to each variable. Following \cite{mccracken2016fred}, $(1)$ implies no transformation, $(5)$ denotes growth rates, defined as log first differences $ln\left(\frac{x_t}{x_{t-1}
}\right)$ and $(7)$ denotes differences in percentage changes with $\Delta \left(\frac{x_t - x_{t-1}}{x_{t-1}
}\right)$. All variables are standardized by subtracting the mean and dividing by the standard deviation. \label{tab:data-descr}}}\end{center}}
\end{table}

For the factor-augmented VAR (\texttt{FA-VAR}) we consider the full data set, compromising $165$ variables. 
For brevity we refer to \cite{mccracken2016fred} for a detailed description and transformation codes. All variables, serving as a basis for the principal components, are transformed to stationarity as suggested in \cite{mccracken2016fred}. 
Finally, we standardise the data by demeaning each variable and dividing through the standard deviation. 
Especially for principal components standardising is important due to the scale variance of the components.
\end{appendices}
\end{document}